\documentclass[ 12pt, article, leqno]{article}
\usepackage[font=small]{caption}
\usepackage[utf8]{inputenc}
\usepackage{amssymb}
\usepackage{natbib}
\usepackage[titletoc]{appendix}
\usepackage{threeparttable,dcolumn}
\usepackage{color,bm}
\usepackage{pdflscape}
\usepackage{booktabs}
\usepackage{graphicx,rotating}
\RequirePackage[colorlinks,citecolor=blue,linkcolor=blue,urlcolor=blue,pagebackref]{hyperref}
\usepackage[table,xcdraw]{xcolor}
\usepackage{amsthm,amsmath,amsfonts,amssymb, xr}
				\RequirePackage{booktabs, adjustbox, graphicx, multirow, xcolor, navigator}
\usepackage{graphics}
\usepackage{epsfig}
\usepackage{booktabs}
\usepackage{array,multirow}
\usepackage{verbatim}
\usepackage{bm}
\usepackage{lscape}
\usepackage{sanitize-umlaut}
\usepackage{diagbox}
\usepackage{latexsym}
\usepackage[bottom]{footmisc}
\usepackage{hyperref}
\hypersetup{
	colorlinks=true,
	linkcolor=blue,
	filecolor=blue,      
	urlcolor=blue,
	citecolor=blue,
}
\usepackage{pdfpages}
\usepackage{lscape}
\usepackage{adjustbox}
\usepackage{amsmath}
\usepackage{amssymb}
\usepackage{enumerate}
\usepackage{float}
\usepackage{setspace}
\usepackage{longtable}
\usepackage{graphics}
\usepackage{epsfig}
\usepackage{amsfonts}
\usepackage{graphicx}
\usepackage{pdflscape}
\usepackage{textcomp}
\usepackage{titling}
\usepackage[resetlabels, labeled]{multibib}%
\providecommand{\U}[1]{\protect\rule{.1in}{.1in}}
\providecommand{\U}[1]{\protect
	\rule{.1in}{.1in}}

\def \1{\hbox{\bf 1}}

\def \beqn{\begin{displaymath}}
	\def \eeqn{\end{displaymath}}

\newcommand{\mockalph}[1]{}

\renewcommand \theequation{\arabic{equation}}

\newtheorem{theorem}{\sc Theorem}
\newtheorem{lemma}{\sc Lemma}
\newtheorem{corollary}{\sc Corollary}
\newtheorem{proposition}[theorem]{\sc Proposition}
\newtheorem{assumption}{\sc Assumption}

\newtheorem{@example}{\sc Example}

\newtheorem{@remark}{\sc Remark}
\newenvironment{remark}{\begin{@remark}\rm}{\end{@remark}}
{\catcode `\@=11 \global \let \AddToReset=\@addtoreset}

\def \D01{\stackrel{\mbox{$\scriptstyle D[0,1]$}}{\longrightarrow}}
\def \L01{\stackrel{\mbox{$\scriptstyle L_2[0,1]$}}{\longrightarrow}}

\def \1{{\bf 1}}

\newcites{OA}{References}

\usepackage[mathscr]{eucal}

\makeatletter
\@addtoreset{theorem}{section}
\makeatother

\usepackage{geometry}
\allowdisplaybreaks
\usepackage{titlesec}
\titleformat{\section}{\large\bfseries}{\thesection}{0.5em}{}

\usepackage{setspace}
\AtBeginDocument{%
	\addtolength\abovedisplayskip{-0.2\baselineskip}%
	\addtolength\belowdisplayskip{-0.2\baselineskip}%
	\addtolength\abovedisplayshortskip{-0.5\baselineskip}%
	\addtolength\belowdisplayshortskip{-0.5\baselineskip}%
}

\usepackage{protecteddef}
\allowdisplaybreaks
\usepackage{authblk}

\usepackage{iftex}
\ifTUTeX
\usepackage{fontspec}
\else
\usepackage[T1]{fontenc}
\usepackage[utf8]{inputenc} 
\DeclareUnicodeCharacter{200B}{{\hskip 0pt}}
\fi

\newcommand\Jel{{\small\textbf{JEL classification: }}}

\newbox\keywbox
\setbox\keywbox=\hbox{\bfseries }%

\newcommand\keywords{%
	\noindent\rule{\wd\keywbox}{0.15pt}\\ \small{\textbf{Keywords: }}}
\makeatletter
\newcommand*{\centerfloat}{%
	\parindent \z@
	\leftskip \z@ \@plus 1fil \@minus \linewidth
	\rightskip\leftskip
	\parfillskip \z@skip}
\makeatother
\newcommand\acknowldgements{{\small\textbf{Acknowledgements: }}}

\newcommand{\email}[1]{\href{mailto:#1}{\nolinkurl{#1}} } 

\begin{document}

	\title{High Dimensional Generalised Penalised Least Squares}
		
		\author[1]{Ilias Chronopoulos
		\thanks{\email{ilias.chronopoulos@essex.ac.uk}}}
		\affil[1]{Essex Business School , University of Essex}
		\author[2]{Katerina Chrysikou 
		\thanks{\email{katerina.chrysikou@kcl.ac.uk}} } 
		\renewcommand\Authands{ and }
		\author[2]{George Kapetanios 
		\thanks{\email{george.kapetanios@kcl.ac.uk}.  Corresponding author}}
		\affil[2]{	King's Business School , 	King's College London}
		\date{{ \today}}
		\maketitle

		\begin{abstract}
			In this paper we develop inference in  high dimensional linear models with serially correlated errors. We examine the \emph{Lasso} estimator  under the assumption of $\alpha$-mixing in the covariates and error processes.  While the \emph{Lasso} estimator performs poorly under such circumstances,  we estimate via \emph{GLS} Lasso the parameters of interest and extend the asymptotic properties of the  \emph{Lasso} under more general conditions.  Our theoretical results indicate that the non-asymptotic bounds for stationary dependent processes are sharper, while the rate of  \emph{Lasso} under general conditions appears slower as $T,p\to \infty$.   Further,  we employ  debiasing methods to perform inference uniformly on the parameters of interest.  Monte Carlo results support the proposed estimator,  {as it has significant efficiency gains over traditional methods.}\\
			
			\noindent	\Jel{C01, C22, C55}
			
		\end{abstract}
	
		\newpage
		
		\section{Introduction}
		
		Research in high-dimensional statistics and econometrics  has witnessed a surge, because  
		the dimensionality of available datasets, models and associated parameter spaces has grown massively, in relation to the sample size.
		After the seminal work of \cite{tibshirani1996regression}, the \emph{Lasso} has become a focus of this continuously growing literature, since it conducts \textcolor{black}{simultaneously} model \textcolor{black}{estimation and} selection. 	More recent work  establishes the asymptotic behaviour of the \emph{Lasso} estimator as well as its model selection consistency, known as the oracle property, see,  for example,    \cite{meinshausen2009lasso}, \cite{raskutti2010restricted}, \cite{van2009conditions}, \cite{van2008high} and \cite{zhang2008sparsity}.

		Although the majority of the theoretical work is centred around the asymptotic behaviour of the \emph{Lasso} and relevant estimators in generalised linear models,  the main interest still lies in simple designs, such as \emph{i.i.d.} or fixed covariates,  while limited work has been done 
		towards the direction of regularised  linear models with time series components. Early work of  \cite{wang2007regression} suggests  a linear model with autoregressive error terms and their resulting \emph{Lasso} estimator satisfies a \cite{knight2000asymptotics}-type asymptotic property, {where} the number of covariates cannot be larger than the sample size.   \cite{hsu2008subset} consider  the \emph{Lasso} estimator under a Vector Autoregressive (VAR) process and derive their asymptotic results following the same setting as the former.   Various other papers derive asymptotic results  ensuring effective model selection for regularised estimators: \cite{nardi2011autoregressive} consider autoregressive structure on the covariates,  \cite{basu2015regularized} consider  stochastic regressions and transition matrix estimation in VAR models, and   \cite{kock2015oracle} consider models with stationary VAR covariates. These papers are indicative of the work in the high-dimensional literature under more general conditions,  but use restrictive assumptions in the error term in order to get  the oracle property.   
		
		There is significant research in high-dimensional econometrics which allows for more relaxed  assumptions:   \cite{kock_2016} shows that  the adaptive \emph{Lasso} is oracle efficient in stationary and non-stationary autoregressions,  \cite{masini2019regularized}  consider  linear time-series models with non-Gaussian errors, and \cite{10.1214/19-AOS1840} consider sparse Gaussian VAR models exploring the efficiency of $\beta$-mixing assumptions,  used to bound the prediction and estimation error of sub-Weibull covariates.  	Another class of papers focuses entirely on high-dimensional financial econometrics: \cite{babii2020machine} and \cite{babii2021machine} consider prediction and 
		now-casting with panel data  and   high-dimensional time series respectively,  sampled at different frequencies using the sparse-group \emph{Lasso}.
		
		Although, the \emph{Lasso} provides an efficient avenue to estimation and variable selection in high-dimensional datasets, one is unable to perform inference on the parameters estimated.   \cite{leeb2005model} have proven that ignoring the model selection step in the \emph{Lasso}, leads to invalid \textcolor{black}{uniform} inference.  Recent developments hinge on \emph{post-selection} inference, see, for example, \cite{berk2013valid} and  \cite{taylor2015statistical}, among others.  The incentive is that \emph{post-selection} methods provide valid  inference on the non-zero coefficients obtained after model selection, which is typically carried out at a first step, using the  \textit{Lasso}.  Although \emph{post-\textit{Lasso}} inference guarantees   valid confidence intervals, it is subject to the model selection made prior to that step, which can be {misleading} when the conditions for a "well-behaved" model are not met, e.g. \emph{i.i.d.} errors and/or covariates.    The latter facilitates the necessity for \textit{post-Lasso} inference to allow for more relaxed assumptions, as well as uniformity in the limit theory for general penalised models. A way to relax the \textcolor{black}{\textit{i.i.d.}/Gaussianity} assumption has been proposed by \cite{tian2017asymptotics} and \cite{tibshirani2018uniform}, who consider a bootstrap approach for asymptotically valid testing.   However,  \emph{i.i.d.} conditions on the covariates and errors need to be  
		{assumed} in order examine the large sample properties of the method.
		
		Alternative approaches that allow inference on the true parameters without the limitation of a "well-behaved" model, have  been developed.     These are based on \emph{debiased} or \emph{desparsified} versions of the \emph{Lasso}, see, for example, \cite{javanmard2014confidence}, \cite{van2014asymptotically},  \cite{zhang2014confidence} and on certain assumptions on the covariance matrix, such as fixed design, {\textit{i.i.d}}-ness or  sub-Gaussianity.  Further, extensions to a time series framework have been introduced in the literature, see, for example, \cite{wang2020lasso}  consider the covariates and error terms of a high-dimensional model to be temporally and cross-sectionally dependent and apply bootstrap methods towards estimation and inference of the true parameters.  \cite{babii2020inference} use \emph{debiased } sparse-group \emph{Lasso} for inference in a lower dimensional group of parameters.  \cite{kock2016oracle}  studies high-dimensional correlated random effects panel data models, where they  allow for correlation between time invariant covariates  and unobserved heterogeneity, as under fixed effects.  \textcolor{black}{Finally,  \cite{adamek2020lasso}	extend the \emph{desparsified Lasso}  to a time series setting, using near-epoch dependence assumptions, allowing for non-Gaussian, serially correlated and heteroscedastic processes.   Notice though that throughout the literature, no link has been made on the use of Generalised Least Squares (\emph{GLS}) type of inference  to account for non-spherical errors, which is the main focus of this paper. }
		
		Inference on \emph{Ridge} regression instead of the \emph{Lasso} has been considered by \cite{zhang2020ridge}, where they propose a wild bootstrap algorithm to construct confidence regions and perform hypothesis testing for a linear combination of parameters.  A similar approach has been considered  by \cite{zhang2021debiased}, allowing for inference under non-stationary and heteroscedastic errors.
		%

		In this paper, we contribute to the ongoing literature of high-dimensional inference under general  conditions.  We propose  a penalised  \emph{GLS}-type  estimator 
		which utilizes estimated autocovariances of the residual in a linear regression, where the  residual  is allowed to follow a general autoregressive process.   \emph{Lasso}  works as a preliminary estimator and is shown to be asymptotically consistent under mild  assumptions on the covariates  and error processes.
		We  perform uniform  inference via the \emph{debiased} \emph{GLS Lasso},  imposing mild restrictions on the autocorrelation of the error term.  In addition, we relax   assumptions on the error and most importantly the  covariates processes, commonly used in the existing literature,  for example fixed design, $i.i.d.$, sub-Gaussianity,  by allowing them to be  stationary  $\alpha$-mixing processes.

		The remainder of this paper is organised as follows.  Sections \ref{theoreticalconsiderations} -- \ref{debiased section} 
		present  the model, proposed methodology and theoretical results.   Section \ref{CV} describes the regularisation parameter tuning, while Section \ref{MCsection} contains the Simulation Study.    Section \ref{discussion} concludes.   Proofs and additional simulation results are relegated to the Supplementary Material.

		\section*{Setup and notation}\label{notation}
		For any vector $\boldsymbol{x}\in \mathbb{R}^{n}, $ we denote the $\ell_{p}$-, $\ell_{\infty}$- and $\ell_0$- norms, as  $\left\lVert \boldsymbol{x} \right\rVert_{p}=\left(\sum_{i=1}^{n}  | x_{i}  |^p\right)^{1/p}$, $\lVert   \boldsymbol{x} \rVert_{\infty} = \max_{i=1,\ldots,n}|x_{i}|$,  $\lVert  \boldsymbol{x}\rVert_{0}= \sum_{i=1}^{n}\boldsymbol{\mathrm{1}}_{(x_{i}\neq 0)} = \rm{supp}(\boldsymbol{\beta})$,  respectively, and $\rm{sign}$$(\boldsymbol{x})$ denotes the sign function applied element-wise on  $\boldsymbol{x}$.  We  use "$\to_{P}$" to denote convergence in probability.  For two deterministic sequences $a_n$ and $b_n$ we define asymptotic proportionality, "$\asymp$",  by writing  $a_n \asymp b_n$ if there exist constants $0 < a_1 \leq a_2$ such that $a_1b_n \leq a_n \leq a_2b_n$ for all $n \geq 1$.   For any set $A$, $ |A|$ denotes its cardinality,  while $A^c$ denotes its complement.   For a real number $a$, $\lfloor a  \rfloor$ denotes the largest integer no greater than $a$.

		\section{Theoretical considerations}\label{theoreticalconsiderations}
		We consider the following high-dimensional linear regression model, 
		\begin{equation}\label{model}
			y_{t}=\boldsymbol{x}_{t}'\boldsymbol{\beta}+u_{t},\quad \;t=1,\ldots, T,
		\end{equation}
		where  $\{\boldsymbol{x}_t\} = \left \{ \left(x_{t,1},\ldots,x_{t,p} \right)' \right \}$   is a $p$-dimensional vector-valued stationary time series,
		$ \boldsymbol{\beta}$ is a $p\times 1$  vector of  unknown parameters and  $u_{t}$ is a general second order stationary process. 
		We make the following assumption on the processes  $\left \{u_{t} \right \}$ and $\left \{ \boldsymbol{x}_t \right  \}$:
		\begin{assumption}\label{errors} 
			$\left \{ u_t \right \}$ is a stationary ergodic $\alpha$-mixing sequence, with mixing coefficients $\alpha_k\leq c\phi^{k}$, $k\geq 1$, for some $0<\phi<1$ and $c>0$. 
		\end{assumption}
	\noindent 	Under Assumption \ref{errors}, $\{u_t\}$ admits an $AR\left(q\right)$ representation, which is a more general setting compared to the standard one, e.g. $u_{t}\sim i.i.d. $  We propose to estimate the unknown parameters associated with \eqref{model} feasibly within the scope of a high-dimensional model.  More specifically 
		$	\left \{u_{t} \right \}$ is a second-order stationary process satisfying
		\begin{align}
			u_{t} = \sum_{j=1}^{q}\phi_{j}u_{t-j}+\varepsilon_t, \quad t = q+1, \ldots, T, \quad q<\infty\label{arerrors}.
		\end{align}	 %
		Notice that a stationary  $AR(p)$ process with \emph{i.i.d.} increments such as \eqref{arerrors}, in which the marginal distribution of $\{ \varepsilon_{t} \}$ has unbounded support satisfies the $\alpha$-mixing assumption.  To further specify the conditions under which $\{ u_{t} \}$ is strong mixing, one may refer to theorem 12.4 of \cite{davidson1994stochastic}, describing  random sequences, which by definition are strong mixing and further on Chapter 14 of \cite{davidson1994stochastic}. 
		\begin{assumption}\label{et}
			$\left \{\varepsilon_{t}  \right \}$ is an ergodic sequence of r.v.'s such that $E\left(\varepsilon_{t}| \mathcal{F}_{t-1} \right )= 0$ a.s., $E\left(\varepsilon_t^{2}|\mathcal{F}_{t-1}\right)= \sigma^{2}>0$ $\,$ a.s., $\sup_{t}E\left(|\varepsilon_{t}|^{4} \right)<\infty$, where $\mathcal{F}_{t-1}$ is the information set at time $t-1$. 
		\end{assumption}
	\begin{remark}
	In  Assumption \ref{et} it is explicitly stated that the innovation process $\{\varepsilon_{t}\}$ is a continuously distributed, random variable with unbounded support, which   is a sufficient and necessary condition  for $\{u_{t} \}$ to be $\alpha$-mixing.   The latter is necessary in order to avoid cases where the innovations $\{ \varepsilon_{t}\}$ are drawn from more  general distributions.  For example in the case where $q=1$ and $\{ \varepsilon_{t} \}$ is a Bernouli r.v., where   $0<\phi\leq 1/2$, then by definition $\{u_{t} \}^{\infty}_{-\infty}$ is not $\alpha$-mixing,  since the mixing coefficient $\alpha_{k} \nrightarrow 0$, see e.g. \cite{andrews1984non}.
	\end{remark} 
		To describe the method in detail, let
		\[\widetilde{u }_t = y_t - \boldsymbol{x}_t'\boldsymbol{\widetilde{\beta}} , \quad t=1,\ldots, T,\]
		where $\boldsymbol{ \widetilde{\beta}}$ is a preliminary \textit{Lasso} estimate of $\boldsymbol{\beta}$.   Further, let $\boldsymbol{\widehat{{\phi}}} = (\widehat{\phi}_1,\ldots, \widehat{\phi}_q) '$  be the $q^{\text{th}}$ order OLS estimator of the autoregressive parameters for $\{\widetilde{u}_{t} \}$ obtained as the the solution to the minimisation of \vspace{-2mm}
		\[
		T^{-1}\sum_{t=q+1}^T \left(\widetilde{u}_t - \phi_{1} \widetilde{u}_{t-1} - \cdots - \phi_{q} \widetilde{u}_{t-q}\right)^2,	\vspace{-1mm}
		\]
		over $\boldsymbol{\phi} = (\phi_1, \ldots, \phi_q)'$.   Then, an asymptotically valid feasible and penalised \textit{GLS} estimate of $\boldsymbol{\beta}$ can be obtained as  the solution to the following problem
		\begin{align}\label{penalised}
			\mathcal{L}\left({\boldsymbol{\beta}}\boldsymbol{;}\;\boldsymbol{\widehat{\phi}} \right) &= \underset{\boldsymbol{\beta}\in\mathbb{R}^{p}}{\arg\min} \frac{1}{2T}\left \lVert \widetilde{  \boldsymbol{y}}- \widetilde{  \boldsymbol{X}}\boldsymbol{\beta}	\right\rVert^{2}_{2} 
			+ \lambda\left\lVert\boldsymbol{\beta}\right\rVert_{1}, 
		\end{align}
		where $ \widetilde{\boldsymbol{X}}= \boldsymbol{\widehat{L} X}  $,  $ \widetilde{\boldsymbol{y}}= \boldsymbol{\widehat{L} y} $, $\boldsymbol{y}=\left( y_{1}, \ldots, y_{T} \right)'$ and  $\widehat{\boldsymbol{L}}$ is a $\left(T-q \right)\times T$ matrix defined as:
		\begin{align}\label          {LL}
			\widehat{\boldsymbol{L}}=\left(\begin{array}{ccccccccc}-\widehat{\phi}_{q} & -\widehat{\phi}_{q-1} &\cdots& \cdots & -\widehat{\phi}_{ 1} & 1 & \cdots & 0 & 0 \\ 
				0 & -\widehat{\phi}_{q} & -\widehat{\phi}_{q-1} & \cdots & \cdots & -\widehat{\phi}_{1} & \cdots & 0 & 0 \\
				\vdots & \vdots & \vdots & \vdots & \vdots & \vdots & \vdots & \vdots & \vdots \\ 
				0 & 0 & 0 & \cdots & -\widehat{\phi}_{q} & -\widehat{\phi}_{q-1} & \cdots & -\widehat{\phi}_{1} & 1\end{array}\right),  \quad \boldsymbol{\widehat{\phi}} =\left(\begin{array}{cccccccccccc}
				\widehat{\phi}_{1}  \\
				\vdots \\
				\widehat{\phi}_{q} \\ 
			\end{array}\right).
		\end{align}
		One can use the scalar representation of $\widetilde{\boldsymbol{y}}$ and vector representation of $\widetilde{\boldsymbol{X}}$, to obtain 
		\begin{equation}
			\widetilde{{y}}_t  =  y_{t}- \sum_{j=1}^{q}\widehat{{\phi}}_j {y}_{t-j} , \quad \widetilde{\boldsymbol{x}}_t  = \boldsymbol{x}_{t}-\sum_{j=1}^{q} \widehat{ {\phi}}_{j} {\boldsymbol{x}}_{t-j}, \quad t = q+1, \ldots, T.\nonumber
		\end{equation}
		The loss function in \eqref{penalised}  corresponds to the $\ell_{1}$-penalised loss  function using the estimates of  $\boldsymbol{\widehat{\phi}}$, where the additional penalty is added to the least squares objective.   Note that the asymptotic consistency of $\boldsymbol{\widehat{\beta}}$ can be established even prior to the estimation of $\boldsymbol{{\phi}}$.  We prove this result in  Lemma \ref{consistencyAR,cor}.    \textcolor{black}{Further, following the same argument  that we used to define $\widehat{\boldsymbol{ L }}$ in \eqref{LL}, we define ${\boldsymbol{L}}$, a $\left(T-q \right)\times T$ matrix, with  $\widehat{\phi}_{j}$ replaced by $\phi_{j}$, denoting the matrix of $\boldsymbol{\phi}$ used to derive the infeasible \textit{GLS-Lasso} estimates, when the degree of  autocorrelation in $u_{t}$ is known. } 
		\begin{remark}
			Notice that Yule–Walker or Burg-type estimates (or maximum entropy estimates), see,  for example, \cite{burg1968new}, \cite{brockwell2005modified},  can  be used instead of the OLS estimates to obtain $\boldsymbol{\widehat{ {\phi}}}$, used in the  construction of $\widehat{\boldsymbol{ L }}$, without changing the asymptotic properties of neither the preliminary estimate, $ \boldsymbol{\widetilde{\beta}}$ nor the GLS-type estimate, ${ \boldsymbol{\widehat{\beta}}}$.
		\end{remark}
		We highlight now the Assumptions on the covariates and errors, necessary to derive our theoretical results. 
		\begin{assumption} \label{DGK}  
			\begin{enumerate}
				\item $\{ \boldsymbol{x}_t\}$ is a $p$-dimensional stationary ergodic   $\alpha$-mixing sequence, with mixing  coefficients $\alpha^{(i)}_m\leq c\phi^{m}$, $m\geq 1$, for some $0<\phi<1$ and $c>0$, $\forall \; i= 1,\ldots,p$. \label{A2}
				\item 	$\{u_t\}  \;\text{and}\; \{\boldsymbol{x}_{t}\}$   are mutually independent.  
			\item 	$\{u_t\}  \;\text{and}\; \{\boldsymbol{x}_{t}\}$   have thin-tailed distributions, that \textcolor{black}{satisfy} for $u_{t}$ and uniformly on $\boldsymbol{x}_t$, using some $d, r>0$,  \label{A3}
				\begin{align}\label{thin}
					\sup_{i} \left[		\max_t E\left[\exp \left(d \left| u_t \right|^{r} \right) \right]\right]&<\infty  ,\\
					\sup_{i} \left[		\max_{t} E\left[\exp \left(d \left | x_{t,i} \right |^{r} \right) \right]\right]&<\infty\label{thin2}.
				\end{align}
			\end{enumerate}
		\end{assumption}
		\begin{remark}
			Assumption \ref{DGK}  controls the tail behaviour of the distribution of $\boldsymbol{x}_{t}$ and $u_t$ via \eqref{thin} and \eqref{thin2}, allowing an exponential decay of correlations that both $\{\boldsymbol{x}_{t}\}$ and $\{u_{t}\}$ can exhibit. {Furthermore, the (strong) mixing condition permits heterosedasticity, typically exhibited in empirical data (e.g. financial data), though the latter is not the focus of this paper.}   Notice that we do not impose the assumption of boundedness on the random variables (covariates), $\{\boldsymbol{x}_{t}\}$, which is typically assumed when implementing Bernstein type of  inequalities.  On the contrary, less restrictive assumptions are imposed on  $\{\boldsymbol{x}_{t}\}$ and $\{u_t\}$ respectively, compared to, for example,  \cite{10.1214/19-AOS1840}, who consider general forms of dynamic misspecification, resulting in serially correlated error terms, which are typically difficult to verify. 
		\end{remark} 	
		\begin{remark} \label{heavy_tails}
			In Assumption \ref{DGK} one can allow for heavy tails in the distribution of $\{ \boldsymbol{x}_{t} \}$ and  $\{u_t\}$  such that for some $\theta>2$, we can write a uniform   heavy-tailed distribution property:
\begin{align}
			\sup_i	\left[\max_{t} E\left| 	{u}_{t}	\right|^{\theta} \right]&<\infty,\label{fat1}\quad   \sup_{i}\left[ \max_{t} E\left| 	{x}_{t, i}	\right|^{\theta}\right]<\infty, \quad t=1,\ldots, T. 
			\end{align}
			The properties presented in \eqref{thin} can be generalised.   Then, we have that there exist $c_0, c_{1}>0$ such that for all $\xi>0$ and $r>0$
			\begin{align}\label{probabilityinequality}
				P(\sup_i|x_{t, i}|\geq \xi) =\left\{\begin{aligned}
					&c_{0}\exp(-c_1 \xi ^{r}), \quad \text{ if $x_{t, i}$ is thin-tailed }\\
					&  c_{0}\xi^{\theta}, \quad \quad \quad\quad \quad \text{ if $x_{t, i}$ is heavy-tailed.}
				\end{aligned}\right. 
			\end{align}
			One can, then, modify the probability inequalities of $\{ u_t\}$ from what appears in  \eqref{probabilityinequality}.  It is clear that using heavy-tailed distributions for   $\{ \boldsymbol{x}_t \}$, and $\{ u_{t} \}$  the probability inequalities in \eqref{probabilityinequality} become polynomial rather than exponential.   Allowing then for heavy tails in $\{ \boldsymbol{x}_t \}$, and $\{ u_{t} \}$, we implicitly allow for departures from exponential probability tails,   implying the  requirement that $p$ should be relatively small compared to $T.$
	\end{remark}
\vspace{-2mm}
Our objective is to incorporate a \emph{GLS} estimator to address the issue of serial-autocorrelation in the error term, $u_{t}$, that will enable  sharper inference following the paradigm of the  \emph{debiased}  \emph{Lasso}.   Such methods have been examined in the past in low dimensional cases, where $T>>p$, see,  for example,   \cite{amemiya1973generalized}, and \cite{kapetanios2016semiparametric}.    We extend this framework to allow for $p>>T$, ensuring that  $\widetilde{u}_t$ is asymptotically consistent  using  a penalised estimate of $\boldsymbol{\beta}$, 
while we limit our research to finite autoregressions.  \textcolor{black}{	Note that theoretical results can be established even when $\{u_{t}\}$ admits an $AR(\infty)$  representation, however certain additional assumptions must be made about $\{u_t\}$, and  the theoretical proofs would  come at an increased mathematical cost. }

The use of penalised models highlights the necessity of imposing sparsity conditions on the  parameter set, $\boldsymbol{\beta}$, which in turn allow  for a  degree of misspecification in the model.   The latter can be approximated by a sparse linear model following certain boundedness conditions on the smallest  eigenvalues of  the sample variance-covariance matrix, $\left(\boldsymbol{x}_t\boldsymbol{x}_t'\right)/T$.    

For some $\lambda\geq 0 $, we define the following  index set
\begin{align} 
S_{0} = \{i: \left|   \beta_i \right|>\lambda ; \; i=1,\ldots, p\},\label{active}
\end{align}
with cardinality $s_{0}=\left| S_{0} \right|$.  Under an appropriate choice of the regularisation   parameter, $\lambda$, $S_{0}$ contains all $ \beta_i$ that are "sufficiently large", while for $\lambda=0$,  $s_0$ is $\lVert  \boldsymbol{\beta}\rVert_{0}$.   We make use of this set in the following assumption, which forms the compatibility condition, as seen in \cite{bickel2009simultaneous}, \cite{raskutti2010restricted} and  Chapter 6 of \cite{buhlmann2011statistics}.   We make the following assumption under $|S_{0}|$.
\begin{assumption} \label{rac} 
For $\boldsymbol{\beta}=$
$\left(\beta_{1}, \ldots, \beta_{p}\right)',$ denote $\boldsymbol{\beta}_{s_{0}}:=\beta_{i}\; \mathbf{1}\left\{i \in S_{0};\; i=1, \ldots, p\right\}$,  $ \boldsymbol{\beta}_{s_{0}^{c}}:=\beta_{i} \;\boldsymbol{1}\left\{i \notin S_{0}, i=1, \ldots, p\right\}$, such that  $\boldsymbol{\beta}_{s_{0}}$ has zeroes outside the set $S_{0}$, 
and $\boldsymbol{\beta}=\boldsymbol{\beta}_{s_{0}}+\boldsymbol{\beta}_{s_{0}^{c}} $  such that $\lVert\boldsymbol{\beta}_{s_{0}^{c}}\rVert_{1} \leq 3\|\boldsymbol{\beta}_{s_{0}}\|_{1}$,
and  $\boldsymbol{{\Sigma}}= E({  \boldsymbol{\widetilde{x}}_t  \boldsymbol{\widetilde{x}}}_t')$.  
We define the following compatibility constant 
\begin{align}
	\zeta_{*}^{2}(s_{0}, \boldsymbol{\phi})=
	\underset{  \boldsymbol\beta\in\mathbb{R}^{p} \boldsymbol{\backslash} \{0\}}{\underset{\left\lVert\boldsymbol\beta_{s_{0}^{c}}\right\rVert_{1}
			\leq 3\left\lVert \boldsymbol{\beta}_{s_{0}} \right\rVert_{1}}{ \min}}  \;\; \frac{ |S_0| \boldsymbol{\beta}' {\boldsymbol{\Sigma}}\boldsymbol{\beta}}{\left \lVert\boldsymbol\beta_{s_0} \right \rVert_{2}^{2}} .
\end{align}
Consider $\zeta_{*}^{2}(s_{0}, \boldsymbol{\phi})>0$, then the following condition holds
\begin{align}
	\left \Vert \boldsymbol{\beta}_{s_{0}} \right \rVert_{1}^{2}\leq \left[\boldsymbol{\beta}{'} {\boldsymbol{\Sigma}}\boldsymbol{\beta}\right]\zeta_{*}^{2} \left (s_{0}, \boldsymbol{\phi} \right).
\end{align}
The constant $3$ might appear arbitrary and can be replaced with a number larger than  1, at the cost of changing the lower bound for $\lambda$. 
\end{assumption}
\begin{remark}
Assumption \ref{rac} implies the “restricted” positive definiteness of the variance-covariance matrix, which is valid  only for the vectors satisfying $\lVert\boldsymbol\beta_{s_{0}^{c}}\rVert_{1} \leq 3\|\boldsymbol{\beta}_{s_{0}}\|_{1}$.
Notice that in Assumption \ref{rac} we  present a modified restricted eigenvalue condition, based on the population variance-covariance matrix, ${\boldsymbol{\Sigma}}$.   In  Lemma  10.7 of the Supplementary Material, we   show that the population covariance $E\left(\widetilde{\boldsymbol{x}}_t\widetilde{\boldsymbol{x}}_t'\right)$ can be approximated well by the sample covariance estimate, $\widetilde{\boldsymbol{x}}_t\widetilde{\boldsymbol{x}}_t'/T$, such that
\begin{align}
	\operatorname{P}\left(\max _{0<i, k \leq p}\left|\left(\widetilde{\boldsymbol{X}}' \widetilde{\boldsymbol{X}}\right)_{i, k}-E\left(\widetilde{\boldsymbol{X}}'\widetilde{\boldsymbol{X}}\right)_{i, k}\right|>\nu \right)\to 0, \quad \text{for some \, } \nu>0. \label{vcv}
\end{align}
\end{remark}
\begin{assumption}\label{lasassumption_rate}
\noindent 
\begin{enumerate}
	\item $p=O(T^{\delta})$ for some $\delta >0$. \label{order of p}
	\item   $ \lambda \asymp \log^{1/2}{p}T^{-1/2}$,  as $(p, T) \rightarrow \infty$.
\end{enumerate}
Assumption \ref{lasassumption_rate} introduces the asymptotic rate of the regularisation parameter $\lambda$, based on which the following non-asymptotic  bounds are established. 
\end{assumption} 

\begin{lemma} \label{consistencyAR,cor}
For some regularisation parameter $\ddot{\lambda}\asymp T^{-1/2}\log^{1/2}{p}$, 	let 
\begin{align}\label{lasso_correlation}
	\mathcal{L}\left(\boldsymbol{\beta}\right) = \left \lVert \boldsymbol{y}-\boldsymbol{X}\boldsymbol{\beta} \right  \rVert^{2}_{2}+\ddot\lambda \left \lVert\boldsymbol{\beta} \right \rVert_{1}
\end{align}	
be the  \textit{Lasso}  regression prior to obtaining $\widetilde{u}_t$. Consider $\boldsymbol{\ddot{\Sigma}}= E\left(  \boldsymbol{x}_t  \boldsymbol{x}_t' \right)$ and the following restricted eigenvalue condition,
\begin{align}\label{rac1}
	s_{0} \left \lVert    \boldsymbol\beta_{s_{0}} \right  \rVert^{2}_{2}\leq   \boldsymbol{\beta}'\boldsymbol{\ddot{\Sigma}}  \boldsymbol{\beta} \phi_{0}^{-2},
\end{align}
for some $\phi_{0}^{-2}>0$ and $s_{0}= |S_0|$.  Under Assumptions \ref{errors} -- \ref{DGK}, 
we show that  \eqref{prederror} and \eqref{coeferror1} hold with probability at least $1- cp^{-\epsilon}$  for some positive constants, $c, \epsilon$, 
\begin{align}
	T^{{-1}}\left\lVert   \boldsymbol{X}\left(\boldsymbol{\widetilde{ \beta}-\beta}\right)  \right \rVert_{2}^{2}& \leq 4\ddot\lambda^{2} s_{0}\phi_{0}^{-2}\label{prederror}\\
	\left\lVert\boldsymbol{\widetilde{ \beta}-\beta}\right\rVert_1 &\leq 4\ddot\lambda s_{0}\phi_{0}^{-2},\label{coeferror1}
\end{align}
where $\boldsymbol{\widetilde{\beta}}$ is the  \textit{Lasso}  estimate obtained from the solution of \eqref{lasso_correlation}.    A detailed  proof of Lemma \ref{consistencyAR,cor} can be found in the Supplementary Material.
\end{lemma}

  Further, in Lemma \ref{consistencyAR,cor}, we show that the oracle inequalities hold prior to estimating $\boldsymbol{\widehat{\phi}}$. Notice that in \eqref{rac1} we use the population covariance matrix instead of the sample,  $T^{-1}{\boldsymbol{x}_t\boldsymbol{x}_t'}$,  for which an argument similar to \eqref{vcv} holds and is formalised in the  Supplementary Material, see Lemma 10.7.    Notice that \eqref{rac1} is the restricted eigenvalue condition on the population covariance, $E(\boldsymbol{x}_{t}\boldsymbol{x}_{t}')$, while $\phi_{0}^{-2}$ is the compatibility constant. 
Lemma \ref{consistencyAR,cor} is of paramount  importance and is required in order to establish the following corollary and Theorem \ref{theorem1}.  
\begin{corollary}\label{rho}
Let Assumptions \ref{errors} -- \ref{DGK}, and \ref{lasassumption_rate}   hold.  Then,    $\widehat{\phi}_{j}$ is an asymptotically consistent estimate of the autoregressive parameters, 
\begin{equation}
	\sum_{j=1}^{q}\left(\widehat\phi_j - \phi_j\right)=O_{P}\left(T^{-1/2}\right),\quad j=1,\ldots,q,\;\; q<\infty.
\end{equation} \label{consistency}
\end{corollary}
\vspace{-1cm}
\begin{remark}\label{testing_discussion}
In Corollary \ref{consistencyAR,cor}, the order, $q$,  though finite, is not known.  To address the latter we adopt a sequential testing technique towards the selection of $q$,   similar to \cite{kapetanios2016semiparametric}.   Allowing more flexibility to our model we also allow for this technique to be dependent on the generation of $\{u_t\}$. 
To minimize dependence on these conditions, we consider the following bound on $q$, $1\leq q< q^{*}$, where $q^{*}= \lfloor T^{1/2} \rfloor$, which implies that the case of $u_t\sim i.i.d.$ is not included in our testing procedure.
\end{remark}
\subsection{A feasible penalised \emph{GLS}}\label{feasibleGLS}
We start with the main theorem that provides non- asymptotic guarantees for the estimation and prediction errors of the  \emph{Lasso}  under a modified  compatibility condition,  i.e. Assumption \ref{rac}, which restricts the smallest eigenvalue of the variance-covariance matrix.  \textcolor{black}{The results of the  next theorem are based on the deviation inequality, which illustrates that as long as  the co-ordinates of $\boldsymbol{X'{  \boldsymbol{L}}'}\boldsymbol{Lu}/T$ are uniformly concentrated around zero  and the quantity $\lVert \boldsymbol{X'{  \boldsymbol{L}}'}\boldsymbol{Lu}/T \rVert_{\infty}$ is sharply bounded,  using $\boldsymbol{\phi}$ instead of $\boldsymbol{\widehat{\phi}}$,  \eqref{empproces} and \eqref{coef} hold w.p.a. one.}
\begin{theorem} 
\label{theorem1}
Let Assumptions \ref{errors} -- \ref{lasassumption_rate}, hold.  For  $p>>T$, we introduce the oracle inequalities corresponding to the solution of \eqref{penalised}, $\boldsymbol{\widehat{\beta}},$
\begin{align}
	T^{-1}\left\lVert\widetilde{  \boldsymbol{X}}\left(\boldsymbol{\widehat{ \beta}}-\boldsymbol{\beta}\right)   \right\rVert_{2}^{2}&\leq 4s_{0}\lambda^{2}\zeta_{*}^{-2}\left(s_{0}, \boldsymbol\phi\right)\label{empproces},\\
	\left\lVert \boldsymbol{\widehat{ \beta}-\beta}\right\rVert_{1}&\leq 4s_{0}\lambda\zeta_{*}^{-2}\left(s_{0},\boldsymbol \phi\right)\label{coef},
\end{align}
which hold with probability at least $1- c p^{1-\frac{c_{4}}{(12C_{0})^{2}}}\label{tailprobability tilde}$, for some $c, C_{0}>0$ and $c_{4}>0$, a large enough constant.    Therefore \eqref{empproces} -- \eqref{coef} hold a.s.,  for $p,T$ sufficiently large.
\end{theorem}
\begin{remark}	
Notice that the non-asymptotic bounds of the $\ell_2$-prediction and $\ell_{1}$-estimation errors of the parameter vector are similar to the bounds of a high-dimensional regression with $i.i.d.$ covariates.  This occurs  because $\zeta_{*}^{2}(s_{0}, \boldsymbol\phi)$ captures the autocorrelation structure of the errors by considering the transformed samples of $\widetilde{  \boldsymbol{x}}_t =    (\widetilde{  \boldsymbol{x}}_{1t}, \ldots, \widetilde{  \boldsymbol{x}}_{pt}    )'$ and bounding the smallest eigenvalue of ${\boldsymbol{\Sigma}}$, rather than $\boldsymbol{\ddot{\Sigma}}$, where ${\boldsymbol{\Sigma}} =  E(\boldsymbol{\widetilde{x}}_t\boldsymbol{\widetilde{x}}_t')$ and $\boldsymbol{\ddot{\Sigma}}=E(\boldsymbol{x}_t\boldsymbol{x}_t')$.   This ensures fast convergence rates of \emph{GLS Lasso}  under high-dimensional scaling.
\end{remark}
The following Corollary serves as a consequence of Theorem \ref{theorem1},  providing asymptotic rates for the bounded processes in \eqref{empproces} and \eqref{coef}.
\begin{corollary}\label{corolaryrate}
Let  Assumptions \ref{errors} -- \ref{lasassumption_rate} hold. Then
\begin{align}
	\left\lVert	\widetilde{\boldsymbol{X}} \left(\boldsymbol{\widehat{\beta}}-{\boldsymbol{\beta}} \right)\right\rVert_{2}^{2} = O_{P}\left(s_{0}\frac{\log p}{T}\right), \quad 
	\left	\lVert \widehat{{\boldsymbol{\beta}} } - {\boldsymbol{\beta}}\right\rVert_{1} = O_{P}\left(s_{0}\sqrt{\frac{\log{p}}{T}}\right)\label{coefe},
\end{align}
hold as $T, p\to \infty$.
\end{corollary}
{	\section{Point-wise valid  inference  based on the \emph{GLS Lasso}} \label{pointwiseSection}
A natural avenue to inference, having obtained the  estimates of $\boldsymbol{\beta}$ from \eqref{penalised}  is to re-estimate the parameters of the estimated active set, $\widehat{S}_{0}= \{ i : \; \widehat\beta_{i}\neq 0 \}$, via OLS.   	
Formally, let $\boldsymbol{\ddot{\beta}}_{S_0}$ be the vector whose $i^{th}$ element equals the least square re-estimate for all $i \in \widehat{S}_{0}$ and zero otherwise, while  $\boldsymbol{\widehat{\beta}}_{S_{0}}$  denotes the oracle assisted least squares estimates only including the relevant variables, those indexed by ${S}_{0}$. 
The following theorem shows that this indeed leads to point-wise valid confidence bands for the non-zero entries of $\boldsymbol{\beta}$, i.e. those indexed by $S_{0} $, since $\widehat{S} _{0}=S_{0}$ asymptotically.  
\begin{theorem}\label{pointwise}
	Let the  Assumptions of Theorem \ref{theorem1} hold, and assume that ${\boldsymbol{\widetilde{X}}'_{S_0}}\boldsymbol{\widetilde{X}}_{S_0} $ is invertible for $s_0<p$.  Then we have that 
	\begin{align} \label{pointwise th}\sqrt{T}	\left\Vert\left( \boldsymbol{\ddot{\beta}}_{S_0}-\boldsymbol{\beta}_{S_0}\right)- \left(\boldsymbol{\widehat{\beta}}_{S_{0}} -\boldsymbol{\beta}_{S_0}\right)\right\Vert_1=o_{p}\left(1\right). 
	\end{align}
\end{theorem}
\begin{remark}
	Theorem \ref{pointwise}  illustrates that performing least squares after model selection leads to inference that is asymptotically equivalent to inference based on least squares only including the covariates in $S_0$. 	However, it is important to note that such inference is of a point-wise nature and it is not uniformly valid. 
	This non-uniformity is visible  in the  confidence intervals,  that could occasionally undercover the true parameter.  This remark serves as a warning in applying  point-wise inference after $\ell_{1}$-regularisation, similar to the discussion in  \cite{leeb2005model}.
\end{remark}
\section{Uniformly valid inference based on the \emph{GLS Lasso}}\label{debiased section}
We define the following notation:
Let $$\boldsymbol{\Sigma}_{xu} = E[(\boldsymbol{{X}}'\boldsymbol{L}'\boldsymbol{Lu})(\boldsymbol{{X}}'\boldsymbol{L}'\boldsymbol{Lu})'], \quad 
\boldsymbol{\widehat{\Sigma}}_{xu} =(\boldsymbol{\widetilde{X}}'\boldsymbol{\widehat{L}}\boldsymbol{u})(\boldsymbol{\widetilde{X}}'\boldsymbol{\widehat{L}}\boldsymbol{u})'/T.$$

We extend the \emph{debiased Lasso} estimator,
introduced in  \cite{javanmard2014confidence}, and \cite{van2014asymptotically} \textcolor{black}{to accommodate non-spherical errors while relaxing the Assumptions made for $\boldsymbol{x}_t$, to perform valid inference in high-dimensional linear models  via a penalised \textit{GLS}-type estimator, introduced in \eqref{penalised}.}
The key idea of the  \emph{debiased Lasso}   is to invert the Karush-Kuhn-Tucker (KKT) conditions of the \emph{Lasso} solution  defined in  \eqref{lasso_correlation}.   To proceed under correlated errors we invert the KKT conditions of \eqref{penalised}. Then, 
\begin{align}
	-	\frac{1}{T}\widetilde{\boldsymbol{X}}'\left(\widetilde{\boldsymbol{y}} - \widetilde{\boldsymbol{X}}\boldsymbol{\widehat{\beta}}\right) + \lambda\widehat{\boldsymbol{\kappa}}=0 ,\quad \widehat{\beta}_{i}\in S_0,  \label{kkt1}
\end{align}
where $\widehat{\boldsymbol{\kappa}}$ is obtained  from the subgradient of the $\ell_{1}$-norm at $\boldsymbol{\widehat{\beta}}$, the last term of \eqref{penalised}, such that
\begin{align}
	\widehat{\boldsymbol{\kappa}} &= \textrm{sign}(\boldsymbol{\widehat{\beta}}),   \;\; \lVert \widehat{ \boldsymbol{\kappa}}\rVert_{\infty}\leq 1, \;\;\forall\;\; \widehat{\beta}_{i}\neq 0, \; \text{where}\;  \textrm{sign}(\widehat{{\beta}}_i)=\left\{\begin{aligned}
		1, &\quad \widehat{\beta}_{i}>0 \\
		[-1,1], & \quad\widehat{\beta}_{i}=0.\\
		-1, &\quad \widehat{\beta}_{i}<0
	\end{aligned}\right.
\end{align} 
The solution of the normal equations, can be further written as:
\begin{align}
	\widehat{ {\boldsymbol{\Sigma}}}\left(\boldsymbol{\widehat{\beta}} - \boldsymbol{\beta}\right) +\lambda\widehat{\boldsymbol{\kappa}} & =  \frac{\widetilde{\boldsymbol{X}}'\boldsymbol{\widehat{L}}\boldsymbol{u}}{T}. \label{normaleq}
\end{align}

If $p>T$, then $\widehat{\boldsymbol{\Sigma}}$ is not  invertible, so  we seek a different avenue on approximating its inverse.    Let $\widehat{\boldsymbol{\Theta}}$ be an approximate inverse of $\widehat{\boldsymbol{\Sigma}}$.  Then, \eqref{normaleq} can be rewritten as
\begin{align}
	\left(\boldsymbol{\widehat{\beta}} - \boldsymbol{\beta}\right)+ \widehat{\boldsymbol{{\Theta}}} \lambda\widehat{\boldsymbol{\kappa}}  + \frac{\boldsymbol{\delta}}{\sqrt{T}}&= \frac{\widehat{\boldsymbol{{\Theta}}}  \widetilde{\boldsymbol{X}}'\boldsymbol{\widehat{L}}\boldsymbol{u}}{T} , \quad 
	\boldsymbol{\delta}= \sqrt{T} \left(\widehat{  \boldsymbol{{\Theta}}}\widehat{\boldsymbol{{\Sigma}} } - \boldsymbol{I}_{(p\times p)}\right) \left(\boldsymbol{\widehat{\beta}} - \boldsymbol{\beta}\right) \label{lasso_b},
\end{align}
where $\boldsymbol{\delta}$ is  the error resulting from the approximation of $\widehat{\boldsymbol{{\Theta}}}$, which is shown to be asymptotically negligible, in  Theorem \ref{normalityDebiased}.
We use the fact that $	{T}^{-1}\widetilde{\boldsymbol{X}}'(\widetilde{\boldsymbol{y}} - \widetilde{\boldsymbol{X}}\boldsymbol{\widehat{\beta}}) = \lambda\widehat{\boldsymbol{\kappa}}$ and  define the \emph{debiased} \emph{GLS Lasso} estimator 
\begin{align}
	\widehat{ \boldsymbol{b}} = \boldsymbol{\widehat{\beta}} +\frac{1}{T}{ \widehat{  \boldsymbol{{\Theta}}} \widetilde{  \boldsymbol{X}}' \left(\widetilde{  \boldsymbol{y}} - \widetilde{  \boldsymbol{X}}\boldsymbol{\widehat{\beta}}\right)}. \label{debLaso}
\end{align}
In essence, \eqref{debLaso} is the same as in equation (5) of  \cite{van2014asymptotically}, but using 
$(\widetilde{  \boldsymbol{y}},\widetilde{  \boldsymbol{X}})$ instead of $(\boldsymbol{y}, \boldsymbol{X})$. 
\textcolor{black}{Notice that, with \eqref{lasso_b}, we obtain the asymptotic pivotal quantity, $	({\boldsymbol{\widehat{b  }}} - \boldsymbol{\beta})$, since  $\boldsymbol{\delta}=o_{P}(1)$, arriving at the following expression} 
\textcolor{black}{
	\begin{align}
		\sqrt{T}\left(\widehat{\boldsymbol{b}}-\boldsymbol{\beta}\right) = \frac{1}{\sqrt{T}} \widehat{\boldsymbol{\Theta}}\widetilde{\boldsymbol{X}}{'}{\boldsymbol{\widehat{L}}\boldsymbol{u} }+ {\boldsymbol{\delta}}\label{debiased}.
	\end{align} 
	The error $(\widehat{\boldsymbol{\Theta}}\widetilde{\boldsymbol{X}}{'}{\boldsymbol{\widehat{L}}\boldsymbol{u}})/\sqrt{T}$ }is asymptotically Gaussian and $\boldsymbol{\delta}$ is asymptotically negligible, as shown in Theorem \ref{normalityDebiased}. 
\subsection{Construction of $\widehat{\boldsymbol{\Theta}}$}\label{Thetasection}
The most common practice of constructing the approximate inverse of $\widehat{\boldsymbol{\Sigma}}$, 
is given by node-wise regressions on the design matrix  $\widetilde{\boldsymbol{X}}$.  The main idea is similar to \cite{van2014asymptotically}, but the asymptotic properties of the method change in the presence of serially correlated errors.    Let $\widetilde{\boldsymbol{X}}$ be a $T\times p$ design matrix, and $\widetilde{\boldsymbol{X}}_{-i}$ a design sub-matrix, missing the $i^{th}$ column, for $i=1,\ldots,p$.   

We first consider the  population nodewise regressions, obtained by the  linear projections
\begin{align}\label{nodewiseregression}
	{{x}}^{\star}_{t,i} &= {\boldsymbol{{x}}^{\star}_{t, -i}}'\boldsymbol{\gamma}_{i}+ \upsilon^{\star}_{t,i}, \\
{	\boldsymbol{{\gamma}}^{\star}_{i}}&=\underset{\boldsymbol{\gamma}\in \mathbb{R}^{p-1}}{\arg\min} \left \{{E}\left[(	{x}^{\star}_{t,i} -{	\boldsymbol{{x}}^{\star}_{t,-i}}' \boldsymbol{\gamma}_{i}	)^{2}\right]	\right	\},\label{gamma0}
\end{align}
where $\boldsymbol{x}_{t}^{\star} = \boldsymbol{x}_t-\sum_{j=1}^q \phi_j\boldsymbol{x}_{t-j}$, and $(	\tau^{\star}_{i})^{2}= E({{\upsilon_{i}^{\star}}})^{2} \label{ti2}$.  	Under Assumption \ref{DGK}	and by consequence of Theorem 14.1 of  \cite{davidson1994stochastic},   we make the following Assumption  for $ j=1,\ldots, p-1,\; i=1,\ldots,p$:
\begin{assumption}\label{Ass_mixingARinf}
		\begin{enumerate}
		\item 	
		$\{	\boldsymbol{x}_{t}^{\star} \}$ is a $p$-dimensional stationary ergodic   $\alpha$-mixing sequence satisfying \eqref{thin}, with mixing coefficients $\alpha^{(j)}_m\leq c\phi^{m}$, $m\geq 1$, for some $0<\phi<1$ and $c>0$, $j= 1,\ldots,p-1$.\label{AA1}
		\item $\{\upsilon^{\star}_{t,i} \}$  is an ergodic sequence of r.v.'s such that $E\left(\upsilon^{\star}_{t, i}| \mathcal{F}_{t-1} \right )= 0$ a.s., $E\left((\upsilon^{\star}_{t, i})^{2}|\mathcal{F}_{t-1}\right)= \sigma_i^{2}>0$ $\,$ a.s., $\sup_i\sup_{t}E\left(|\upsilon^{\star}_{t,i}|^{4} \right)<\infty$, where $\mathcal{F}_{t-1}$ is the information set at time $t-1$, $i=1,\ldots,p$.\label{AA2}
		\item $E[	\upsilon^{\star}_{t,i} ]=0, \; E[	\upsilon^{\star}_{t,i} 	\boldsymbol{x}_{k, t}^{\star}	]= 0$ for all $t, \; i\neq  k$. 
		 \item  $\{ \boldsymbol{x}_{t}^{\star}  \}$   and  $\{ {\upsilon}_{t,i}^{\star}  \}$  have thin-tailed distributions, that satisfy for $\upsilon_{t,i},\;{x}_{t, i}^{\star} ,\; 1\leq i\leq p, \; 1\leq t\leq T$, using some $m,\; s>0$,  \label{A34}
		\begin{align}
		\sup_i	\left[\max_t E\left[\exp \left(m \left | \upsilon^{\star}_{t,i} \right |^{s} \right) \right]\right]&<\infty, \; \text{and }
		\sup_i	\left[\max_{t} E\left[\exp \left(m \left | 	{x}_{t, i}^{\star}\right |^{s} \right) \right]\right]<\infty \label{tails}.
		\end{align}
	\end{enumerate}
\end{assumption}
\begin{remark}
	Assumption \ref{Ass_mixingARinf} complements Assumption \ref{DGK}, where the transformed processes,  	$\{ {	\boldsymbol{x}}^{\star}_{t} \}, \; \{ \upsilon^{\star}_{it} \}$, by consequence  of Theorem 14.1 of \cite{davidson1994stochastic} are  $\alpha$-mixing  series  with  properties noted in \eqref{tails}.    
The main argument in Assumption  \ref{Ass_mixingARinf} is that the errors of each of the node-wise regressions, defined in \eqref{nodewiseregression}, is a $p$-dimensional martingale difference sequence (m.d.s.). The proof of this statement is a direct application of Lemma 8.1 of the Supplementary Material. 
\end{remark}
A feasible way to estimate $\boldsymbol{\gamma}_{i} $  would be to use $\boldsymbol{\widetilde{	\boldsymbol{x}}}_t$ instead of $\boldsymbol{x}^{\star}_{t}$ in \eqref{gamma0}. Let  $\widehat{\boldsymbol{\gamma}}_{i}$ be the estimates of the node-wise regressions, obtained as the solution of the following problem:
\begin{equation}\label{node-wise}
	\widehat{\boldsymbol{\gamma}}_{i} = \underset{\boldsymbol{\boldsymbol{\gamma}_{i}}\in\mathbb{R}^{p-1}}{\arg\min}\frac{1}{T}\left\lVert \widetilde{\boldsymbol{x}}_{i} - \widetilde{\boldsymbol{X}}_{-i}\boldsymbol{\gamma}_{i}\right\rVert^{2}_{2} + 2\lambda_{i}\lVert\boldsymbol{\gamma}_{i}\rVert_{1}, 
\end{equation}
with components $\widehat{\boldsymbol{\gamma}}_{i}=\left\{ \widehat{{\gamma}}_{i,\kappa},\; \kappa\neq i,\; \kappa=1,\ldots,p  \right \}$.  Using the latter we define the following  $p\times p$ matrix:
\begin{equation}\label{C}
	\widehat{\boldsymbol{C}}:=\left[\begin{array}{cccc}
		1 & -\widehat{\gamma}_{1,2} & \cdots & -\widehat{\gamma}_{1, p} \\
		-\widehat{\gamma}_{2,1} & 1 & \cdots & -\widehat{\gamma}_{2, p} \\
		\vdots & \vdots & \ddots & \vdots \\ 
		-\widehat{\gamma}_{p, 1} & -\widehat{\gamma}_{p, 2} & \cdots & 1
	\end{array}\right],
\end{equation}
which is the solution to the series of the $p$ node-wise regressions, each  of them tuned with a different regularisation parameter $\lambda_{i}\asymp \sqrt{\log{p}/T}$.
We define  $\widehat{\boldsymbol{\Theta}}$ as:
\begin{align}
	\widehat{\boldsymbol{\Theta}} = 	\boldsymbol{\widehat{M}}^{-2}\widehat{\boldsymbol{C}}, \label{THETA}
\end{align}
where
\begin{align}
	\boldsymbol{\widehat{M}}^{2}& = \text{diag}(\widehat\tau_{1}^{2},\ldots, \widehat\tau_{p}^{2}), \quad 
	\widehat{{\tau}}_{i}^{2}={T}^{-1}\lVert \boldsymbol{\widetilde{x}}_{i} - \widetilde{\boldsymbol{X}}_{-i}{\widehat{\boldsymbol{\gamma}}}_{i}\rVert^{2}_{2} + \lambda_{i}\lVert{\widehat{\boldsymbol{\gamma}}}_{i}\rVert_{1}, \quad 	\forall\; i=1,\ldots,p.
\end{align}
It remains to show that $\widehat{\boldsymbol{\Theta}}$ is a suitable approximation of the inverse of $\boldsymbol{\widehat{\Sigma}}$, which we explore in Proposition 10.1 of the Supplementary Material.   Similarly to \cite{van2014asymptotically} and \cite{kock2016oracle}, we define the $i^{th}$ row of $\boldsymbol{\widehat{\Theta}}$ as $\boldsymbol{\widehat{\Theta}}_{i}$, a $1\times p$ vector and analogously $\widehat{\boldsymbol{C}}_i$.  Hence, $\boldsymbol{\widehat{\Theta}}_i= \widehat{\boldsymbol{C}}_i\;\widehat{\tau}_{i}^{-2}$.     The first order (KKT) conditions for the node-wise \emph{Lasso} regressions in \eqref{node-wise}, imply that 
\begin{align}
	\left\lVert\widehat{\boldsymbol{\Theta}}'_{i}\widehat{\boldsymbol{\Sigma}}-\boldsymbol{e}_{i}\right\rVert_{\infty}\leq \lambda_{i}\widehat\tau_{i}^{-2}, \quad  i=1,\ldots,p,\label{thetanorm}
\end{align}
where $\boldsymbol{e}_{i}$ is a $p\times 1$ unit vector.   Notice that the bound in \eqref{thetanorm} is controlled by $\widehat{\tau}_i^{2}$, which depends explicitly on the solutions of the node-wise lasso regressions, $\widehat{\boldsymbol{\gamma}}_i$.  In Lemma 10.2 of the Supplementary Material, we show  that $\widehat{\boldsymbol{\gamma}}_i$ is consistently estimated, and $\lVert\boldsymbol{\gamma}_i-\widehat{\boldsymbol{\gamma}}_i\rVert_{\ell}, \; \ell=\{1,2\}$, attains a non-asymptotic bound.
We further show that \eqref{thetanorm} holds a.s.,  in Proposition 10.1 of the Supplementary Material.
\subsubsection{Asymptotic properties of $\widehat{\boldsymbol{\Theta}}$}\label{AsymptTheta}
In order to show that $\sqrt{T}{\boldsymbol{\Theta}}\widetilde{\boldsymbol{X}}{'}{\boldsymbol{\widehat{L}u}} $ is asymptotically Gaussian, where  $\widetilde{\boldsymbol{X}}$ is defined prior to  \eqref{LL}, one needs to explore the limiting behaviour of  the approximate inverse $\widehat{ {\boldsymbol{\Theta}}}$.   Consider ${\boldsymbol{\Sigma}} = E( \boldsymbol{\widetilde{x}}_t \boldsymbol{\widetilde{x}}_t')$ and  $\boldsymbol{\Theta} $ its approximate inverse. Following a similar partition of $\boldsymbol{\Sigma}^{-1}$ as \cite{yuan2010high}, we show that $\widehat{\boldsymbol{\Theta}}$ is asymptotically equivalent to ${\boldsymbol{\Theta}}$ under certain sparsity assumptions (formalised  in Lemma 10.4 of the Supplementary Material).   We write 
\begin{align}
	\Theta_{i,i}             & = \left[		{\Sigma}_{i,i} - \boldsymbol{\Sigma}_{i,-i}\boldsymbol{\Sigma}_{-i,-i}^{-1}\boldsymbol{\Sigma}_{-i,i}	\right]^{-1}, 
	\quad 
	\boldsymbol{\Theta}_{i,-i} = -\Theta_{i,i}\;\boldsymbol{\Sigma}_{i,-i}\;\boldsymbol{\Sigma}_{-i,-i}^{-1}\label{thetai-i},
\end{align} 
where $\Theta_{i,i} $ is the $i^{th}$ element of the diagonal of $\boldsymbol{\Theta}$, $\Sigma_{i,i}$  is the $i^{th}$ element of the diagonal of $\boldsymbol{\Sigma}$, $\boldsymbol{\Sigma}_{i,-i}$ is the $1\times (p-1)$ vector of the design, obtained by removing the $i^{th }$ column of the $i^{th}$ row, $\boldsymbol{\Sigma}_{-i,-i}$ is the sub-matrix of $\boldsymbol{\Sigma}$ resulting from the removal of the $i^{th}$ row and column, $\boldsymbol{\Sigma}_{-i,i}	$ is the $(p-1)\times 1$ vector of the design, obtained by removing the $i^{th}$ row of the same column and $\boldsymbol{\Theta}_{i,-i}$ is the $1\times (p-1)$ vector of $\boldsymbol{\Theta}$,  obtained by removing the $i^{th }$ column of the $i^{th}$ row.   \\ \indent 
Let \eqref{gamma0} be the  $p$ node-wise regressions.  By linear projections, let its first order conditions be 
\begin{align}\label{kktgamma}
	\boldsymbol{\widetilde{\gamma}}_i =E\left[ \left(	\widetilde{\boldsymbol{X}}_{-i}{'}\widetilde{\boldsymbol{X}}_{-i}	\right)^{-1}\left(\widetilde{\boldsymbol{X}}_{-i}{'}\widetilde{	\boldsymbol{x}}_{i}\right)\right] =\boldsymbol{\Sigma}_{-i,-i}^{-1}\;\boldsymbol{\Sigma}_{-i,i}, \quad   i=1,\ldots,p.
\end{align}
Replacing $\widetilde{\boldsymbol{X}}_{-i,-i}^{-1}\;\widetilde{\boldsymbol{x}}_{-i,i}$ of  \eqref{kktgamma} into \eqref{thetai-i}, we have that
\begin{align}\boldsymbol{\Theta}_{i,-i} = -\Theta_{i,i} {\boldsymbol{\widetilde{\gamma}}_{i}},
\end{align}
which illustrates that the $i^{{th}}$ row of $\boldsymbol{\Theta}$ is sparse, if and only if, \textcolor{black}{${\boldsymbol{\widetilde{\gamma}}_{i}}$} is sparse.  Furthermore, let the error, $\boldsymbol{\psi}_{i}= \widetilde{\boldsymbol{x}}_{i}-\widetilde{\boldsymbol{X}}_{-i}\textcolor{black}{\boldsymbol{\widetilde{\gamma}}_{i}}$, then 
\begin{align}	
	\boldsymbol{\widetilde{x}}_{i} = 	\widetilde{\boldsymbol{X}}_{-i}\textcolor{black}{\boldsymbol{\widetilde{\gamma}}_{i}}+\boldsymbol{\psi}_{i}, \quad  i=1,\ldots, p.
\end{align}
Then, using Assumption \ref{Ass_mixingARinf}, and the  definition of $\textcolor{black}{\boldsymbol{\widetilde{\gamma}}_{i}}$	as a penalised  quadratic objective function we have that
\begin{align}
	\frac{1}{T}E\left( \boldsymbol{\widetilde{X}}_{-i}'\boldsymbol{\psi_{i}}\right)\leq c,
\end{align}
for some small positive constant $c$. 
In light of Theorem \ref{theorem1}, it is sensible to note that $\widehat{\boldsymbol{\gamma}}_{i}$, defined in \eqref{node-wise} is close to its population counterpart, $\textcolor{black}{\boldsymbol{\widetilde{\gamma}}_{i}}$.
We then can  write 
\begin{align}
	\tau_{i}^{2} &= E\left[\left(	\widetilde{\boldsymbol{x}}_{i} - \widetilde{\boldsymbol{X}}_{-i}\textcolor{black}{\boldsymbol{\widetilde{\gamma}}_{i}}	\right)^{2}\right]
	= \Sigma_{i,i} -\boldsymbol{\Sigma}_{i,-i}\boldsymbol{\Sigma}_{-i,-i}^{-1}\boldsymbol{\Sigma}_{-i,i}={\Theta_{i,i}}^{-1}.
\end{align}
Notice that $\Theta_{i,-i}=-\textcolor{black}{\boldsymbol{\widetilde{\gamma}}_{i}}\tau_{i}^{-2}$.  Therefore, replacing the terms in \eqref{THETA}, with their population counterparts, we can write $\boldsymbol{{M}}^{2} = \text{diag}(\tau_{1}^{2},\ldots, \tau_{p}^{2})$, where matrices $\boldsymbol{T}, \;\boldsymbol{C}$ are defined similarly to $\widehat{\boldsymbol{T}},\; \widehat{\boldsymbol{C}}$ replacing $\widehat{\boldsymbol{\gamma}}_{i}$ with $\textcolor{black}{\boldsymbol{\widetilde{\gamma}}_{i}}$, which in turn is asymptotically close to the true/population value of $\boldsymbol\gamma_i$ as it is shown in Lemma 10.4 of the Supplementary Material.    Further, we  illustrate in Lemma  10.4 of the Supplementary Material that  $(\widehat{\tau}^{2}_i- \tau_{i}^2)\to_{P} 0$, such that,   $\widehat{\boldsymbol{\Theta}}\to_{P} \boldsymbol{\Theta}$ a.s. for $p\to \infty$.	

\subsubsection{Asymptotic normality}\label{DebiasedSection}
We explore the asymptotic normality of the debiased  \emph{GLS Lasso}, and establish uniformly valid confidence intervals based on the theoretical results considered throughout Section \ref{theoreticalconsiderations}.
\begin{theorem}\label{normalityDebiased}  		
	Let Assumptions \ref{errors} -- \ref{Ass_mixingARinf}, sparsity assumptions on $s_{0}=o(\sqrt{T}/
	\log{p})$  hold.  Further,  consider the linear model in \eqref{model}, the  \textit{GLS Lasso} estimator in \eqref{penalised}, let  $\widehat{\boldsymbol{\Theta}}$  be a  suitable approximation of  $\boldsymbol{\widehat{\Sigma}}^{-1}$, for a proper selection of $\lambda_{i} \asymp \sqrt{T^{-1}\log{p}}\,$  for the \textit{Lasso} for node-wise regressions in \eqref{node-wise},  
	where ${\boldsymbol{\widehat{\Sigma}}} = (\widetilde{  \boldsymbol{x}}_t\widetilde{  \boldsymbol{x}}_{t}')/T$.  Then 
	\begin{align}
		\sqrt{T}(\widehat{ \boldsymbol{b}} - \boldsymbol{\beta}) & =  \boldsymbol{z} + {\boldsymbol{\delta}}\label{rootTbeta}, 
		\quad 		\boldsymbol{z}  = \frac{1}{\sqrt{T}}{ {\boldsymbol{\widehat{\Theta}}}}\widetilde{\boldsymbol{X}}{'}{\boldsymbol{\widehat{L}}\boldsymbol{u}}\sim \mathcal{N}\left(0,{\sigma}_{u}^2{\boldsymbol{\Theta}}\boldsymbol{{\Sigma}}_{xu}'{\boldsymbol{\Theta}}\right),\quad
		{\boldsymbol{\delta}} = o_{P}(1), 
	\end{align}\label{normality}
	where $\widehat{{\boldsymbol\Theta}} $ is defined in \eqref{THETA},  ${\boldsymbol{\delta}}:={\sqrt{T}}(\boldsymbol{ \widehat{\Theta}} {\widehat{\boldsymbol\Sigma}}-\boldsymbol{I}_{(p\times p)})(\boldsymbol{\widehat{\beta}}-\boldsymbol{\beta})$.
\end{theorem}
\begin{remark}
	Theorem  \ref{normalityDebiased}  holds uniformly,  thus both  confidence intervals and tests based on these statements are honest, see,  for example,   the discussion in \cite{li1989honest}.   More importantly, the non-uniformity of  the limit theory for regularised estimators as it was described in  \cite{leeb2005model}, does not exist for  $\boldsymbol{ \widehat{b}}$.   However, we treat that in a more general manner, by using an asymptotic pivotal quantity,   $(\boldsymbol{\widehat{b}} - \boldsymbol{\beta})$ which is less restrictive, see \cite{van2014asymptotically}.
\end{remark}
Following the results of Theorem \ref{normalityDebiased}, we introduce  asymptotic point-wise confidence intervals for the true parameter, $\beta_{i}$, $1\leq i\leq p$, given by  
\begin{equation}\label{confint}
	\text{CI}\left(\alpha \right)=\left[\widehat{b}_{i}\pm z_{\alpha/2}\sqrt{\widehat{\sigma}_{u}^2({\widehat{  \boldsymbol{\Theta}}}{'}\widehat{{\boldsymbol\Sigma}}_{xu}'\widehat{  \boldsymbol{\Theta}})_{i,i}/T}\right],
\end{equation}
$\boldsymbol{\widehat{\Sigma}}_{xu} =T^{-1} (\boldsymbol{\widetilde{X}}'\boldsymbol{\widehat{L}}\boldsymbol{u})(\boldsymbol{\widetilde{X}}'\boldsymbol{\widehat{L}}\boldsymbol{u})'$,  $z_{\alpha/2}:=\Phi^{-1}\left(1-\alpha/2\right)$,   $\alpha $ is the confidence level,  and $\Phi(\cdot)$ is the standard normal CDF, }such that
\begin{align}\underset{  \boldsymbol{\beta}\in \mathcal{B}(s)}{\sup}\left| P \left(\beta_{i}	\in \text{CI}\left(\alpha \right)\right)- (1-\alpha) \right|= o_P(1),\;\forall\; i=1,\ldots,p\label{confidence}\end{align}
and $\widehat{\sigma}_{u}^2$ is a consistent estimate of the variance of  ${u}_{t}$.
Thus, one can perform inference on $\beta_{i}$, for some $z\in \mathbb{R}$, using the following form
\begin{align}
P\left(	\frac{\sqrt{T} (\widehat{b}_i -{\beta}_i)  }{\widehat{\sigma}_{u}\sqrt{	\left({\widehat{  \boldsymbol{\Theta}}}\widehat{{\boldsymbol\Sigma}}_{xu}'\widehat{  \boldsymbol{\Theta}}\right)_{i,i}}}   \leq z\right) - \Phi(z)
=o_{p}(1), 		\label{confidence2}\end{align} for any $z\in \mathbb{R}$.  
Note that \eqref{confidence2} is  implied in Theorem \ref{normalityDebiased} for a single-dimensional component, ${\beta}_i$.  It can be generalised to hold trivially for a group $G$,  of components,  $S_G\subseteq\{1,\ldots,p\}$ which can be  allowed to be large, see,  for example,   \cite{van2014asymptotically}.  

\section{Selection of the optimal regularisation parameter}\label{CV}
The selection of the regularisation parameter used to obtain  $\boldsymbol{ \widehat{\beta}}$, $\boldsymbol{\widetilde{\beta}}$, and $\boldsymbol{\widehat{\gamma}}_i$,  defined  in \eqref{penalised},  \eqref{lasso_correlation} and \eqref{node-wise} respectively, is carried out by a $k$-fold-cross validation scheme, where each fold can be considered as a block.   Due to the autocorrelation present in the errors, we  do not randomly shuffle the data within the blocks, retaining the structure of dependence intact.   Each of the blocks $B_{1}, \ldots, B_{k}$ are  non-overlapping and consequent to each other.

However, different types of cross validation schemes can be considered.   \cite{racine2000consistent} proposed a  cross-validation approach carried out in blocks and highlighting near-independence of the training and validation sub-samples, as an extension to the $h$-\emph{block} method of \cite{burman1994cross}, which proved consistent for general stationary processes.  The former  encompasses our method of cross-validation, the \emph{leave-one-out} method and the $h$-\emph{block }method of \cite{burman1994cross}, and can be an alternative in  selecting the optimal regularisation parameter in our framework as well.

\section{Simulation study}\label{MCsection}
We use simulations to verify the theoretical properties of the proposed methodology in finite samples.
We assess the performance of our method against the  \emph{debiased} \emph{Lasso}. 
We generate an array of samples from the following model	
\begin{align}\label{mod}
	y_{t}= \boldsymbol{x}_{t}'\boldsymbol{\beta} + u_t, \quad
	u_t={\phi} u_{t-1 } + \varepsilon_{t},
\end{align}
where  $\boldsymbol{x}_{t}, \; \varepsilon_{t}\sim\rm{i.i.d.} \, \mathcal{N}(0,1)$  and the sample sizes $T,p=\{100, 200, 500\}$. The active set has cardinality $s_{0} = |S_{0}|= 3$, where $S_{0}=\{  i:\beta_{i}\neq0, \; i=1,\ldots,p \}$, while $S_{0}^{c}= \{i:\beta_{i}=0, \; i=s_0+1,\ldots, p \}$.  We choose the sparsity level $s_0$, according to the results of \cite{van2014asymptotically}, which are indicative of good performance for the \emph{debiased Lasso}.  Simulation results for $s_0=7$ are relegated to the Supplementary Material.
Each of the sets, $S_{0}, \;S_{0}^{c}$ assume the form:
$
S_{0} = \{1,2,\ldots,s_{0}\}\equiv\{v_{1},\ldots,v_{s_{0}}\},
$
where $v_{1},\ldots,v_{s_{0}}$ is a realisation of random draws of $S_{0}$ without replacement from $\{1,\ldots,p\}$.  The parameters, $\beta_{i}\in S_0$ are simulated 
from the  $U[0,1]$ distribution  at  each replication, while the autoregressive parameter $\phi$ used for the simulation of $u_{t}$, takes values $\phi=[0,0.5,0.8,0.9]$, where $\phi=0$ indicates that    $u_{t}\sim\rm{i.i.d.} \, \mathcal{N}(0,1)$.

We base our findings on 1000 Monte Carlo replications, and evaluate the following performance measures:
\begin{equation}
	\text{AvgCov}   =  \frac{1}{z} \sum_{i \in {S}} P(\beta_{i}\in \text{CI}_{i}), \quad 	\text{AvgLength}  =  \frac{1}{z} \sum_{i \in {S}} \;\text{length}(\text{CI}_{i})\label{measures} ,  
\end{equation}
where $z=s_{0}, \; S= S_{0}$ when $\beta_{i}\in {S_{0}}$, $z=p-s_{0},\; S=S_{0}^{c}$, when $\beta_{i}\in {S_{0}^{c}}$, $\forall\; i=1,\ldots,p$ and 
CI$_{i}$ is a two-sided confidence interval for either $\beta_{i}\in {S_{0}}$ or $\beta_{i}\in {S_{0}^{c}}$, denoted in \eqref{confint}.  Further, 
we test the null hypothesis  $H_{0} : \beta_i = 0$, using the following t-statistic
\begin{align}\label{statistic}
	\widehat{S}_i= {T}^{1/2}\frac{\widehat{b}_{i}-\beta_{i| H_0}}{\widehat{\sigma}^{2}_{u}\sqrt{\left(\widehat{\boldsymbol{\Theta}}\widehat{\boldsymbol{\Sigma}}\widehat{\boldsymbol{\Theta}}\right)_{i,i}}},
\end{align}	
where,  for all $i=1,\ldots, p$,  $\widehat{S}_i\sim t^{(T-1)}$, and $ t^{(T-1)}$ the student's-$t$ distribution with $T-1$ degrees of freedom, $\beta_{i|H_0}=0$ under the null,  $\widehat{b}_{i}$  is defined in \eqref{debLaso},  $\widehat{\boldsymbol{\Sigma}}={T}^{-1}\widetilde{  \boldsymbol{x}}_t\widetilde{  \boldsymbol{x}}_t'$   is the  sample covariance matrix, and   $\boldsymbol{\widehat{\Theta}}$ is  defined in Section \ref{Thetasection}.    Note that in the case where $i\in S_{0}$, we report the empirical  power, while for  $i\in S_{0}^{c}$, we report the empirical size

All tests are carried out at  $\alpha=5\%$ significance level and confidence intervals are at the $1-\alpha=95\%$ level.  	 Prior  to the estimation of $\widehat{b}_{i}$,  the estimate of  ${\widehat{\phi}}$, and a value for the regularisation parameter are required.  In Section \ref{theoreticalconsiderations}, we describe the method used for the estimation of  ${\widehat{\phi}}$, along with the testing scheme used for  the selection of the  optimal lag order, $q$, of the $AR(q)$ model.   Further, we use   the $k$-fold cross-validation scheme described in Section \ref{CV},  where $k=10$, to select the optimal  regularisation parameter, $\lambda$ to obtain $\boldsymbol{ \widehat{\beta}}$,  $\boldsymbol{\widetilde{\beta}}$   and $\boldsymbol{\widehat{\gamma}}_i$ in \eqref{penalised}, \eqref{lasso_correlation} and \eqref{node-wise},  respectively. 
The results are displayed in Table  \ref{Avg_cov_03}.

We also examine the performance of the  \emph{Lasso}, \emph{GLS Lasso}, \emph{debiased Lasso} and  \emph{debiased \emph{GLS} Lasso} estimators in terms of average \emph{root mean squared error} (RMSE) throughout 1000 replications of model \eqref{mod}.  	The results are displayed in Table  \ref{rmse1}.

The first panel of Table \ref{rmse1} reports the ratio of the average RMSE of the \emph{Lasso} estimator  over the RMSE of the \emph{GLS Lasso}. The second panel of  Table \ref{rmse1}, reports the ratio of the average   RMSE of the \emph{debiased Lasso} estimator over 
the RMSE of \emph{debiased \emph{GLS} Lasso}.   Entries larger than 1 indicate superiority of the competing model (\emph{GLS} \emph{Lasso}).   Highlighted are the entries corresponding to the RMSE of the   {\emph{GLS} Lasso} in Panel I and  the \emph{debiased \emph{GLS} Lasso} in Panel II.  

The evidence is compelling, since both the \emph{GLS Lasso} and  \emph{debiased GLS Lasso } have, in general,  the smallest  RMSE compared with their counterpart respectively. The most pronounced cases of improvement are when the autocorrelation is strong, i.e. $\phi=0.9$ for all possible sample sizes, where the RMSE of the proposed method is almost (and /or more) two times smaller than the one recorded by the benchmark,  \emph{Lasso}.  Notice that in the case of $\phi=0$ \emph{GLS Lasso}	and \emph{Lasso} perform equally well.

In the first panel of  Table \ref{Avg_cov_03} we present the  average coverage rates (as $\text{AvgCov}$ $S_0$ or $S_{0}^{c}$) and average length of the confidence intervals, of the  \emph{  debiased GLS Lasso} compared to the  \emph{debiased Lasso}. Notice that $1-\text{AvgCov} S_0^{c} $ is the size of the test in \eqref{statistic}.   In the second panel of Table \ref{power_03} we facilitate the comparison of the two methods by reporting the size adjusted power  of the test.  By size-adjusted power, we mean that if the test is found to reject at a rate $\widehat{\alpha}>0.05$ under the null,  then the power of the test is adjusted by finding the $0.95$ empirical quantile of the statistic in  \eqref{statistic}.  We then report the rate that exceeds this quantile.  Alternatively, if $\widehat\alpha\leq 0.05$, then no adjustment is needed.

It is noticeable that when autocorrelation is present in the error term, \emph{debiased Lasso} underperforms severely in terms of all measures reported.  More specifically, as  $T$ increases the benefits of \emph{debiased GLS Lasso } are more clear, the coverage rates reported for our method are approaching the desirable rate of $0.95$, while for the \emph{debiased Lasso} there are noticeable deviations from the nominal rate.    The latter is more severe for higher serial autocorrelation reported, e.g. $\phi=[0.8,0.9]$.  

Further,  \emph{debiased GLS Lasso} reports narrower confidence intervals, indicating that the variance of the proposed estimator is significantly smaller compared to the competing method.  In addition, our method appears  to be more powerful and correctly sized throughout the  different cases reported. The best performing  case reported is  in Table \ref{Avg_cov_03}, for $(p,T)=(200,500)$ and $\phi=0.9$, where the size is $0.05$ and for $(p,T)= (500,500)$,  using the same $\phi$, the reported size is $0.049$.   

\begin{table}[h!]
	\caption{Relative to \emph{GLS Lasso} \text{RMSE}.   }
	\label{rmse1}
	\resizebox{\linewidth}{!}{%
		\begin{tabular}{llllllllllllll}
			\toprule
			\multicolumn{14}{l}{\textbf{Panel I}}                                                                                           \\ \midrule \midrule
			& $p/T$ & \multicolumn{4}{c}{100}       & \multicolumn{4}{c}{200}       & \multicolumn{4}{c}{500}       \\ \midrule 
			$\phi$         &       & 0     & 0.5   & 0.8   & 0.9   & 0     & 0.5   & 0.8   & 0.9   & 0     & 0.5   & 0.8   & 0.9   \\ \midrule 
			\emph{Lasso}/GLS \emph{Lasso}          & 100   & 0.996 & 1.237 & 1.798 & 2.234 & 0.994 & 1.257 & 2.049 & 2.608 & 0.998 & 1.241 & 1.890 & 2.788 \\
			\rowcolor[HTML]{C0C0C0} 
			GLS \emph{Lasso}      &       & 0.042 & 0.039 & 0.036 & 0.034 & 0.029 & 0.026 & 0.023 & 0.023 & 0.019 & 0.017 & 0.015 & 0.014 \\ \midrule 
			\emph{Lasso}/GLS \emph{Lasso}           & 200   & 0.998 & 1.236 & 1.934 & 2.574 & 0.993 & 1.275 & 2.010 & 2.518 & 0.998 & 1.236 & 2.036 & 2.945 \\
			\rowcolor[HTML]{C0C0C0} 
			GLS \emph{Lasso}      &       & 0.033 & 0.031 & 0.028 & 0.028 & 0.022 & 0.020 & 0.018 & 0.018 & 0.015 & 0.013 & 0.011 & 0.011 \\ \midrule 
			\emph{Lasso}/GLS \emph{Lasso}           & 500   & 0.998 & 1.273 & 2.249 & 3.184 & 1.001 & 1.256 & 1.928 & 2.641 & 0.997 & 1.253 & 2.188 & 3.016 \\
			\rowcolor[HTML]{C0C0C0} 
			GLS \emph{Lasso}      &       & 0.023 & 0.022 & 0.020 & 0.020 & 0.016 & 0.015 & 0.013 & 0.013 & 0.010 & 0.009 & 0.008 & 0.007 \\ \midrule 
			\multicolumn{14}{l}{\textbf{Panel II}}                                                                                          \\ \midrule \midrule
			Debiased \emph{Lasso}/Debiased GLS  & 100   & 1.014 & 1.258 & 1.928 & 2.515 & 1.005 & 1.284 & 2.068 & 2.847 & 1.001 & 1.287 & 2.087 & 2.943 \\
			\rowcolor[HTML]{C0C0C0} 
			Debiased GLS   &       & 0.094 & 0.086 & 0.077 & 0.075 & 0.069 & 0.062 & 0.054 & 0.052 & 0.044 & 0.040 & 0.035 & 0.033 \\
			Debiased \emph{Lasso}/Debiased GLS  & 200   & 1.015 & 1.237 & 1.816 & 2.299 & 1.007 & 1.285 & 2.072 & 2.855 & 1.002 & 1.288 & 2.092 & 2.946 \\
			\rowcolor[HTML]{C0C0C0} 
			Debiased GLS   &       & 0.090 & 0.083 & 0.078 & 0.079 & 0.067 & 0.060 & 0.053 & 0.051 & 0.044 & 0.039 & 0.034 & 0.033 \\
			Debiased \emph{Lasso}/Debiased GLS & 500   & 1.017 & 1.128 & 1.228 & 1.257 & 1.005 & 1.291 & 2.073 & 2.734 & 1.003 & 1.293 & 2.088 & 2.986 \\
			\rowcolor[HTML]{C0C0C0} 
			Debiased GLS   &       & 0.081 & 0.079 & 0.082 & 0.088 & 0.064 & 0.057 & 0.051 & 0.050 & 0.044 & 0.039 & 0.034 & 0.033 \\ \bottomrule\\
		\end{tabular}%
	}
\end{table}
\begin{table}[!h]
	\caption{	Average coverage rates, lengths of $CI$, size-adjusted power and size of \emph{debiased} estimates.}	
	\label{Avg_cov_03}
	\resizebox{\linewidth}{!}
	{%
		\begin{tabular}{lllllllllllllll}
			\toprule
			&                            & p/T                 & \multicolumn{4}{c}{100}       & \multicolumn{4}{c}{200}       & \multicolumn{4}{l}{500}       \\ \midrule 
			
			$\phi$ &&
			&
			\multicolumn{1}{c}{0} &
			\multicolumn{1}{c}{0.5} &
			\multicolumn{1}{c}{0.8} &
			\multicolumn{1}{c}{0.9} &
			\multicolumn{1}{c}{0} &
			\multicolumn{1}{c}{0.5} &
			\multicolumn{1}{c}{0.8} &
			\multicolumn{1}{c}{0.9} &
			\multicolumn{1}{c}{0} &
			\multicolumn{1}{c}{0.5} &
			\multicolumn{1}{c}{0.8} &
			\multicolumn{1}{c}{0.9} \\ \midrule
			\emph{Debiased Lasso} & AvgCov $S_0$               & 100                     & 0.796 & 0.758 & 0.694 & 0.660 & 0.818 & 0.775 & 0.680 & 0.619 & 0.846 & 0.802 & 0.679 & 0.594 \\
			& AvgCov $S_0^{c}$ &                         & 0.864 & 0.831 & 0.753 & 0.709 & 0.868 & 0.830 & 0.730 & 0.654 & 0.879 & 0.836 & 0.713 & 0.613 \\
			& AvgLength                  &                         & 0.306 & 0.314 & 0.348 & 0.394 & 0.224 & 0.228 & 0.248 & 0.278 & 0.150 & 0.151 & 0.157 & 0.170 \\
			& AvgLength$^{c}$  &                         & 0.306 & 0.313 & 0.348 & 0.394 & 0.224 & 0.228 & 0.248 & 0.278 & 0.150 & 0.151 & 0.157 & 0.170 \\
			\emph{Debiased GLS}   & AvgCov $S_0$               &                         & 0.789 & 0.811 & 0.845 & 0.857 & 0.811 & 0.854 & 0.889 & 0.892 & 0.842 & 0.872 & 0.909 & 0.916 \\
			& AvgCov $S_0^{c}$ &                         & 0.868 & 0.894 & 0.923 & 0.929 & 0.870 & 0.902 & 0.932 & 0.938 & 0.879 & 0.913 & 0.943 & 0.950 \\
			& AvgLength                  &                         & 0.305 & 0.308 & 0.313 & 0.314 & 0.224 & 0.227 & 0.229 & 0.229 & 0.150 & 0.152 & 0.154 & 0.154 \\
			& AvgLength$^{c}$  &                         & 0.305 & 0.308 & 0.312 & 0.313 & 0.224 & 0.227 & 0.229 & 0.229 & 0.150 & 0.152 & 0.154 & 0.154 \\\midrule
			\emph{Debiased Lasso} & AvgCov $S_0$               & 200                     & 0.771 & 0.743 & 0.667 & 0.621 & 0.810 & 0.778 & 0.691 & 0.640 & 0.840 & 0.799 & 0.680 & 0.605 \\
			& AvgCov $S_0^{c}$ &                         & 0.870 & 0.847 & 0.782 & 0.742 & 0.875 & 0.840 & 0.745 & 0.678 & 0.878 & 0.835 & 0.721 & 0.633 \\
			& AvgLength                  &                         & 0.302 & 0.312 & 0.352 & 0.400 & 0.223 & 0.228 & 0.250 & 0.286 & 0.148 & 0.150 & 0.158 & 0.175 \\
			& AvgLength$^{c}$  &                         & 0.301 & 0.312 & 0.352 & 0.400 & 0.223 & 0.227 & 0.250 & 0.286 & 0.148 & 0.150 & 0.158 & 0.175 \\
			\emph{Debiased GLS}   & AvgCov $S_0$               & \multicolumn{1}{l}{}    & 0.759 & 0.780 & 0.810 & 0.822 & 0.804 & 0.839 & 0.873 & 0.879 & 0.839 & 0.872 & 0.911 & 0.923 \\
			& AvgCov $S_0^{c}$ & \multicolumn{1}{l}{}    & 0.873 & 0.897 & 0.918 & 0.916 & 0.877 & 0.906 & 0.935 & 0.940 & 0.878 & 0.912 & 0.943 & 0.950 \\
			& AvgLength                  & \multicolumn{1}{l}{}    & 0.301 & 0.303 & 0.308 & 0.310 & 0.223 & 0.225 & 0.227 & 0.227 & 0.148 & 0.150 & 0.152 & 0.152 \\
			& AvgLength$^{c}$  & \multicolumn{1}{l}{}    & 0.300 & 0.303 & 0.308 & 0.310 & 0.223 & 0.225 & 0.227 & 0.227 & 0.148 & 0.150 & 0.152 & 0.152 \\\midrule
			\emph{Debiased Lasso} & AvgCov $S_0$               & \multicolumn{1}{l}{500} & 0.720 & 0.662 & 0.537 & 0.468 & 0.796 & 0.766 & 0.680 & 0.616 & 0.834 & 0.791 & 0.689 & 0.624 \\
			& AvgCov $S_0^{c}$ & \multicolumn{1}{l}{}    & 0.887 & 0.868 & 0.863 & 0.870 & 0.887 & 0.853 & 0.761 & 0.707 & 0.876 & 0.837 & 0.737 & 0.651 \\
			& AvgLength                  & \multicolumn{1}{l}{}    & 0.299 & 0.308 & 0.349 & 0.391 & 0.219 & 0.224 & 0.251 & 0.287 & 0.146 & 0.148 & 0.160 & 0.182 \\
			& AvgLength$^{c}$  & \multicolumn{1}{l}{}    & 0.299 & 0.308 & 0.348 & 0.390 & 0.219 & 0.224 & 0.250 & 0.287 & 0.146 & 0.148 & 0.160 & 0.182 \\
			\emph{Debiased GLS}   & AvgCov $S_0$               & \multicolumn{1}{l}{}    & 0.717 & 0.735 & 0.779 & 0.776 & 0.794 & 0.818 & 0.853 & 0.854 & 0.832 & 0.867 & 0.903 & 0.904 \\
			& AvgCov $S_0^{c}$ & \multicolumn{1}{l}{}    & 0.892 & 0.904 & 0.903 & 0.888 & 0.888 & 0.916 & 0.939 & 0.938 & 0.877 & 0.913 & 0.943 & 0.949 \\
			& AvgLength                  & \multicolumn{1}{l}{}    & 0.300 & 0.302 & 0.307 & 0.309 & 0.219 & 0.221 & 0.224 & 0.224 & 0.146 & 0.148 & 0.149 & 0.149 \\
			& AvgLength$^{c}$  & \multicolumn{1}{l}{}    & 0.299 & 0.302 & 0.307 & 0.309 & 0.219 & 0.221 & 0.224 & 0.224 & 0.146 & 0.148 & 0.149 & 0.149 \\ \bottomrule
		\end{tabular}%
	}
	\label{power_03}
	\resizebox{\linewidth}{!}{%
		\begin{tabular}{@{}lcllllllllllll@{}}
			
			\multicolumn{14}{c}{Size-adjusted power}                                                                                                               \\ \midrule
			& p/T                  & \multicolumn{4}{c}{100}       & \multicolumn{4}{c}{200}       & \multicolumn{4}{c}{500}       \\ \midrule 
			$\phi$ &
			&
			\multicolumn{1}{c}{0} &
			\multicolumn{1}{c}{0.5} &
			\multicolumn{1}{c}{0.8} &
			\multicolumn{1}{c}{0.9} &
			\multicolumn{1}{c}{0} &
			\multicolumn{1}{c}{0.5} &
			\multicolumn{1}{c}{0.8} &
			\multicolumn{1}{c}{0.9} &
			\multicolumn{1}{c}{0} &
			\multicolumn{1}{c}{0.5} &
			\multicolumn{1}{c}{0.8} &
			\multicolumn{1}{c}{0.9} \\ \midrule   
			Debiased \emph{Lasso} &       100                   & 0.789 & 0.765 & 0.669 & 0.578 & 0.865 & 0.836 & 0.764 & 0.692 & 0.916 & 0.902 & 0.865 & 0.817 \\
			Debiased \emph{GLS}   &                     & 0.787 & 0.807 & 0.820 & 0.829 & 0.868 & 0.870 & 0.952 & 0.897 & 0.916 & 0.919 & 0.930 & 0.932 \\
			Debiased \emph{Lasso} &      200                   & 0.780 & 0.748 & 0.649 & 0.556 & 0.852 & 0.838 & 0.780 & 0.706 & 0.913 & 0.901 & 0.857 & 0.802 \\
			Debiased \emph{GLS}   &                      & 0.776 & 0.793 & 0.807 & 0.809 & 0.855 & 0.865 & 0.878 & 0.884 & 0.913 & 0.920 & 0.929 & 0.933 \\
			Debiased \emph{Lasso} &       500                  & 0.776 & 0.749 & 0.640 & 0.525 & 0.824 & 0.836 & 0.752 & 0.704 & 0.913 & 0.901 & 0.857 & 0.804 \\
			Debiased \emph{GLS}   &                      & 0.775 & 0.782 & 0.792 & 0.789 & 0.850 & 0.868 & 0.881 & 0.884 & 0.913 & 0.920 & 0.929 & 0.933 \\ 
			\bottomrule
		\end{tabular}%
	}
\end{table}

\begin{figure}[h!]	
	\centerfloat
	\includegraphics[width=\linewidth]{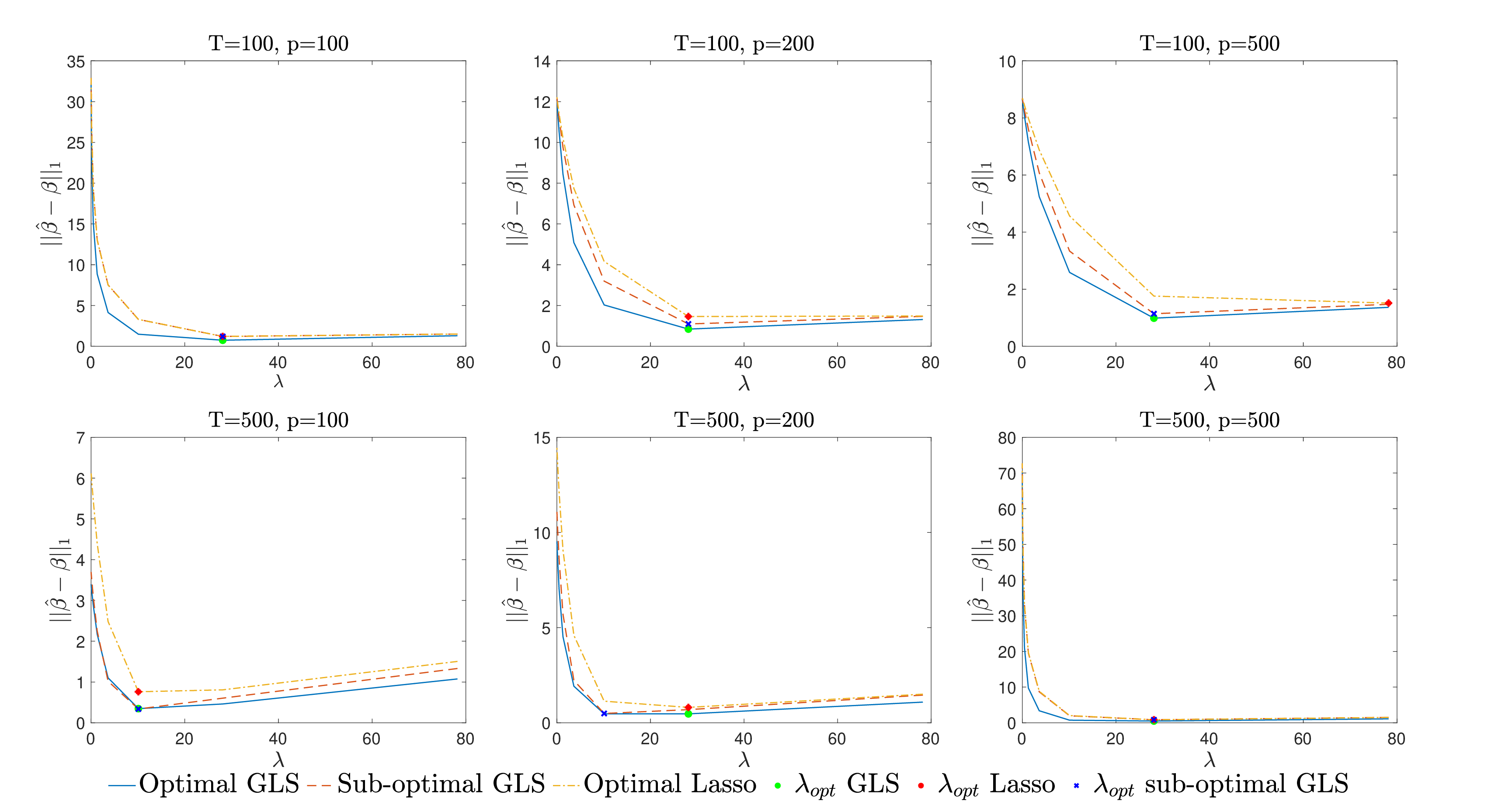}
	\caption{Estimation loss, $ \Vert\boldsymbol{\widehat{\beta}} - \boldsymbol{\beta}\Vert_1$.  Optimal \emph{GLS} corresponds the \emph{GLS Lasso} defined in \eqref{penalised} when the preliminary \emph{Lasso} estimator, $\boldsymbol{\widetilde{ \beta}}$ in \eqref{lasso_correlation} is obtained according to the optimal selection of the regularisation parameter via the cross-validation scheme highlighted in Section \ref{CV}, while \emph{Sub-optimal GLS} corresponds the \emph{GLS Lasso} as in \eqref{penalised} when the preliminary \emph{Lasso} estimator, $\boldsymbol{\widetilde{ \beta}}$ is not obtained according to the maximisation of \eqref{cvloss}.   Further,   \emph{Optimal Lasso } corresponds to \eqref{lasso_correlation} selecting the optimal $\lambda$ through the cross-validation scheme highlighted in Section 5, and  $\lambda_{opt}$ corresponds to the optimal selection of $\lambda$ for each of the methods. }
	\label{fig:cv_bias}
\end{figure}

\begin{figure}[h!]
	\centerfloat
	\includegraphics[width=\linewidth]{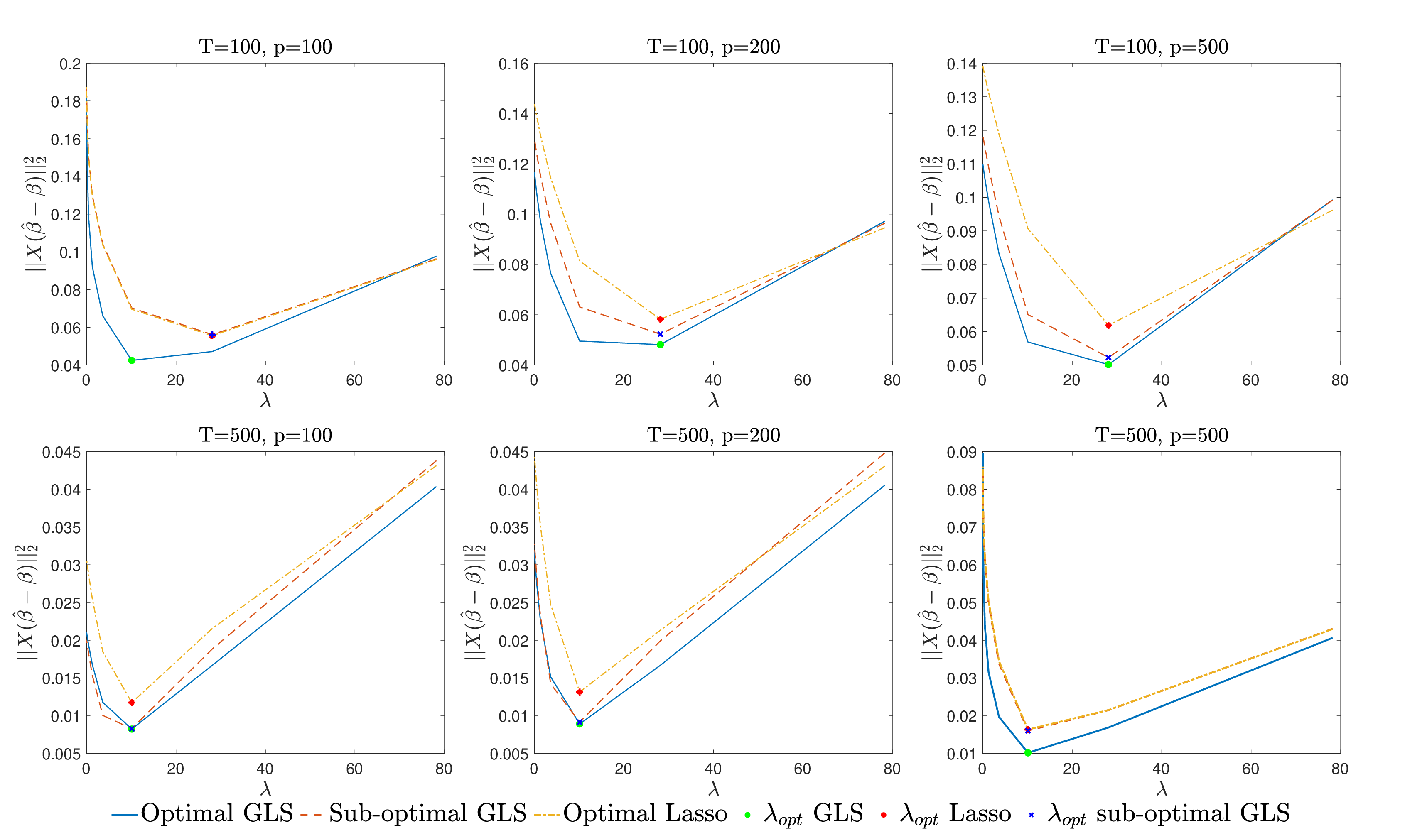}
	\caption{Prediction loss, $\Vert \boldsymbol{ X } (\boldsymbol{ \widehat{\beta}} - \boldsymbol{\beta})\Vert_{2}^{2}$.   Look at description of Figure \ref{fig:cv_bias}}
	\label{fig:cv}
\end{figure}

In Figures \ref{fig:cv_bias}--\ref{fig:cv} we graph  the  estimation and prediction error rates of \emph{GLS Lasso}, defined in \eqref{coef} and \eqref{empproces}  respectively, of  \emph{Lasso}, defined in \eqref{prederror} and \eqref{coeferror1}, and of the \emph{sub-optimal GLS Lasso} considering a $\lambda$, in the preliminary estimate $\boldsymbol{\widetilde{\beta}}$ which maximizes rather than minimising the following test-loss function w.r.t. the regularisation parameter, $\lambda$, for sample sizes $T=[100,500], \; p=[100,200,500]$ and $u_{t} = 0.9 u_{t-1}+\varepsilon_{t},$ 
\begin{align}
	\mathcal{L}_{CV}(\lambda)  = \frac{1}{T} \sum_{i=1}^{k} \sum_{j\in B_{k}} \left(	\boldsymbol{{y}}_{j} - 	\boldsymbol{\widetilde{f}}_{\lambda}^{-i}\left(\boldsymbol{{X}}_{j}\right)	\right)^2,\label{cvloss}
\end{align}
where $B_k$ is one block (fold) for $k=1,\ldots,10$,  $\boldsymbol{\widetilde{f}}_{\lambda}^{-i}\left(\boldsymbol{{X}}_{j}\right) =\boldsymbol{X}_{j} \boldsymbol{\widetilde{\beta}}$ excluding block $i$.

Notice that the prediction and estimation error  for the \emph{GLS Lasso} are bounded from below as Theorem \ref{theorem1} indicates, while a lower bound for the \emph{Lasso}  is provided in Lemma \ref{consistencyAR,cor}.     In empirical problems, cross-validation does not always yield an  optimal selection of $\lambda$, because one has to select between true model recovery (interpretable model) and  a parsimonious (highly regularised) model.  The latter  can potentially be an issue because the  selection of $\lambda$  controls the lower bound  of both the prediction and estimation error, see,  for example,   \eqref{empproces}--\eqref{coef}.   To that end, \emph{GLS Lasso} provides an asymptotic guarantee that even when the selection of $\lambda$ is sub-optimal, it is preferable to using \emph{Lasso}. Indicative cases of such behaviour are $T=100,p=200, 500$ and $T=100, p=100, 200, 500$.  Furthermore, in the remaining sets of cases, of a sub-optimal selection of $\lambda$, \emph{GLS Lasso} still reports smaller errors than \emph{Lasso} but  by a smaller margin.

As a summary of Figure \ref{fig:cv}, \emph{GLS Lasso} can be more   attractive under  relaxed assumptions due to the fact that neither the estimation, nor the prediction error deteriorate severely with different selections of $\lambda$.  Further, even when the selection of $\lambda$ for the preliminary estimate is sub-optimal, \emph{GLS Lasso} appears to be, in the majority of the cases,  the best method to use as a preliminary vehicle to inference.  

\begin{figure}[!ht]
	\includegraphics[width=1\linewidth]{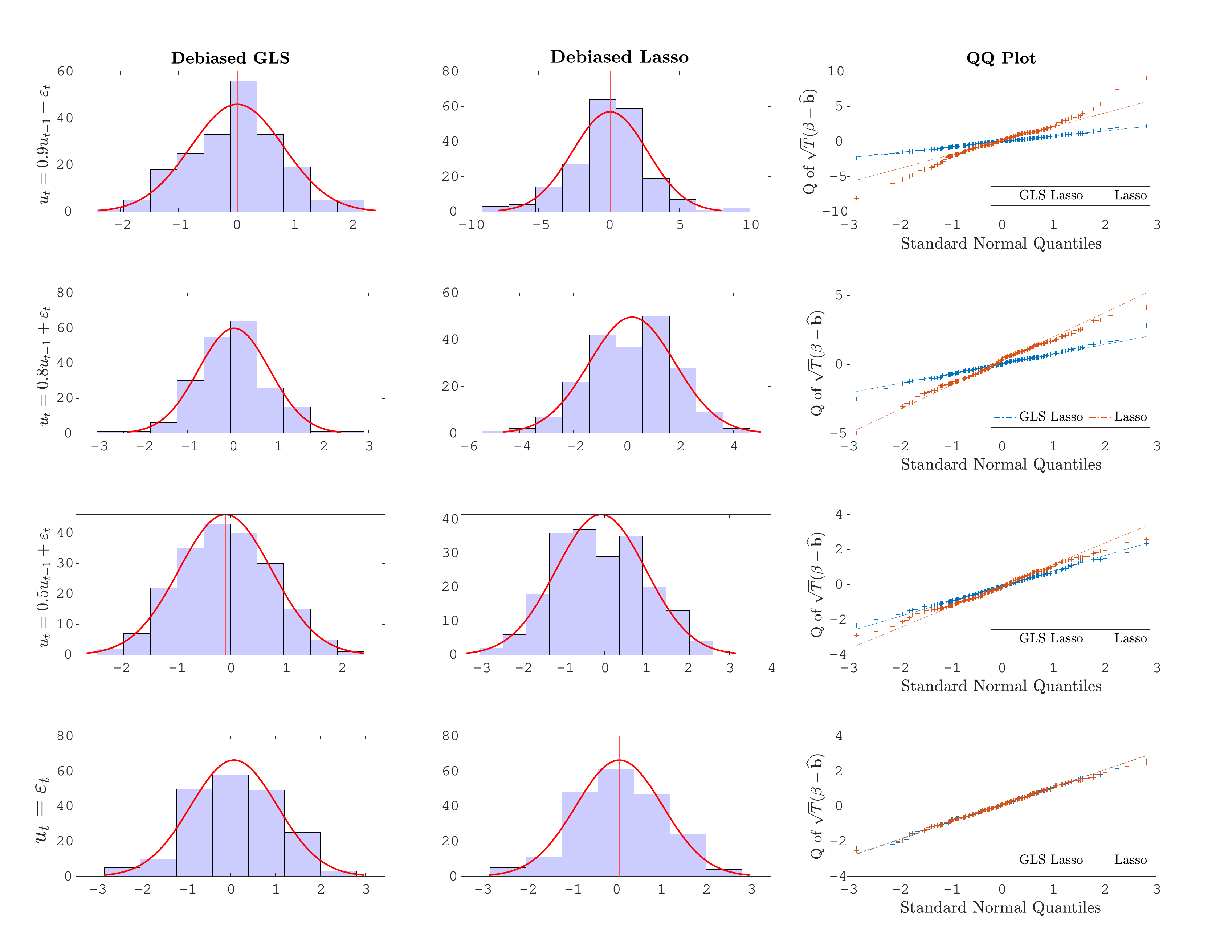}\vspace{-8mm}
	\caption{Histogram and QQ-plot of $\sqrt{T}(\boldsymbol{\beta}-\boldsymbol{\widehat{{b}}} )$, where $\boldsymbol{\widehat{b }}$ corresponds to the \emph{debiased GLS Lasso} estimator defined in \eqref{debLaso},   in the first column of  Figure \ref{fig1} and to the \emph{debiased Lasso} estimator in the second  column of  Figure \ref{fig1}, for the case of $T=500$, $p=200$ and all cases of $\phi$ and degree of sparsity $s_{0}= 3$,  as they are reported in Tables \ref{rmse1}--\ref{Avg_cov_03}. }
	\label{fig1}
\end{figure}

In Figure \ref{fig1} we graph  the density estimates of $\sqrt{T}(\boldsymbol{\beta} - \boldsymbol{\widehat{{b}}} )$ corresponding to \emph{debiased GLS Lasso} starting from the left panel, the second panel corresponds to \emph{debiased Lasso} and the third panel illustrates the quantile-quantile (QQ) plots of the aforementioned.  Figure \ref{fig1} corresponds to the case where the sample size is the largest, i.e. $T=500$ and the number of regressors is  $p=200$ for all $\phi= [0, 0.5, 0.8, 0.9]$.  A full set of graphs considering all the cases that are reported in the Tables \ref{rmse1}--\ref{Avg_cov_03} are available upon request.

In the first row of graphs in Figure \ref{fig1} where $\phi=0.9$,  it is observed that extreme dependence between $u_t$ and $u_{t-1}$ is related to overdispersion of the distribution of $\sqrt{T}(\boldsymbol{\beta} - \boldsymbol{\widehat{{b}}} )$,  corresponding to the \emph{debiased Lasso} estimator,  while in the same case,  departures from the standard normal are observed, with the empirical data showing evidence of heavy tails.    On the contrary, the empirical distribution of the \emph{debiased GLS Lasso} estimator is closer to the Standard Normal, as it is indicative from the QQ-plot of the same row.  However, this effect fades as $\phi$ takes values closer to $0$, indicative of the case where  $u_{t}\sim \; i.i.d.\mathcal{N}(0,1).  $  In the second row of Figure \ref{fig1}, \emph{debiased Lasso } appears slightly skewed left with heavy tails to persist indicative of departures from the standard normal distribution, while \emph{debiased GLS Lasso}  appears to approach the standard normal distribution, similar behaviour for both methods is reported in the third row.  Finally, in the case where $\phi= 0 $ the two methods coincide supporting the evidence in  Tables \ref{rmse1}--\ref{Avg_cov_03}.

\section{Discussion}\label{discussion}
This paper provides a complete inferential methodology for  high dimensional linear regressions with serially correlated error terms.   We  propose \emph{GLS}-type of estimators in the high-dimensional framework, derive a  \emph{debiased  GLS Lasso} estimator, and establish its asymptotic normality.  Theoretical results lead to valid confidence intervals regardless of the dimensionality of the parameters and the degree of autocorrelation,  under stationarity of the error and covariate processes.       
We derive non-asymptotic bounds using Bernstein-type of inequalities derived in  \cite{DGK}, allowing for a degree of  flexibility in our model compared to what classical assumptions typically allow.   
Finally, our method explores the underlying structure of the  process in the error term,  introducing a flexible way of dealing with autocorrelation in high-dimensional models, without a prior knowledge of neither  its existence nor its structure.   We support this statement both theoretically and with our  simulation results, where we show that even when autocorrelation is absent the \emph{debiased Lasso} and \emph{debiased GLS Lasso} are asymptotically equivalent.

There are a number of interesting avenues for future work.  For example, a natural extension would be to generalise our method using not only $\ell_1$-regularised models or $\ell_2$ (e.g. \cite{zhang2021debiased}), but a combination of the two,  e.g. the elastic net, covering a wider class of penalised  models.    Further, since our assumptions allow for a wide class of dependent stochastic processes, one could consider heteroskedastic error processes, see,  for example,   the framework in \cite{cck2022}, where
\begin{align}
	u_{t} = h_t\epsilon_t, \quad t=1,\ldots,T,
\end{align}
where $h_t$ is a persistent scaling factor, smoothly varying in time, while $\epsilon_t\sim\; i.i.d.$	
\bigskip 

\noindent\acknowldgements{For discussions and/or comments we thank  Jianqing Fan, Dacheng Xiu, Zacharias Psaradakis, Tassos Magdalinos and Yunyi Zhang.} 

		\bibliography{bib_file_02_08_20}
		\bibliographystyle{econometrica}
		\newpage
	\renewcommand{\theequation}{\thesection.\arabic{equation}}
		\setcounter{page}{1}
				\counterwithin*{equation}{section}
			\counterwithin{table}{section}
				\emptythanks
		\title{Supplement to "High Dimensional Generalised Penalised Least Squares"}
		\date{ }
		\maketitle\emptythanks
		\vspace{-25mm}
		\section*{}
		This Supplement provides proofs of the theoretical results given in the text of the main paper. It is organised as follows: Section  \ref{oracleinequalities} provides proofs of Lemma \textcolor{red}{1}, Corollary \textcolor{red}{1},  Theorem \textcolor{red}{1} and Corollary \textcolor{red}{2} of the main paper. Section \ref{appendixb} provides the proofs of Theorem \textcolor{red}{2} and Theorem \textcolor{red}{3} of the main paper. Section \ref{secondary res} provides  proofs of auxiliary technical lemmas.   Section \ref{simulations} contains supplementary simulation material. 
		
		Formula numbering in this  supplement includes the section number, e.g. $(8.1)$,
		and references to lemmas
		are  signified as ``Lemma 8.\#",  ``Lemma 9.\#",  e.g. Lemma 8.1.  Formula numbering in the Notation section is signified with the letter N.\#, e.g. (N.1). 
		Theorem references to the main paper  are signified, e.g. as   Theorem $ \textcolor{red}{1}$,  while equation references are signified as, e.g.  \textcolor{red}{(1)}, \textcolor{red}{(2)} (denoted in red).  
		
		In the proofs, $C, d, e$ stand for  generic positive constants which may assume different values in different contexts.

		\vspace{3mm}
		{
			\textbf{Notation:} \hspace{2mm}
			For any vector $\boldsymbol{x}\in \mathbb{R}^{n}, $ we denote the $\ell_{p}$-, $\ell_{\infty}$- and $\ell_0$- norms, as  $\left\lVert \boldsymbol{x} \right\rVert_{p}=\left(\sum_{i=1}^{n}  | x_{i}  |^p\right)^{1/p}$, $\lVert   \boldsymbol{x} \rVert_{\infty} = \max_{i=1,\ldots,n}|x_{i}|$ and $\lVert  \boldsymbol{x}\rVert_{0}= \sum_{i=1}^{n}\boldsymbol{\mathrm{1}}_{(x_{i}\neq 0)} = \rm{supp}(\boldsymbol{\beta})$,  respectively.   Throughout this supplement, $\mathbb{R}$ denotes the set of  real numbers. We denote the cardinality of a set $S_0$ by $s_{0}=|S_0|$, while $S_{0}^c$ denotes its complement.  For any $\boldsymbol{x}\in \mathbb{R}^{n}$, $\rm{sign}(x)$ denotes the sign function applied to each component of $\boldsymbol{x}$, where $\mathrm{sign}(x)=0$ if $x=0$.   We  use "$\to_{P}$" to denote convergence in probability.  For two deterministic sequences $a_n$ and $b_n$ we define asymptotic proportionality, "$\asymp$",  by writing  $a_n \asymp b_n$, if there exist constants $0 < a_1 \leq a_2$ such that $a_1b_n \leq a_n \leq a_2b_n$ for all $n \geq 1$.      For a real number $a$, $\lfloor a  \rfloor$ denotes the largest integer no greater than $a$.    { For a symmetric  matrix $\boldsymbol{B}$, we denote its minimum and  maximum eigenvalues by $\lambda_{\min}\left(	\boldsymbol{B}	\right)$ and $\lambda_{\max}\left(	\boldsymbol{B}	\right)$  respectively.   For a general (not necessarily square)  $n\times m$ matrix $\boldsymbol{B}=(b_{ij})$, $\Vert \boldsymbol{B}\Vert_1$, $\Vert \boldsymbol{B}\Vert_F$, $\Vert \boldsymbol{B}\Vert_{\infty}$ are its $\ell_{1}$-, $\ell_{F}$-, $\ell_{\infty}$-norm respectively.  In particular, $\Vert \boldsymbol{B}\Vert_1= \max_j\sum_{i}|b_{ij}|$,
				$\Vert \boldsymbol{B}\Vert_F = [\rm{tr}(\boldsymbol{B}\boldsymbol{B}')]^{1/2}$,
				$\Vert \boldsymbol{B}\Vert_{\infty} = \max_i\sum_{j} |b_{ij}|$ and $\rm{tr}(\boldsymbol{B}) $$= \sum_{i=1}^{n}b_{ii}$.} 
			We further write that, $\widetilde{{\boldsymbol{\beta}}}$ is the \textit{Lasso} estimator obtained from the solution of \textcolor{red}{(13)} in the main paper,  $\boldsymbol{\widehat{\beta}}$ is the \emph{GLS Lasso } estimator obtained from the solution of \textcolor{red}{(3)}, given some estimates of the autoregressive parameters ${ \boldsymbol{\widehat{\phi}}} = ( \widehat{\phi}_{1}, \ldots, \widehat{\phi}_{q} )'$  and $\boldsymbol{\beta}$ is a $p\times 1$ parameter vector containing the true parameters.    
		
			We define  $\widetilde{\boldsymbol{x}}_t= \boldsymbol{x}_t- \sum_{ j=1}^{q}\widehat\phi_{j}\boldsymbol{x}_{t-j}$, and its matrix counterpart, $ \widetilde{\boldsymbol{X}}= \boldsymbol{\widehat{L} X}  $, where $\widehat{\boldsymbol{L}}$ is a $\left(T-q \right)\times T$ matrix defined as:
			\begin{align}\label          {L}
				\tag{N.1}
				\widehat{\boldsymbol{L}}=\left(\begin{array}{ccccccccc}-\widehat{\phi}_{q} & -\widehat{\phi}_{q-1} &\cdots& \cdots & -\widehat{\phi}_{ 1} & 1 & \cdots & 0 & 0 \\ 
					0 & -\widehat{\phi}_{q} & -\widehat{\phi}_{q-1} & \cdots & \cdots & -\widehat{\phi}_{1} & \cdots & 0 & 0 \\
					\vdots & \vdots & \vdots & \vdots & \vdots & \vdots & \vdots & \vdots & \vdots \\ 
					0 & 0 & 0 & \cdots & -\widehat{\phi}_{q} & -\widehat{\phi}_{q-1} & \cdots & -\widehat{\phi}_{1} & 1\end{array}\right),  \quad \boldsymbol{\widehat{\phi}} =\left(\begin{array}{cccccccccccc}
					\widehat{\phi}_{1}  \\
					\vdots \\
					\widehat{\phi}_{q} \\ 
				\end{array}\right).
			\end{align}
			Further, following the same argument that we used to define $\widehat{\boldsymbol{ L }}$, we define ${\boldsymbol{L}}$, a $\left(T-q \right)\times T$ matrix, with  $\widehat{\phi}_{j}$ replaced by $\phi_{j}$, denoting the autoregressive parameters, $\forall\; j=1,\ldots,q$.  As a consequence, we define ${\boldsymbol{X} }^{\star} = {\boldsymbol{L}}\boldsymbol{ X }$. 
			
			We shall frequently refer to the $\alpha$-mixing Assumption \textcolor{red}{2} and its property $\alpha_{k}\leq c_{*}\phi^{k}, \; k\geq 1$ of $\{\boldsymbol{x}_{t}\}$ and $\{u_{t } \}$ of Section \textcolor{red}{2} of the main paper.   The r.v.'s $\boldsymbol{x}_{t}, \; {u}_{t}$ have  thin-tailed distributions with properties  denoted in \textcolor{red}{(5)}--\textcolor{red}{(6)}   of the main paper.

			\section{Proofs of  Lemma 1, Corollary 1,  Theorem 1 and Corollary 2 of the main paper} \label{oracleinequalities}
			This section contains the proofs of the results of  Section \textcolor{red}{2} of the main paper.    

				\begin{proof}[\unskip\nopunct]\textbf{Proof of Lemma \textcolor{red}{1}.} In Lemma \textcolor{red}{1} we analyse the asymptotic properties of the \emph{Lasso} estimator under mixing Assumptions, e.g. Assumption \textcolor{red}{1} and offer non-asymptotic bounds for the empirical process and prediction errors, as seen in \textcolor{red}{(15)} and \textcolor{red}{ (16)} of the main paper. 
				Re-arranging the basic inequality, as in Lemma 6.1 of \cite{buhlmann2011statistics}, we obtain
				\begin{align}\label{firstlem}
					\frac{1}{T}\left\lVert \boldsymbol{X} \left(\widetilde{{\boldsymbol{\beta}}} - {\boldsymbol{\beta}} \right) \right\rVert^2_{2}&\leq \frac{2}{T} \boldsymbol{u}'\boldsymbol{X}  \left(\widetilde{{\boldsymbol{\beta}}} - {\boldsymbol{\beta}} \right)+ \lambda (\lVert {\boldsymbol{\beta}}\rVert-\lVert\widetilde{{\boldsymbol{\beta}}}\rVert).
				\end{align}
				The  "empirical process" part of the right hand side of \eqref{firstlem}, $2 \boldsymbol{u}'\boldsymbol{X}  \left(\widetilde{{\boldsymbol{\beta}}} - {\boldsymbol{\beta}} \right)/T$, can be bounded further in terms of  the $\ell_{1}$-norm, such that, 
				\begin{equation}\label{maxfirstElem}
					\frac{1}{T}\left|\boldsymbol{u}'\boldsymbol{X} \left(\widetilde{{\boldsymbol{\beta}}}-{\boldsymbol{\beta}}\right)\right| \leq \frac{2}{T} \left\lVert\boldsymbol{u}'\boldsymbol{X} \right\rVert_{\infty}\left\lVert \widetilde{{\boldsymbol{\beta}}}-{\boldsymbol{\beta}}\right\rVert_{1}.
				\end{equation}
				The regularisation parameter $\lambda$ is chosen such that $2\left\lVert\boldsymbol{u}'\boldsymbol{X}\right\rVert_{\infty}/T\leq \lambda$. Hence, we introduce the event 
				\begin{equation}\label{event1}
					\mathcal{E}:= \left\{ {T^{-1}}\left\lVert\boldsymbol{u}'\boldsymbol{X}\right\rVert_{\infty}\leq \frac{\lambda_{0}}{2} \right\},
				\end{equation}
				which needs to hold with high probability, where $\lambda_{0}= \sqrt{{\log{p}}/{T}}\leq\lambda/2$.    	
				We proceed to illustrate the former. Using the identity $P(\mathcal{E}^{c}) = 1-P(\mathcal{E})$ and  the union bound we obtain:
				\begin{align}
					P(\mathcal{E}^{c})&=1-P\left(	\max_{i} \left|\frac{1}{{T}}\sum_{t=1}^{T}{x}_{t,i}u_t\right|\leq \frac{\lambda_{0}}{2}	\right)
					\leq  \sum_{i}P\left(	\frac{1}{{\sqrt{T}}} \left|\sum_{t=1}^{T}{x}_{t,i}u_t\right|> \frac{\lambda_{0}\sqrt{T}}{2}	\right)\nonumber.
				\end{align}
				Notice that $\left\{  \boldsymbol{x}_{t}u_{t} \right\}$ is $\alpha$-mixing as a product of two $\alpha$-mixing  sequences,  $\forall\; j=1,\ldots, q, \, i = 1, \ldots, p$. Then by direct application of Lemma 1 of \cite{DGK},  we obtain
				\begin{align}
					\sum_{i} P\left(	\frac{1}{{\sqrt{T}}} \left|\sum_{t=1}^{T}{x}_{t,i}u_t\right|> \frac{\lambda_{0}\sqrt{T}}{2}	\right)
					&\leq  p c\left\{ 	\exp \left(-c_{2}\left(\frac{\sqrt{T}\lambda_{0}}{4}\right) ^{2}\right)\right. \nonumber\\
					&\left. \quad \quad + \exp\left({-c_{3}\frac{\sqrt{T\log{(p)}}}{4\log^{2}T}}\right)^{\zeta}	\right\}\nonumber\\
					& = A_{1} +A_{2}, \label{lemma21}
				\end{align}
				for some $c_{2}, \, c_{3}, \,\zeta>0$.
				It is sufficient to bound $A_{1}$, then for a proper selection of $c_{2}>0$, we have that $(c_{2}/2)^2>1+\epsilon$, $\epsilon>0 $,  and $c>0$ we have that 
				\begin{align}\label{firstpartexponent}
					A_{1}
					=p c\exp\left(-c_{2}\left(\frac{\sqrt{\log{p}}}{4\sqrt{T}}\sqrt{T}\right)^{2} \right) 
					&= pc\left(\frac{1}{\exp(\log{(p)}(1+\epsilon))}\right)=cp^{-\epsilon}. 
				\end{align}
				Let Assumptions \textcolor{red}{1 -- 3,  5} and the conditions supporting Lemma \textcolor{red}{1}  hold, then Theorem 6.1 of \cite{buhlmann2011statistics} implies 
				\begin{align}
					\left	\lVert \boldsymbol{X}\left(\widetilde{\boldsymbol{\beta}}-{{\boldsymbol{\beta}}} \right)\right\rVert_{2}^{2}  \leq 4{\frac{\log{p}}{T}}s_{0}\phi_{0}^{-2}, \quad
					\left\lVert\widetilde{{\boldsymbol{\beta}}}-{\boldsymbol{\beta}}\right\rVert_1\leq 4\sqrt{\frac{\log{p}}{T}} s_{0}\phi_{0}^{-2} , \nonumber 
				\end{align}
				with probability at least $1- cp^{-\epsilon}$, for some positive values of  $\epsilon$ and  $\phi_{0}^{-2}\leq \frac{s_{0} \boldsymbol{\beta}'{E\left(\boldsymbol{x}_t\boldsymbol{x}_t' \right)}\boldsymbol{\beta} }{\left\lVert \boldsymbol\beta_{s_{0}} \right\rVert^{2}_{1}}$. 
			\end{proof}
			
			\begin{proof}[\unskip\nopunct]
				\textbf{Proof of Corollary \textcolor{red}{1}}. 		\textcolor{black}{In Corollary \textcolor{red}{1}  we  provide an asymptotic rate of convergence for the error between the estimated autoregressive parameters, $\boldsymbol{\widehat{\phi}}$ utilising $\widetilde{\beta}$ and the true autoregressive parameters, $\boldsymbol{\phi}$.   First we define the following 
					\begin{align}
						\begin{array}{l}
							\boldsymbol{u} = \left(u_{q+1}, \ldots, u_{T}\right)', \quad \label{uhat}
							\boldsymbol{ U} =\left[\begin{array}{cccc}
								u_{q}& u_{q-1} & \cdots & u_{1} \\
								u_{q+1}& u_{q} & \cdots & u_{2} \\
								\vdots & \vdots & \vdots & \vdots \\
								u_{T-1} & u_{T-2} & \cdots & u_{T-q}
							\end{array}\right]
						\end{array},
					\end{align}
					where $\boldsymbol{u}$ is a $\left(T-q \right)\times 1$ vector and $\boldsymbol{ U}$ is a $\left(	T-q	\right)\times q$ design matrix.  
					Note that  $\boldsymbol{\widetilde{\phi}} = (	\widetilde{\phi}_{1}, \ldots, \widetilde{\phi}_{q}	)'$ denotes the OLS estimate of the regression coefficients in the following regression: 
					\begin{align}
						\boldsymbol{		u} = \boldsymbol{	U} \boldsymbol{	\widetilde{\phi }} +\boldsymbol{	\varepsilon},
					\end{align}
					where $\boldsymbol\varepsilon = (\varepsilon_1, \ldots, \varepsilon_{T-q})'$,   follows Assumption \textcolor{red}{2}  and $q<\infty$. 
					Similarly to \eqref{uhat}, we define $	\boldsymbol{ \widehat{u}} $ and $	\boldsymbol{ \widehat{U}}$, and consider  $\boldsymbol{\widehat{\phi}} = (	{\widehat{\phi}} _{1},\ldots, {\widehat{\phi}} _{q}	)'$, the OLS estimates of an AR(q) regression,  using  $	(\boldsymbol{ \widehat{u}} ,	\boldsymbol{ \widehat{U}})$ instead of $(\boldsymbol{ {u}} ,{\boldsymbol{U}})$ and $\boldsymbol{\widehat{\phi}} $ instead of $ \boldsymbol{\widetilde{\phi}} $.}
				
				Then, we can write that 
				\begin{align}
					\boldsymbol{\widehat{\phi}} = \left(\boldsymbol{\widehat{U}}'  \boldsymbol{\widehat{U}}\right)^{-1}\boldsymbol{\widehat{U}}'\boldsymbol{\widehat{u}}  %
					, \quad \boldsymbol{\widetilde{\phi}} =  \left(\boldsymbol{{U}}'  \boldsymbol{{U}}\right)^{-1}\boldsymbol{{U}}'\boldsymbol{{u}} %
					,\quad {\boldsymbol{\phi}} = E\left[\left(\boldsymbol{{U}}'  \boldsymbol{{U}}\right)^{-1}\boldsymbol{{U}}'\boldsymbol{{u}} \right]%
					,\label{phis}
				\end{align}
				and \begin{align}\widehat{\sigma}^{2} =  \left(	\boldsymbol{\widehat{u}}	- \boldsymbol{\widehat{U}}\boldsymbol{\widehat{\phi}}	\right)'&\left(	\boldsymbol{\widehat{u}}	- \boldsymbol{\widehat{U}}\boldsymbol{\widehat{\phi}}	\right),\quad  \widetilde{\sigma}^{2} =  \left(	\boldsymbol{{u}}	- \boldsymbol{U}\boldsymbol{\widetilde{\phi}}	\right)' \left(	\boldsymbol{{u}}	- \boldsymbol{U}\boldsymbol{\widetilde{\phi}}	\right),\nonumber\\
					\sigma^{2}&= E\left[ \left(	\boldsymbol{{u}}	- \boldsymbol{U}\boldsymbol{{\phi}}	\right)' \left(	\boldsymbol{{u}}	- \boldsymbol{U}\boldsymbol{{\phi}}	\right)	\right].\nonumber
				\end{align}
				Notice that $\boldsymbol{\widehat{\phi}}-\boldsymbol{\phi} = \left(	\boldsymbol{\widehat{\phi}} - \boldsymbol{\widetilde{\phi}} \right)+\left(\boldsymbol{\widetilde{\phi}} - {\boldsymbol{\phi}}\right)$. 
				We are interested to show that the following holds
				\begin{align}
					\boldsymbol{\widehat{\phi}} -{\boldsymbol{\phi}}  = O_{P}\left(\frac{1}{\sqrt{T}}\right) . \label{statement}
				\end{align}
				To show \eqref{statement},  it is sufficient to show that 
				\begin{align}
					\boldsymbol{\widehat{\phi}} - \boldsymbol{\widetilde{\phi}}  = O_{P}\left(		s_{0}\frac{\log{p}}{T}\right)&, \quad \boldsymbol{\widetilde{\phi}} - {\boldsymbol{\phi}}= O_{P}\left(		T^{-1/2}\right), 	 \label{statement1}  \\
					T^{-1}\widehat{\sigma}^{2}\to {\sigma}^{2}&, \quad T^{-1}\widetilde{\sigma}^{2}\to {\sigma}^{2}	\label{statement2} .
				\end{align}
				\textit{Proof of \eqref{statement1}}. 
				To prove  the first two statements in \eqref{statement1} it is sufficient to show that
				\begin{align}
					T^{-1}\left(\sum_{t=T-k-j+1}^{T}\left(\widehat{u}_{t-j}\widehat{u}_t- {u}_{t-j}{u}_t\right)\right)&= O_{P}\left(		s_{0}^{2}\frac{{\log{p}}}{T}\right),\label{state3} \\
					T^{-1}\left(\sum_{t=T-k-j+1}^{T}(\widehat{u}^{2}_{t-j}- {u}^{2}_{t-j}) \right)&= O_{P}\left(  	s_{0}\frac{{\log{p}}}{T}\right)\label{state34}.
				\end{align}

				\noindent \textit{Proof of \eqref{state3}}.   	
				\eqref{state3} is bounded by $s_{1}+s_{2}+s_{3}$, where 
				\begin{align}
					s_{1}  & = 	T^{-1}\sum_{t=T-k-j+1}^{T}\left(\widehat{u}_t- {u}_t\right)  \left(\widehat{u}_{t-j}- {u}_{t-j}\right),\\
					s_{2} & =T^{-1}\sum_{t=T-k-j+1}^{T}\left(\widehat{u}_t- {u}_t\right) {u}_{t-j}, \quad 
					s_{3}  =	T^{-1}\sum_{t=T-k-j+1}^{T}{u}_t\left(\widehat{u}_{t-j}- {u}_{t-j}\right).
				\end{align}
				Notice that $	\widehat{{u}}_{t} ={u}_{t} + \boldsymbol{x}_t' \left({\widetilde{\boldsymbol{\beta}}}-{\boldsymbol{\beta}}\right)$, and  by Assumption \textcolor{red}{3}, $\{\boldsymbol{x}_t\}$, $\{u_{t}	\}$ are stationary, ergodic and  mutually independent,  $\alpha$-mixing series. Hence,  $E(\boldsymbol{x}_tu_{t}) = E(\boldsymbol{x}_t)E(u_{t})$ and  the following hold
				\begin{equation}
					\frac{1}{T}\sum_{t=T-k-j}^{T}\boldsymbol{x}_{t}\boldsymbol{x}_{t-j}' = O_{P}(1), \quad   \frac{1}{\sqrt{T}}\sum_{t=T-k-j}^{T}    u_{t}\boldsymbol{x}_{t-j}  =O_{P}(1), \quad \frac{1}{\sqrt{T}}\sum_{t=T-k-j}^{T} \boldsymbol{x}_{t}u_{t-j}  =O_{P}(1),\label{11}
				\end{equation}
				where $k=0,1,2,...$ and $j=1,\ldots, q.$ 
				Further, by Lemma \textcolor{red}{1}, we have that 
				\begin{align}
					\left\Vert{\boldsymbol{\widetilde{\beta}}}-{\boldsymbol{\beta}} \right\Vert_{1}  = O_{P}\left(	s_{0}\sqrt{\frac{\log{p}}{T}}	\right).\label{lem1Res}
				\end{align}
				Then by the Cauchy-Schwartz inequality and  substituting \eqref{11} and \eqref{lem1Res} in  \eqref{state3}, we obtain 
				{		\small	\begin{align}
						s_{1} &=  \left(\boldsymbol{\widetilde{\beta}} - \boldsymbol{ \beta}\right)'\left( T^{-1}  \sum_{t=T-k-j+1}^{T}\boldsymbol{x}_{t}\boldsymbol{x}_{t-j}'\right)\left(\boldsymbol{\widetilde{\beta}} - \boldsymbol{ \beta}\right)\nonumber\\
						&\leq		 \left\Vert \boldsymbol{\widetilde{\beta}} - \boldsymbol{ \beta}\right\Vert_{1}^{2}{\left( T^{-1}  \sum_{t=T-k-j+1}^{T}\boldsymbol{x}_{t}\boldsymbol{x}_{t-j}' \right)}
						=O_{P}\left(	s^{2}_{0}\frac{{\log{p}}}{T}\right).\nonumber 
				\end{align}}
				Next,
				\begin{align}
					s_{2} &= \left(\boldsymbol{\widetilde{\beta}} - \boldsymbol{ \beta}\right) \left(T^{-1}\sum_{t=T-k-j+1}^{T}\boldsymbol{x}_{t}u_{t-j}			\right)\nonumber\\
					&	\leq  \left\Vert \boldsymbol{\widetilde{\beta}} - \boldsymbol{ \beta}\right\Vert_{1} \left(T^{-1}\sum_{t=T-k-j+1}^{T}\boldsymbol{x}_{t}u_{t-j}			\right) = O_{P}\left(	s_{0}\frac{\sqrt{\log{p}}}{T}\right). \nonumber
				\end{align}			 
				Similarly, $s_{3}$ is bounded using the same arguments with $s_{2}$,
				which completes the proof of the first part of \eqref{statement1}.  			
				In view of the results in \eqref{state3} -- \eqref{state34} and the analysis in [8.3.17]-- [8.3.19] of \cite{hamilton2020time}, we can conclude that the second part of \eqref{statement1} holds. 
				
				\textit{Proof of \eqref{state34}}.  Using \eqref{11}--\eqref{lem1Res}, obtain  
				\begin{align}
					T^{-1}\left( {\sum_{t=T-k-j+1}^{T}\widehat{u}^{2}_{t-j}}\right)&= T^{-1}\Bigg( \sum_{t=T-k-j+1}^{T} {u}^{2}_{t-j} +  \left({\widetilde{\boldsymbol{\beta}}}-{\boldsymbol{\beta}}\right) '\boldsymbol{x}_{t-j}	\boldsymbol{x}_{t-j}' 	\left({\widetilde{\boldsymbol{\beta}}}-{\boldsymbol{\beta}}\right)\nonumber\\
					&	\quad \quad\quad\quad+2\boldsymbol{x}_{t-j}{u}_{t-j} \left({\widetilde{\boldsymbol{\beta}}}-{\boldsymbol{\beta}}\right)  \Bigg)
					\nonumber \\
					T^{-1}\left(\sum_{t=T-k-j+1}^{T} \widehat{u}^{2}_{t-j} -{u}^{2}_{t-j} \right)&\leq 2 \left\Vert{\widetilde{\boldsymbol{\beta}}}-{\boldsymbol{\beta}} \right\Vert_{1} ' \left(T^{-1}	\sum_{t=T-k-j+1}^{T} {u}_{t-j}\boldsymbol{x}_{t-j} \right)\nonumber\\
					& + \left\Vert \left(  T^{-1}\sum_{t=T-k-j}^{T}\boldsymbol{x}_{t-j}\boldsymbol{x}_{t-j}'\right)\left(	{\widetilde{\boldsymbol{\beta}}}-{\boldsymbol{\beta}}\right) \right\Vert_{2}^{2}\nonumber\\
					&= O_{P}\left(	s_{0}\frac{{\log{p}}}{T}\right). \nonumber
				\end{align}
				Together with \eqref{state3}--\eqref{state34}, we obtain 
				\begin{align}
					\boldsymbol{\widehat{\phi}} - \boldsymbol{\widetilde{\phi}}  = O_{P}\left(		s_{0}^{2}{\frac{\log{p}}{T}}	\right).
				\end{align}
				\textit{Proof of \eqref{statement2}}.    The convergence of both statements in \eqref{statement2} follows directly from the definitions of $\widehat{\sigma}^{2}$ and $\widetilde{\sigma}^{2}$ through statements \eqref{state3}, \eqref{state34} and the definitions of  $\boldsymbol{\widehat{\phi}},  \; \boldsymbol{\widetilde{\phi}}$, and $\boldsymbol{\phi}$. 
				
				Together with \eqref{statement1} and \eqref{statement2}, we obtain 
				\begin{align}
					\boldsymbol{ \widehat{\phi}} - \boldsymbol{\phi}=  O_{P}\left(	\frac{1}{\sqrt{T}}\right) \nonumber . 
				\end{align}
				For the latter to hold, it sufficient to note that $s_{0}^{2}{\frac{\log{p}}{T}}\lesssim T^{-1/2}$ , as $p,T\to \infty$. 
			\end{proof}
			\begin{corollary}\label{probabilisticRates}
				Under Assumptions \textcolor{red}{1 -- 3, 5}   the following holds
				\begin{align}
					P\left(\left\Vert   \boldsymbol{\widehat{\phi}} - \boldsymbol{\phi}\right\Vert_{1}>C	\right)
					&= o(1),\label{Ls}
				\end{align}
				for some  finite large enough constant $C>0$, where $\boldsymbol{\widehat{\phi}}$, $\boldsymbol{\phi}$ are defined in \eqref{phis}.
			\end{corollary}
			\begin{remark}
				Corollary \ref{probabilisticRates} is an implication of Corollary \textcolor{red}{1} and is useful to bound different quantities throughout the Supplement.  
			\end{remark}
			
			\begin{proof}[\unskip\nopunct]\textbf{Proof of Corollary \ref{probabilisticRates}}. 	
				Notice that $\boldsymbol{\widehat{\phi}}-\boldsymbol{\phi} = \left(	\boldsymbol{\widehat{\phi}} - \boldsymbol{\widetilde{\phi}} \right)+\left(\boldsymbol{\widetilde{\phi}} - {\boldsymbol{\phi}}\right)$, where $\boldsymbol{\widetilde{\phi}}, \; \boldsymbol{\widehat{\phi}}$ and $\boldsymbol{\phi}$ are defined in \eqref{phis}, then 
				\begin{align}
					P\left(\sqrt{T}\left\Vert   \boldsymbol{\widehat{\phi}} - \boldsymbol{\phi}\right\Vert_{1}>C	\right)\leq P\left(\sqrt{T}\left\Vert   \boldsymbol{\widehat{\phi}} - \boldsymbol{\widetilde{\phi}}\right\Vert_{1}>\frac{C}{2}\right) +
					P\left(\sqrt{T}\left\Vert   \boldsymbol{\widetilde{\phi}} - \boldsymbol{\phi}\right\Vert_{1}>\frac{C}{2}	\right) .
				\end{align}
				To show \eqref{Ls}, it is sufficient to show that the following holds: 
				\begin{align}
					P\left(\sqrt{T}\left\Vert   \boldsymbol{\widehat{\phi}} - \boldsymbol{\widetilde{\phi}}\right\Vert_{1}>\frac{C}{2}\right) =o(1), \; \text{and }\; 
					P\left(\sqrt{T}\left\Vert   \boldsymbol{\widetilde{\phi}} - \boldsymbol{\phi}\right\Vert_{1}>\frac{C}{2}	\right) 			=o(1). \label{LS2}
				\end{align} 
				
				Consider $\boldsymbol{		u}, \; \boldsymbol{		\widehat{u}}, \; \boldsymbol{	U}, \; \boldsymbol{		\widehat{U}}$,  and $\boldsymbol{	\widehat{\phi }}, \; \boldsymbol{	\widetilde{\phi }}, \; \boldsymbol{	{\phi }}$ defined in \eqref{uhat} -- \eqref{phis}.  
				In the remainder of our analysis we use the following matrix inequalities from Chapter 8 of \cite{lutkepohl1997handbook};  $\Vert \boldsymbol{A}\Vert_1\leq \sqrt{m}\Vert \boldsymbol{A}\Vert_F$, $\Vert \boldsymbol{AB}\Vert_F \leq \Vert \boldsymbol{A}\Vert_F\Vert \boldsymbol{B}\Vert_F\leq \Vert \boldsymbol{A}\Vert_F\Vert \boldsymbol{B}\Vert_1$ since  $\Vert \boldsymbol{B}\Vert_F\leq \Vert \boldsymbol{B}\Vert_1$,
				for an $m\times n$ matrix $\boldsymbol{A}$ and an $n\times {l}$ matrix $\boldsymbol{B}$, where $\lambda_{\min}(\boldsymbol{A}'\boldsymbol{A}),\; \lambda_{\min}(\boldsymbol{B}'\boldsymbol{B})>0$, $(\boldsymbol{A}'\boldsymbol{A})^{-1}, \; (\boldsymbol{B}'\boldsymbol{B})^{-1}$ exist.   We start by showing the first part of \eqref{LS2}: 
				{\small	\begin{align}
						P\left(\sqrt{T}\left\Vert   \boldsymbol{\widehat{\phi}} - \boldsymbol{\widetilde{\phi}}\right\Vert_{1}>\frac{C}{2}	\right)	& \leq 	P\left(\sqrt{T}\left\Vert \left(\frac{\boldsymbol{\widehat{U}}'  \boldsymbol{\widehat{U}}}{T}\right)^{-1}\frac{\boldsymbol{\widehat{U}}'\boldsymbol{\widehat{u}} }{T}- \left(\frac{\boldsymbol{{U}}'  \boldsymbol{{U}}}{T}\right)^{-1}\frac{\boldsymbol{{U}}'\boldsymbol{{u}} }{T}\right\Vert_{F}>{C}_{T}\right) \nonumber\\
						&\leq  	P\left(\sqrt{T}\left\Vert  \left(\frac{\boldsymbol{{U}}'  \boldsymbol{{U}}}{T}\right)^{-1} \frac{\left(\boldsymbol{\widehat{U}} -\boldsymbol{{U}}\right) '  \boldsymbol{{u}}}{T}\right\Vert_F	>\frac{C_T}{5}	\right)\\
						&+  P\left(\sqrt{T}\left\Vert 	\left[ 	 \left(\frac{\boldsymbol{\widehat{U}}'  \boldsymbol{\widehat{U}}}{T}\right)^{-1}- 	\left(\frac{\boldsymbol{{U}}'  \boldsymbol{{U}}}{T}\right)^{-1} \right]\frac{ \left(\boldsymbol{\widehat{U}}-\boldsymbol{{U}} \right) '(\boldsymbol{\widehat{u}}- \boldsymbol{u})}{T}\right\Vert_F>\frac{C_T}{5}\right)\\
						&+P\left(\sqrt{T}\left\Vert 	\left[ 	 \left(\frac{\boldsymbol{\widehat{U}}'  \boldsymbol{\widehat{U}}}{T}\right)^{-1}- 	\left(\frac{\boldsymbol{{U}}'  \boldsymbol{{U}}}{T}\right)^{-1} \right] \frac {\boldsymbol{{U}} '(\boldsymbol{\widehat{u}}- \boldsymbol{u})}{T}\right\Vert_F>\frac{C_T}{5}\right)\\
						&+  P\left(	\sqrt{T}\left\Vert		\left(\frac{\boldsymbol{{U}}'  \boldsymbol{{U}}}{T}\right)^{-1} \frac{\left(\boldsymbol{\widehat{U}} -\boldsymbol{{U}}\right) '(\boldsymbol{\widehat{u}}- \boldsymbol{u})}{T}	\right\Vert_F	>\frac{C_T}{5}\right)\\
						&+P\left(\sqrt{T}\left\Vert 	\left(\frac{\boldsymbol{{U}}'  \boldsymbol{{U}}}{T}\right)^{-1}  \frac{\boldsymbol{{U}}  '(\boldsymbol{\widehat{u}}- \boldsymbol{u})}{T}\right\Vert_F>\frac{C_T}{5}\right)=\sum_{i=1}^5 A_i ,
				\end{align}}
				for some $C_T = \frac{C}{2}(T-q)^{-1/2}$, $0<q<\infty$ and $C$ a large enough positive and finite constant.
				It is sufficient to show that $A_1, \; A_3$ and $A_5$ are sufficiently small for some large generic positive constant $C_T$. We start with $A_1$:
				\begin{align}
					A_1& =	P\left(\sqrt{T}\left\Vert  \left(\frac{\boldsymbol{{U}}'  \boldsymbol{{U}}}{T}\right)^{-1} \frac{\left(\boldsymbol{\widehat{U}} -\boldsymbol{{U}}\right) '  \boldsymbol{{u}}}{T}\right\Vert_F	>\frac{C_T}{5}	\right)\nonumber\\
					&\underset{(1)}{\leq}P\left(\sqrt{T}\left\Vert \boldsymbol{\widehat{\Sigma}}_{U}^{-1}\right\Vert_{F} \left\vert \frac{\left(\boldsymbol{\widetilde{\beta}} -\boldsymbol{\beta}\right)'\boldsymbol{X}' \boldsymbol{{u}}}{T}\right\vert		> \frac{C_T}{5}	\right)\nonumber\\
					&\underset{(2)}{\leq}  P\left(\sqrt{T}\left\Vert \boldsymbol{\widehat{\Sigma}}_{U}^{-1}\right\Vert_{F} \left\Vert \boldsymbol{\widetilde{\beta}} -\boldsymbol{\beta}\right\Vert_{1} \left\Vert \frac{\boldsymbol{X}' \boldsymbol{{u}}}{T}\right\Vert_\infty		> \frac{C_T}{5}	\right)\nonumber\\
					&\leq P\left(\left\Vert \boldsymbol{\widehat{\Sigma}}_{U}^{-1}\right\Vert_{F} >\frac{C_T}{5C_1}\right) + P\left(\sqrt{T}\left\Vert \boldsymbol{\widetilde{\beta}} -\boldsymbol{\beta}\right\Vert_{1} \left\Vert \frac{\boldsymbol{X}' \boldsymbol{{u}}}{T}\right\Vert_\infty		> C_1	\right)
					, \label{1}
				\end{align}
				where  $\boldsymbol{\widehat{\Sigma}}_{U} = \boldsymbol{{U}}'  \boldsymbol{{U}}/T$,  $\boldsymbol{\widehat{\Sigma}}_{U}^{-1}$ exists and $\boldsymbol{{U}}'  \boldsymbol{{U}}$ is positive definite.  Step $(1)$ results by using  (B.60) in Lemma A11  of \cite{chudik2018one}, while  $(2)$ results from the inequality where $\vert \boldsymbol{a}'\boldsymbol{b}\vert \leq \Vert \boldsymbol{a}\Vert_1\Vert \boldsymbol{b}\Vert_\infty$, for $\boldsymbol{a}, \; \boldsymbol{b}$ two $m\times 1$ vectors.     
				
				It is of interest to show that 
				\begin{align}
					P\left(\left\Vert  \boldsymbol{\widehat{\Sigma}}^{-1}-  \boldsymbol{\Sigma}_{U}^{-1}\right\Vert_{F} >\frac{C_T}{5C_1} \right)=o(1),
				\end{align}
				where $\boldsymbol{\Sigma}_{U}^{-1} = E[\boldsymbol{{U}}'\boldsymbol{{U}}]$.
				Note that  due to positive definiteness of $\boldsymbol{{U}}'  \boldsymbol{{U}}$ and consequent existence of $\boldsymbol{\widehat{\Sigma}}_{U}^{-1}$, it is equivalent to show that 
				\begin{align}
					P\left(
					\lVert	\boldsymbol{\widehat{\Sigma}}^{-1}\rVert_{F}
					\left\Vert   \boldsymbol{\widehat{\Sigma}} -  \boldsymbol{\Sigma}_{U}\right\Vert_{F} \lVert  \boldsymbol{\Sigma}_{U} ^{-1}\rVert_{F} >\frac{C_T}{5C_1} \right)=o(1),\label{sigmaU}
				\end{align}
				since 
				$
				\Vert  \boldsymbol{\widehat{\Sigma}}^{-1}-  \boldsymbol{\Sigma}_{U}^{-1}\Vert_{F} \leq 
				\lVert	\boldsymbol{\widehat{\Sigma}}^{-1}\rVert_{F}
				\Vert   \boldsymbol{\widehat{\Sigma}} -  \boldsymbol{\Sigma}_{U}\Vert_{F} \lVert  \boldsymbol{\Sigma}_{U} ^{-1}\rVert_{F}
				$
				holds.   To show $\eqref{sigmaU}$, it sufficient to show that 
				\begin{align}
					&P\left(
					\left\Vert   \boldsymbol{\widehat{\Sigma}} -  \boldsymbol{\Sigma}_{U}\right\Vert_{F} >\frac{C_T}{5C_1} \right)=o(1),
				\end{align}
				which holds, following a similar analysis to Lemma \ref{vcov}. 
				
				Further, to show that $A_1=o(1)$ it is  sufficient to show that 
				\begin{align}
					P\left(\sqrt{T} \left\Vert \frac{ \boldsymbol{{u}} '\boldsymbol{X}}{T} \right\Vert_\infty	\left\Vert \boldsymbol{\widetilde{\beta}} -\boldsymbol{\beta}\right\Vert_{1} 	> C_1	\right)=o(1),\label{btildeXu}
				\end{align}
				for a large enough $C_1>0.$ Note that 
				\begin{align}
					P\left(\sqrt{T}\left\Vert \boldsymbol{\widetilde{\beta}} -\boldsymbol{\beta}\right\Vert_{1} \left\Vert \frac{\boldsymbol{X}' \boldsymbol{{u}}}{T}\right\Vert_\infty		> C_1	\right)\underset{(1)}{\leq} P\left(  \left\Vert  \boldsymbol{\widetilde{\beta}} -\boldsymbol{\beta}\right\Vert_1> \frac{C_1}{C_2}\right) +    P\left(\frac{1}{\sqrt{T}}  \left\Vert \boldsymbol{X} '\boldsymbol{{u}} \right\Vert_\infty>C_2\right) ,
				\end{align}
				where $(1)$ results by using  (B.60) in Lemma A11  of \cite{chudik2018one}, for some $C_2>0.$ By the arguments of Lemma  \textcolor{red}{1}, \eqref{btildeXu} holds,  if $C_2\leq \frac{\lambda_0}{2}$ and $\frac{C_1}{C_2}\leq T^{-3/2}\sqrt{\log{p}}$, for some $\lambda_{0}={T^{-1/2}\log^{1/2}{p}}$.
				
				We proceed to analyse $A_3$: 
				\begin{align}
					A_3&=	P\left(\sqrt{T}\left\Vert 	\left[ 	 \left(\frac{\boldsymbol{\widehat{U}}'  \boldsymbol{\widehat{U}}}{T}\right)^{-1}- 	\left(\frac{\boldsymbol{{U}}'  \boldsymbol{{U}}}{T}\right)^{-1} \right] \frac {\boldsymbol{{U}} '(\boldsymbol{\widehat{u}}- \boldsymbol{u})}{T}\right\Vert_F>\frac{C_T}{5}\right) \\
					& \leq  P\left(\sqrt{T}\left\Vert 	\left[		 \left(\frac{\boldsymbol{\widehat{U}}'  \boldsymbol{\widehat{U}}}{T}\right)^{-1}- 	\left(\frac{\boldsymbol{{U}}'  \boldsymbol{{U}}}{T}\right)^{-1} \right] \right\Vert_{F}   \left\Vert  \frac{\boldsymbol{{U}} '\boldsymbol{X}  \left(\boldsymbol{\widetilde{\beta}} -\boldsymbol{\beta}\right) }{T}    \right\Vert_1>\frac{C_T}{5}\right) \\
					&\leq  P\left(\sqrt{T}\left\Vert 	 \left(\frac{\boldsymbol{\widehat{U}}'  \boldsymbol{\widehat{U}}}{T}\right)^{-1}- 	\left(\frac{\boldsymbol{{U}}'  \boldsymbol{{U}}}{T}\right)^{-1} \right\Vert_{F} >\frac{C_T}{5C_0}\right) \\
					&+  
					P\left(  \sqrt{T}\left\Vert\frac{\boldsymbol{{U}} '\boldsymbol{X}  \left(\boldsymbol{\widetilde{\beta}} -\boldsymbol{\beta}\right)    }{T}  \right\Vert_1>C_0\right) = A_{3,1}+A_{3,2}.
				\end{align}
				To show that $A_3=o(1),$  it is sufficient to show that $A_{3,1}=o(1)$ and $A_{3,2}=o(1)$, for some $C_0>0$.  
				Considering  Lemma A16 of \cite{chudik2018one} and the fact that $\boldsymbol{{U}}'  \boldsymbol{{U}}$ is positive definite,  it is sufficient to bound the following quantity as a lower bound for $A_{3,1}$
				\begin{align}
					A_{3,1}&=P\left(\sqrt{T}\left\Vert 	 \left(\frac{\boldsymbol{\widehat{U}}'  \boldsymbol{\widehat{U}}}{T}\right)^{-1}- 	\left(\frac{\boldsymbol{{U}}'  \boldsymbol{{U}}}{T}\right)^{-1} \right\Vert_{F} >\frac{C_T}{5C_0}\right)\nonumber\\
					& \leq
					P\left(\sqrt{T}\left\Vert \frac{\boldsymbol{\widehat{U}}'\left( \boldsymbol{\widehat{U}}- 	\boldsymbol{{U}}\right) }{T}
					+
					\frac{\boldsymbol{{U}}'\left( \boldsymbol{\widehat{U}}- 	\boldsymbol{{U}}	\right)}{T} \right\Vert_{F} >\frac{C_T}{5C_0}\right) \nonumber \\
					&\leq  
					P\left(\sqrt{T}\left\Vert \frac{\boldsymbol{\widehat{U}}'\left( \boldsymbol{\widehat{U}}- 	\boldsymbol{{U}}\right) }{T} \right\Vert_{F}>\frac{C_T}{10C_0}\right) 
					+P\left(\sqrt{T}\left\Vert \frac{\boldsymbol{{U}}'\left( \boldsymbol{\widehat{U}}- 	\boldsymbol{{U}}	\right)}{T}  \right\Vert_{F} >\frac{C_T}{10C_0}\right).\nonumber 
				\end{align}
				To show that $A_{3,1}=o(1)$, it is sufficient to show that 
				$P\left(\sqrt{T}\Vert  \boldsymbol{{U}}'( \boldsymbol{\widehat{U}}- 	\boldsymbol{{U}}	)/T \Vert_{F} >\frac{C_T}{10C_0}\right) = o(1)$: 
				
				\begin{align}
					P\left(\sqrt{T}\left\Vert  \frac{\boldsymbol{{U}}'\left( \boldsymbol{\widehat{U}}- 	\boldsymbol{{U}}	\right) }{T}\right\Vert_{F} >\frac{C_T}{10C_0}\right)
					&\underset{}{\leq }P\left(  \left\Vert  \boldsymbol{\widetilde{\beta}} -\boldsymbol{\beta}\right\Vert_1>\frac{C_T}{10C_1C_0} \right)\nonumber\\
					&\quad +  \max_{j=1,\ldots,q}   P\left(  \left\Vert\boldsymbol{{u}}_j '\boldsymbol{X}    \right\Vert_\infty>C_1\sqrt{T}\right). \label{analysisa32}
				\end{align}
				The inequality holds  by using  (B.60) in Lemma A11  of \cite{chudik2018one} for some positive constants, $C_0,C_1.$  By the arguments of Lemma \textcolor{red}{1},  $A_{3,2}=o(1)$, if $C_1\leq \frac{\lambda_0}{2}$ and $\frac{{C}_T}{10C_1C_0} < T^{-3/2}\sqrt{\log{p}}$.  Therefore  $A_3=o(1)$.  
				
				Finally, we analyse $A_5$: 
				\begin{align}
					A_5&=P\left(\sqrt{T}\left\Vert 	\left(\frac{\boldsymbol{{U}}'  \boldsymbol{{U}}}{T}\right)^{-1}  \frac{\boldsymbol{{U}}  '(\boldsymbol{\widehat{u}}- \boldsymbol{u})}{T}\right\Vert_F>\frac{C_{T}}{5}\right)\\
					& \leq  P\left(\left\Vert  \boldsymbol{\Sigma}_{U}^{-1}\right\Vert_{F} >\frac{C_T}{5C_2}\right) + \max_{j=1,\ldots,q}P\left(\left\Vert \boldsymbol{\widetilde{\beta}} -\boldsymbol{\beta}\right\Vert_{1} \left\Vert \boldsymbol{X}' \boldsymbol{{u}}_j \right\Vert_\infty		> {C_2}{\sqrt{T}}	\right),
				\end{align}
				for some $C_2>0.$  \textcolor{black}{The analysis follows similarly to \eqref{1} using the results of  \eqref{analysisa32}, showing that $A_5=o(1)$, which completes the proof of the first part of \eqref{LS2}.   The proof of the second part follows a similar line of arguments, therefore is not included. }
			\end{proof}
			
			\begin{proof}[\unskip\nopunct] 
				\textbf{Proof of Theorem \textcolor{red}{1}} 
				In this Theorem, we illustrate that the feasible \emph{GLS Lasso} estimator, attains similar non-asymptotic bounds to the \emph {Lasso}, both in terms of prediction and estimation errors. The proof follows closely  the steps in Chapter 6 of \cite{buhlmann2011statistics}. We consider the feasible  GLS corrected model:
				\begin{align*}
					{\widetilde{y}_t} = \boldsymbol{\widetilde{x}}_t'{\boldsymbol{\beta}}+\widehat{\varepsilon}_t, \quad \text{where} 
				\end{align*}
				\begin{equation*}
					\widetilde{{y}}_t  =  y_{t}- \sum_{j=1}^{q}\widehat{\phi}_j y_{t-j}, \quad \widetilde{\boldsymbol{x}}_t  = \boldsymbol{x}_{t} - \sum_{j=1}^{q} \widehat{\phi}_j\boldsymbol{x}_{t-j} , \quad  \widehat{\varepsilon}_t  = u_{t}- \sum_{j=1}^{q}\widehat{\phi}_ju_{t-j}, \quad t = q+1, \ldots, T,
				\end{equation*}
				where $\widehat{\phi}_{j}$ is the OLS estimate obtained from an $AR(q)$ regression on the residuals $\widehat{u}_{t}$, where $\widehat{u}_t = y_t - \boldsymbol{x}'_t \boldsymbol{\widetilde{\beta}}$,   $\boldsymbol{\widetilde{\beta}}$ the solution to the original \emph{Lasso} problem, using $\left(\boldsymbol{y},\boldsymbol{X}\right)$. Recall from Corollary \textcolor{red}{1}  that 	$\sum_{j=1}^{q}(\widehat\phi_j - \phi_j)=O_{P}(T^{-1/2})$.			Re-arranging the basic inequality, as in Lemma 6.1 of \cite{buhlmann2011statistics}, we obtain
				\begin{align}
					\frac{1}{T}\left\lVert\boldsymbol{\widetilde{X}}\left(\boldsymbol{\widehat{\beta}}-{{\boldsymbol{\beta}}}\right)	\right\rVert^2_2 \leq \frac{2}{T}{\widehat{\boldsymbol{\varepsilon}}'} \boldsymbol{\widetilde{X}}\left(\boldsymbol{{\widehat{\beta}} }-{\boldsymbol{\beta}}\right) + \lambda \left( \lVert{\boldsymbol{\beta}}\rVert_{1}  -  \lVert\boldsymbol{\widehat{\beta}}\rVert_{1} \right). \label{errorTilde}
				\end{align}
				Define ${2}{\widehat{\boldsymbol{\varepsilon}}'} \boldsymbol{\widetilde{X}}\left(\boldsymbol{\widehat{\beta}}-{\boldsymbol{\beta}}\right)/T$ as the "empirical process".
				Notice that the latter can  be bounded further in terms of  the $\ell_{1}$-norm, such that, 
				\begin{equation}\label{maxfirstElem2}
					\frac{1}{T}\left|{\widehat{\boldsymbol{\varepsilon}}'} \boldsymbol{\widetilde{X}}\left(\boldsymbol{\widehat{\beta}}-{\boldsymbol{\beta}}\right)\right| \leq \frac{2}{T}\left\lVert{\widehat{\boldsymbol{\varepsilon}}'} \boldsymbol{\widetilde{X}}\right\rVert_{\infty}\left\lVert \boldsymbol{\widehat{{\beta}}}-{\boldsymbol{\beta}}\right\rVert_{1}.
				\end{equation}
				The regularisation parameter, $\lambda$ is chosen such that ${T^{-1}}\left\lVert{\widehat{\boldsymbol{\varepsilon}}'} \boldsymbol{\widetilde{X}}\right\rVert_{\infty}\leq \lambda$. Hence, we introduce the following event 
				\begin{equation}\label{event2}
					\widetilde{\mathcal{E}}:= \left\{ {T^{-1}}\left\lVert{\widehat{\boldsymbol{\varepsilon}}'} \boldsymbol{\widetilde{X}}\right\rVert_{\infty}\leq \frac{\lambda_{0}}{2} \right\},
				\end{equation}
				which needs to hold with high probability, where $\lambda_{0}= \sqrt{\frac{\log{p}}{T}}\leq\lambda/2$.      To illustrate the latter, we  use the vector representation of the process ${\widehat{\boldsymbol{\varepsilon}}'} \boldsymbol{\widetilde{X}}$, and  proceed with the following steps:
				{\small	\begin{align}
						\frac{2}{T} \sum_{t=1}^{T}{\widehat{\varepsilon}}_t \boldsymbol{\widetilde{x}}_t'\left(\boldsymbol{\widehat{\beta}}-{{\boldsymbol{\beta}} } \right)&=\frac{2}{T}\sum_{t= 1}^{T}\left(u_{t}-\sum_{j=1}^{q}\widehat{\phi}_{j}u_{t-j}\right)\widetilde{\boldsymbol{x}}_{t}'\left(\boldsymbol{\widehat{\beta}}-{{\boldsymbol{\beta}} } \right) \nonumber \\
						&= \frac{2}{T}\sum_{t=1}^{T}\left[u_{t}\widetilde{\boldsymbol{x}}_{t}' \left(\boldsymbol{\widehat{\beta}}-{{\boldsymbol{\beta}} }\right) +\widetilde{\boldsymbol{x}}_{t}'\left(\boldsymbol{\widehat{\beta}}-{{\boldsymbol{\beta}} } \right) \sum_{j=1}^{q}\widehat{\phi}_{j}u_{t-j}\right]\nonumber\\
						&=\frac{2}{T}\sum_{t=1}^{T}\left[(\varepsilon_{t}+\sum_{j=1}^{q}\phi_{j} u_{t-j} ) \widetilde{\boldsymbol{x}}_{t}'\left(\boldsymbol{\widehat{\beta}}-{{\boldsymbol{\beta}} }\right) +\widetilde{\boldsymbol{x}}_{t}' \left(\boldsymbol{\widehat{\beta}}-{{\boldsymbol{\beta}} } \right)\sum_{j=1}^{q}\widehat{\phi}_{j}u_{t-j} \right]\nonumber \\
						&= \frac{2}{T}\sum_{t=1}^{T} \left[ \left(\widetilde{\boldsymbol{x}}_{t}\sum_{j=1}^{q}\phi_{j} u_{t-j} + \varepsilon_{t} \widetilde{\boldsymbol{x}}_{t}+ \widetilde{\boldsymbol{x}}_{t}\sum_{j=1}^{q}\widehat{\phi}_{j}u_{t-j}\right)\left(\boldsymbol{\widehat{\beta}}-{{\boldsymbol{\beta}} }\right)\right]\nonumber\\
						&	\leq \underset{i=1,\ldots,p}{\max} \left|\frac{2}{T}\left[\sum_{t=1}^{T}\left(\widetilde{{x}}_{t,i}\sum_{j=1}^{q}\phi_{j} u_{t-j} + \varepsilon_{t} \widetilde{{x}}_{t,i}+ \widetilde{{x}}_{t,i}\sum_{j=1}^{q}\widehat{\phi}_{j}u_{t-j}\right)\right]\right| \left\lVert \boldsymbol{\widehat{\beta}}-{\boldsymbol{\beta}}\right\rVert_{1} \label{empProcessTilde}.
				\end{align}}
				Then, we create the set $\widetilde{\mathcal{E}}$, under which \eqref{maxfirstElem2} holds.
				\begin{align}
					\widetilde{\mathcal{E}}& =\left\{  \underset{i=1,\ldots,p}{\max} \left|\frac{1}{T}\sum_{t=1}^{T}\left[\widetilde{{x}}_{t,i}\sum_{j=1}^{q}\phi_{j} u_{t-j} + \varepsilon_{t} \widetilde{{x}}_{t,i}+ \widetilde{{x}}_{t,i}\sum_{j=1}^{q}\widehat{\phi}_{j}u_{t-j}\right]\right|\leq \frac{\lambda_{0}}{2}\right\}\\
					&=  \left\{\overset{p}{\underset{i=1}{\bigcup}}\left| \frac{1}{T}\left[\sum_{t=1}^{T}\widetilde{{x}}_{t,i}\sum_{j=1}^{q}\phi_{j} u_{t-j} + \sum_{t=1}^{T}\varepsilon_{t} \widetilde{{x}}_{t,i}+ \sum_{t=1}^{T}\widetilde{{x}}_{t,i}\sum_{j=1}^{q}\widehat{\phi}_{j}u_{t-j}\right]\right|\leq \frac{\lambda_{0}}{2}\right\}. \label{ev}
				\end{align}
				Using the identity $P(\widetilde{\mathcal{E}}^{c}) = 1-P(\widetilde{\mathcal{E}})$ and the union bound, we get 
				\begin{align}
					P(\widetilde{\mathcal{E}} ^{c}) &= 1-	P\left(\overset{p}{\underset{i=1}{\bigcup}} \left|\frac{1}{T}\left[\sum_{t=1}^{T}\widetilde{{x}}_{t,i}\sum_{j=1}^{q}\phi_{j} u_{t-j} + \sum_{t=1}^{T}\varepsilon_{t} \widetilde{{x}}_{t,i}+ \widetilde{{x}}_{t,i}\sum_{j=1}^{q}\widehat{\phi}_{j}u_{t-j}\right]\right|\leq \frac{\lambda_{0}}{2}	\right)\nonumber\\
					&\leq 1- \sum_{i}P \left( \left| \frac{1}{T}\left[ \sum_{t=1}^{T}\widetilde{{x}}_{t,i}\sum_{j=1}^{q}\phi_{j} u_{t-j} +\sum_{t=1}^{T} \varepsilon_{t} \widetilde{{x}}_{t,i}+\sum_{t=1}^{T} \widetilde{{x}}_{t,i}\sum_{j=1}^{q}\widehat{\phi}_{j}u_{t-j}\right]\right|\leq \frac{\lambda_{0}}{2} \right)\nonumber\\
					&\leq \sum_{i}\left\{P \left(  \left|\frac{1}{T}\sum_{t=1}^{T}{x}_{t,i}\sum_{j=1}^{q} \left(\widehat{\phi}_{j}- {\phi}_j\right) u_{t-j} \right|>\frac{\lambda_{0}}{6}\right)+ P \left(  \frac{1}{T}\left|\sum_{t=1}^{T} \varepsilon_{t}{x}_{t,i}\right| >\frac{\lambda_{0}}{6}\right) \right.\nonumber\\
					&\left.\quad  \quad \quad+ P \left(  \left|\frac{1}{T}\sum_{t=1}^{T}\sum_{j=1}^{q}\widehat{\phi}_{j}{x}_{t-j,i}\left(\widehat{\phi}_{j}- {\phi}_j\right)u_{t-j}\right| >\frac{\lambda_{0}}{6}\right)\right\}= (I)+(II)+(III)\label{expoall}.  
				\end{align}
				We proceed to analyse each term.   Define $\{ \boldsymbol{z}_{t}\} = \{\boldsymbol{x}_{t}\sum_{j=1}^{q} u_{t-j} \}$ which is a $p$-dimensional, zero-mean, stationary and ergodic $\alpha$-mixing series,  as a product of two $\alpha$-mixing series by Assumption \textcolor{red}{3}  and using Theorem 14.1 of \cite{davidson1994stochastic},   with properties similar to \textcolor{red}{(5)}.  For $(I)$ and $i=1,\ldots,p$, we have: 
				\begin{align}
					&\sum_i	P \left(  \left| \frac{1}{T}\sum_{t=1}^{T}{x}_{t, i}\sum_{j=1}^{q}\left(\widehat{\phi}_{j}- {\phi}_j\right) u_{t-j} \right|>\frac{\lambda_{0}}{6}\right)\nonumber\\
					& \leq \sum_i	 P \left(  \left|\frac{1}{\sqrt T}\sum_{t=1}^{T} z_{t, i}\right|>\frac{\lambda_{0}{\sqrt{T}}}{6C}\right) 
					+ P \left(  \left|  \sum_{j=1}^{q} \left(\widehat{\phi}_{j}- {\phi}_j\right) \right|>C\right) \nonumber \\
					&	\leq \sum_i	P \left(  \left|\frac{1}{\sqrt T}\sum_{t=1}^{T}{z}_{t,i} \right|>\frac{\sqrt{\log{p}}}{6C}\right)+o(1)\nonumber\\
					&\leq  p c\left\{ \exp	\left(-c_{2}\frac{\log{p}}{36C^{2}}\right) +  \exp\left({-c_{3}\frac{\sqrt{\log{p}}}{6C\log^{2}T}}\right)^{\zeta}	\right\}
					= I_1+I_2 \label{(I)}, 
				\end{align}
				where we use Lemma 1 of \cite{DGK} to obtain  \eqref{(I)}.    It is then sufficient to bound $I_{1}$: 
				\begin{align}
					I_1=pc\exp	\left(-c_{2}\frac{\log{p}}{36C^{2}}\right) &=c p^{1-\left(\frac{c_{2}}{36C^{2}}\right)} ,
				\end{align}
				for some large enough constant, $c_{2}>0$,  and positive constants, $c, C$. 
				For $i=1,\ldots,p$, we have,  $$\sum_i	P \left(  \left|\frac{1}{\sqrt T}\sum_{t=1}^{T}{z }_{t, i} \right|>\frac{\sqrt{\log{p}}}{6C}\right)\leq c p^{1- \left(\frac{c_{2}}{36C^{2}}\right)} = o\left(p^{1-\left(\frac{c_{2}}{36C^{2}}\right)}\right).$$
				We proceed to  analyse $(II)$.  Define $\{\boldsymbol{v}_{t} \} = \{\boldsymbol{x}_{t}\varepsilon_t \}$ as a m.d.s.  by Lemma \ref{mds}. By direct application of  Lemma A3 of  \cite{chudik2018one}  we obtain 
				\begin{align}
					\sum_i		P \left(  {T^{-1}}\left|\sum_{t=1}^{T} {v}_{t,i}\right| >\frac{\lambda_{0}}{6}\right)&\leq p  \exp\left[	-	c_0\left(\frac{\sqrt{T\log{p}}}{6}\right)^{\frac{s}{(s+1)}}	   \right]\\
					&{\leq}p \exp \left[  - \frac{c_0}{\sqrt{6}}  p^{\frac{1}{4\delta}} \log^{\frac{1}{4}}\left(p \right)\right]\leq p^{1-\left(\frac{c_{2}}{36C^{2}}\right)},
				\end{align} 
				where	$s=1$ and $\delta>0$, as $p\to \infty$.  
				
				To  analyse $(III)$, recall that
				$\{\sum_{j=1}^{q}{\phi}_{j}\boldsymbol{x}_{t-j}u_{t-j}\}$ is $\alpha$-mixing as  a product of  two $\alpha$-mixing series, by Theorem 14.1 of \cite{davidson1994stochastic}, then define $\{\boldsymbol{\zeta}_{t} \}  = \{\sum_{j=1}^{q}\boldsymbol{x}_{t-j}u_{t-j}\}$, $ \{\boldsymbol{\eta}_{t} \}  = \{\sum_{j=1}^{q}\phi_{j}\boldsymbol{x}_{t-j}u_{t-j}\}$. Then:
				{\small	\begin{align}
						&\sum_i	P \left(  \left|\frac{1}{T}  \sum_{t=1}^{T}\sum_{j=1}^{q}\widehat{\phi}_{j}{x}_{t-j, i} \left(\widehat{\phi}_{j} - {\phi}_{j} \right)u_{t-j}\right| >\frac{\lambda_{0}}{6}\right)\nonumber  \\ 
						&=\sum_i	P \left(  \left|\frac{1}{T}  \sum_{t=1}^{T}\sum_{j=1}^{q}\left[\left(\widehat{\phi}_{j}- {\phi}_{j}\right){x}_{t-j, i} \left(\widehat{\phi}_{j} - {\phi}_{j} \right)u_{t-j} +\phi_{j}{x}_{t-j, i} \left(\widehat{\phi}_{j} - {\phi}_{j} \right)u_{t-j} \right] \right| >\frac{\lambda_{0}}{6}\right)\nonumber \\
						&=	\sum_i	P \left(  \left|\frac{1}{T}  \sum_{t=1}^{T}\sum_{j=1}^{q}\left[\left(\widehat{\phi}_{j}- {\phi}_{j}\right){x}_{t-j, i} \left(\widehat{\phi}_{j} - {\phi}_{j} \right)u_{t-j} 		\right] \right| >\frac{\lambda_{0}}{12}\right)	\nonumber\\
						&\quad  +\sum_i		P \left(  \left|\frac{1}{T}  \sum_{t=1}^{T}\sum_{j=1}^{q}\left[\phi_{j}{x}_{t-j, i} \left(\widehat{\phi}_{j} - {\phi}_{j} \right)u_{t-j}		\right] \right| >\frac{\lambda_{0}}{12}\right)	 	\nonumber \\
						&\underset{(1)}{\leq }	\sum_i	\left[		P \left(  \left|\frac{1}{T}  \sum_{t=1}^{T}{\zeta}_{t, i} \right| >\frac{\lambda_{0}}{12C_{0}}\right)	P \left(  \left|\frac{1}{T}  \sum_{t=1}^{T}{\eta}_{t, i}	\right| >\frac{\lambda_{0}}{12C_{1}}	\right) \right] \nonumber\\
						&\quad +	P \left(  \left|  \sum_{j=1}^{q}\left(\widehat{\phi}_{j}- {\phi}_{j}\right)^{2}\right|> C_{0}	\right)+ P \left(  \left| \sum_{j=1}^{q}\left(\widehat{\phi}_{j}- {\phi}_{j}\right)\right|> C_{1}	\right)  = B_{1}\times B_{2}+ B_{3}+B_{4}.\label{last1}
				\end{align}}
				Note that $(1)$ results by direct 	application of Lemma A.11, equation (B.60) of \cite{chudik2018one}.  
				Notice that  by Corollary \ref{probabilisticRates} $B_{3}= o(1), \; B_{4}=o(1)$, $B_{1}$ and $B_{2}$  follow similar line of arguments with term $(I)$ of \eqref{expoall}, therefore $$B_{1} = o\left(	p^{-\frac{c_{4}}{144C^{2}_{0}}}	\right), \quad\text{and } \; B_{2} = o\left(	p^{-\frac{c_{5}}{144C^{2}_{1}}}	\right),$$ for some  large enough and positive constants $c_{4}, c_{5}$ and small $C_{0}, C_{1}>0$, as $T\to \infty$.   
				Combining the results from \eqref{(I)}--\eqref{last1}  we have that 
				\begin{align} 
					P(\widetilde{\mathcal{E}} ^{c}) \leq c\left[ 2p^{1-\frac{c_{2}}{(6C)^{2}}} 
					+ 	p^{1-\frac{c_{4}}{(12C_{0})^{2}} }+ p^{1-\frac{c_{5}}{(12C_{1})^{2}} }\right].  \label{prob}
				\end{align}
				Since $2p^{1-\frac{c_{2}}{(6C)^{2}}} 
				\leq p^{1-\frac{c_{4}}{(12C_{0})^{2}} }$ and for some  large enough positive constant, $c_{4}>0$, we obtain 
				\begin{align}
					P(\widetilde{\mathcal{E}} ^{c})& \leq c	 p^{1-\frac{c_{4}}{(12C_{0})^{2}} }\to 0 ,\; 
					P(\widetilde{\mathcal{E}}) = 1- p^{1-\frac{c_{4}}{(12C_{0})^{2}} } \to 1 .  
				\end{align} 
				Note that the "prediction error", 
				$\left  \lVert  \boldsymbol{\widetilde{X}} \left(\boldsymbol{\widehat{\beta}}- {\boldsymbol\beta} \right) \right \rVert_{2}^{2}$ is $\ell_2$-bounded and $\left\lVert\boldsymbol{\widehat{\beta}}- {\boldsymbol\beta}\right\rVert_1$ is $\ell_1$-bounded, then by Corollary 6.2 of \cite{buhlmann2011statistics}
				\begin{align}
					\frac{1}{T}\left\lVert \boldsymbol{\widetilde{X}}(\boldsymbol{\widehat{\beta}}- {\boldsymbol\beta})	\right\rVert^2_2 &\leq \frac{2}{T} {\widehat{\boldsymbol{\varepsilon}}} '\boldsymbol{\widetilde{X}}\left(\boldsymbol{\widehat{\beta}}- {\boldsymbol\beta} \right)	+ \lambda \left( \left \lVert{\boldsymbol{\beta}} \right \rVert_{1}  -   \lVert \boldsymbol{\widehat{\beta}} \rVert_{1} \right), \\
					&\leq \left\lVert{\boldsymbol{\widehat{\beta}}}-{\boldsymbol{\beta}}\right\rVert_{1}+ \lambda \left( \lVert{\boldsymbol{\beta}}\rVert_{1}  -  \lVert \boldsymbol{\widehat{\beta}}\rVert_{1} \right)\label{pred}  .
				\end{align}
				Since $\lambda\geq2\lambda_{0}$ under $\widetilde{\mathcal{E}}$, and by Assumption \textcolor{red}{4}, 
				\begin{align}
					\left	\lVert \boldsymbol{\widehat{\beta} } - \boldsymbol{\beta} \right \rVert_{1} &= \left \lVert \boldsymbol{\widehat{\beta}}_{S_{0}} - \boldsymbol{\beta}_{S_{0}} \right \rVert_{1} + \left \lVert \boldsymbol{\widehat{\beta}}_{S^{c}_{0}} \right \rVert_{1}, \quad  \left \lVert \boldsymbol{\widehat{\beta}}_{S^{c}_{0}} \right \rVert_{1} \leq 3 \left \lVert \boldsymbol{\widehat{\beta}}_{S_{0}} - \boldsymbol{\beta}_{S_{0}} \right \rVert_{1}  \label{b1}\\
					\left  \lVert \boldsymbol{\widehat{\beta}}_{S_{0}} - \boldsymbol{\beta}_{S_{0}} \right \rVert_{1}&\leq\sqrt{s_{0}} \left \lVert \boldsymbol{\widehat{\beta}}_{S_{0}} - \boldsymbol{\beta}_{S_{0}} \right \rVert_{2},\quad
					\left  \lVert\boldsymbol{\widehat{\beta}} \right \rVert_{1}= \left  \lVert \boldsymbol{\widehat{\beta}}_{S_{0}} \right \rVert_{1}+ \left \lVert \boldsymbol{\widehat{\beta}}_{S_{0}^{c}} \right \rVert_{1}\\
					\left \lVert  \boldsymbol{\beta}_{S_{0}} \right \rVert^{2}_{1}&\leq \left(\boldsymbol{\beta}'{\boldsymbol{{\Sigma}}}\boldsymbol{\beta} \right)s_{0}\zeta^{-2}_{*}\left(s_{0}, \boldsymbol{\phi} \right), \label{b2} 
				\end{align}
				where $\boldsymbol{{\Sigma}}= E({  \boldsymbol{\widetilde{x}}_t  \boldsymbol{\widetilde{x}}}_t')$ and $\zeta^{-2}_{*}\left(s_{0}, \boldsymbol{\phi} \right)>0$ defined in Assumption \textcolor{red}{4}.
				Then, substituting  \eqref{b1}--\eqref{b2} in 
				\eqref{pred}, we obtain the following  dual bound 
				\begin{align}
					\left\lVert  \boldsymbol{\widetilde{X}} \left(\boldsymbol{\widehat{\beta}} - \boldsymbol{\beta} \right) \right\rVert_{2}^{2} +\left \lVert \boldsymbol{\widehat{\beta}} - \boldsymbol{\beta}\right\rVert_{1}&\leq \left\lVert  \boldsymbol{\widetilde{X}}\left(\boldsymbol{\widehat{\beta}}- {\boldsymbol\beta} \right)	\right\rVert_{2}^{2}+\left\lVert \boldsymbol{\widehat{\beta}}_{S_{0}} - \boldsymbol{\beta}_{S_{0}}\right\rVert_{1} + 3\left\lVert \boldsymbol{\widehat{\beta}}_{S_{0}} - \boldsymbol{\beta}_{S_{0}}\right\rVert_{1}\nonumber  \\
					&\leq  4s_{0}\zeta^{-2}_{*}(s_{0}, \boldsymbol{\phi})\frac{\log{p}}{T}, \nonumber  \end{align}
				which leads to the result:
				\begin{align}
					\left\lVert  \boldsymbol{\widetilde{X}} \left(\boldsymbol{\widehat{\beta}}- {\boldsymbol\beta} \right)	\right\rVert^{2}_{2} &\leq   4\zeta^{-2}_{*}(s_{0}, \boldsymbol{\phi})s_{0} \frac{\log{p}}{T}, \label{prediction}\\
					\left\lVert \boldsymbol{\widehat{\beta}}- {\boldsymbol\beta}	\right\rVert_{1} & \leq 4\zeta^{-2}_{*}(s_{0}, \boldsymbol{\phi})s_{0}\sqrt{\frac{\log{p}}{T}} \label{coefficient},
				\end{align}
				with probability at least $1- c p^{1-c_{4}/ (12C)^{2}}$, for some $c,C>0$ and a large enough constant $c_{4}>0$,   obtained in \eqref{prob}.
			\end{proof}
			\begin{proof}[\unskip\nopunct] 
				\textbf{Proof of Corollary \textcolor{red}{2} } \\
				The proof follows directly from Theorem 1 of \cite{raskutti2010restricted}. Consider $s_0=o(T\log^{-1}{p})$, then \eqref{prediction}, \eqref{coefficient} hold with probability at least $1- c p^{1-\frac{c_{4}}{ (12C)^{2}}}$, for some $c,C>0$ and a large enough constant $c_{4}>0$, e.g. Theorem \textcolor{red}{1}.
			\end{proof}
		}
		
		\section{Proofs of Theorems 2 and 3 of the main paper}\label{appendixb}
		This Section provides proofs of Theorems \textcolor{red}{2}, and \textcolor{red}{3} of the main paper. 
		\begin{proof} [\unskip\nopunct] \textbf{Proof of Theorem \textcolor{red}{2}}
			Consider Assumptions \textcolor{red}{1 -- 4} to hold.    Then, with a proper selection of $\lambda< \sqrt{T^{-1}\log{p}}$, the weak irrepresentable condition proposed by \cite{zhao2006model}, holds, hence 
			$
			P\left(	\exists \lambda>0:\;\boldsymbol{\widehat{\beta}} = \boldsymbol{\beta}	\right)\to 1
			$, where $S_{0}$ is defined in \textcolor{red}{(9)} of the main paper, and $s_{0}=|S_0|$,  implying that $P\left(\left|\widehat{S}_{0} /\  S_{0} \right|	\right)\to 1$, where ${S}_{0}= \{ i : \; \beta_{i}\neq 0 \},\; \widehat{S}_{0}= \{ i : \; \widehat\beta_{i}\neq 0 \}$.  This can be easily confirmed by directly  applying  Theorem 4 of \cite{zhao2006model}.  For the sake of clarity we  provide the proof below: 
			
			Let $\boldsymbol{\beta}_{s_{0}}:=\beta_{i}\; \boldsymbol{1}\left\{i \in S_{0};\; i=1, \ldots, p\right\}$, and  $ \boldsymbol{\beta}_{s_{0}^{c}}:=\beta_{i} \;\boldsymbol{1}\left\{i \notin S_{0}, i=1, \cdots, p\right\}$.  Denote $\widetilde{\boldsymbol{X}}(S_{0})$,  $\boldsymbol{\widetilde{X}}(S_{0}^{c})$ as the first $s_{0}$ and last $p-s_0$ columns of $\boldsymbol{\widetilde{X}}$ respectively and let $\boldsymbol{\widehat{{\Sigma}}}= \boldsymbol{\widetilde{X}} '\boldsymbol{\widetilde{X}}/T$. By setting $\boldsymbol{\widehat{{\Sigma}}}_{11}=\boldsymbol{\widetilde{X}}(S_0)' \boldsymbol{\widetilde{X}}(S_0)/T,  \boldsymbol{\widehat{{\Sigma}}}_{22}= \boldsymbol{\widetilde{X}}(S_0^{c})'  \boldsymbol{\widetilde{X}}(S_0^{c})/T, \boldsymbol{\widehat{{\Sigma}}}_{12}= \boldsymbol{\widetilde{X}}(S_0)' \boldsymbol{\widetilde{X}}(S_0^{c})/T,$ and $\boldsymbol{\widehat{{\Sigma}}}_{21}= \boldsymbol{\widetilde{X}}(S_0^{c})'  \boldsymbol{\widetilde{X}}(S_0)/T$.   $  \widehat{\boldsymbol\Sigma}$ can then be expressed in a block-wise form as follows:
			$$\boldsymbol{\widehat{{\Sigma}}}=\left[\begin{array}{cc}
				\boldsymbol{\widehat{{\Sigma}}}_{11} & \boldsymbol{\widehat{{\Sigma}}}_{12} \\
				\boldsymbol{\widehat{{\Sigma}}}_{21}& \boldsymbol{\widehat{{\Sigma}}}_{22}
			\end{array}\right].$$
			We then define two distinct events,
			\begin{align}
				A_{T} &=\left\{\left|\left(\boldsymbol{\widehat{\sigma}}_{11, i}\right)^{-1} w_i(S_0)\right|<\sqrt{T}\left(\left|{\beta}_{i}\right|-\frac{\lambda}{2 T}\left|\left(\boldsymbol{\widehat{\sigma}}_{i1,1}\right)^{-1} \operatorname{sign}\left({\beta}_{i}\right)\right|\right)\right\},\quad i=1,\ldots, s_0,  \nonumber \\
				B_{T} &=\left\{\left|\omega_i - w_i(S_0^{c})\right| \leq \frac{\lambda}{2 \sqrt{T}}\right\}, \quad i=s_0+1,\ldots, p,\nonumber 
			\end{align}
			where
			$$
			\boldsymbol{\omega}= \left(\boldsymbol{\widehat{{\Sigma}}}_{21}\left(\boldsymbol{\widehat{{\Sigma}}}_{11}\right)^{-1} \boldsymbol{\widetilde{X}}(S_0)'\widehat{\boldsymbol{\varepsilon}}\right), \quad W(S_0)=\frac{1}{\sqrt{T}} \boldsymbol{\widetilde{X}}(S_0)'\widehat{\boldsymbol{\varepsilon}}, \quad  \frac{1}{\sqrt{T}} W(S_0^{c})=\boldsymbol{\widetilde{X}}(S_0^{c})'\widehat{ \boldsymbol{\varepsilon}}.
			$$
			Event $A_T$ implies that the signs of the active set, $S_0$, are correctly estimated, while $A_T, B_T$ together imply that the signs of the non-active set, $S_{0}^{c}$,  are estimated consistently.
			
			To show $P\left(	\exists \lambda>0:\;\boldsymbol{\widehat{\beta}} = \boldsymbol{\beta}	\right)\to 1$, it is sufficient to show that 
			\begin{align}
				P\left(	\exists \lambda>0:\;\boldsymbol{\widehat{\beta}} = \boldsymbol{\beta}	\right) \geq P\left(	A_{T}	\cap B_T\right). \label{sign}
			\end{align}
			Using the Identity of $1-P\left(	A_{T}	\cap B_T\right)\leq   P\left(	A_{T}\right)^{c}+ 	P\left(B_{T}\right)^{c}$ we have that 
			\begin{align}
				P\left(	A_{T}\right)^{c}+ 	P\left(B_{T}\right)^{c} & \leq 
				\sum_{i\in S_0} P\left(\frac{1}{\sqrt{T}} \left|\left(\boldsymbol{\widehat{\sigma}}_{11, i}\right)^{-1} \boldsymbol{\widetilde{x}}_i'\widehat{\boldsymbol{\varepsilon}}\right|\geq\sqrt{T}\left|{\beta}_{i}\right|-\frac{\lambda}{2 T}\left|\left(\boldsymbol{\widehat{\sigma}}_{11, i}\right)^{-1} \rm{sign}\left({\beta}_{i}\right)\right|\right)\nonumber \\
				&+ \sum_{i\in S_{0}^{c}} P\left(\frac{1}{\sqrt{T}}\left|\omega_{i}-\boldsymbol{\widetilde{x}}_i'\widehat{\boldsymbol{\varepsilon}}\right| \leq \frac{\lambda}{2 \sqrt{T}} \right)= A+B,  \label{sign2}
			\end{align}
			where  $\boldsymbol{\omega}$ is a $(p-s_0)\times 1$ vector. 
			Notice that $\forall \; i, j = 1,2$ and by Assumption \textcolor{red}{4},  $0<\lambda_{\min }\left(\boldsymbol{\widehat{{\Sigma}}}_{ij}\right)\leq \lambda_{\max } \left(	\boldsymbol{\widehat{{\Sigma}}}_{ij}\right)$ holds , 
			hence 
			\begin{align}
				\frac{\lambda}{2 T}\left|\left(\boldsymbol{\widehat{\sigma}}_{11, i}\right)^{-1} \operatorname{sign}\left({\beta}_{i}\right)\right| \leq
				\frac{\lambda}{2c_0 T} \left\lVert     \rm{sign}\left({\beta}_{i}\right) \right\rVert_{2} \leq
				\sqrt{s_0}\frac{\lambda}{2c_0 T},  \nonumber  
			\end{align}
			for some positive constant $c_{0}$.  Denote $ \boldsymbol{ \Sigma }^{\star}  =  \boldsymbol{ X }'\boldsymbol{ L }'  \boldsymbol{ L }\boldsymbol{ X }$, $\widehat{\boldsymbol{\varepsilon} }= \boldsymbol{\widehat{L}}\boldsymbol{u}$, and ${\boldsymbol{\varepsilon} ^{\star}}= \boldsymbol{{L}}\boldsymbol{u}$ where $\boldsymbol{ L }$ is defined in \eqref{L}, then $A$ of \eqref{sign2} becomes: 
			\begin{align}
				A&=\sum_{i\in S_0} P\left(\frac{1}{\sqrt{T}} \left|\left(\boldsymbol{\widehat{\sigma}}_{11, i}\right)^{-1} \boldsymbol{\widetilde{x}}_i'\widehat{\boldsymbol{\varepsilon}}     -  \left(\boldsymbol{\widehat{\sigma}}_{11, i}\right)^{-1} \boldsymbol{\widetilde{x}}_i'{\boldsymbol{\varepsilon}^{\star}}    \right|\geq\sqrt{T}\left|{\beta}_{i}\right|-\frac{\lambda\sqrt{s_0}}{8c_0 T}\right)\nonumber\\
				&\quad + \sum_{i\in S_0} P\left(\frac{1}{\sqrt{T}} \left|\left(\boldsymbol{\widehat{\sigma}}_{11, i}\right)^{-1} \boldsymbol{\widetilde{x}}_i'{\boldsymbol{\varepsilon}}   - 
				\left(\boldsymbol{\sigma}^{\star}_{11, i}\right)^{-1}\boldsymbol{\widetilde{x}}_i'{\boldsymbol{\varepsilon}}^{\star}   
				\right|\geq\sqrt{T}\left|{\beta}_{i}\right|-\frac{\lambda\sqrt{s_0}}{8c_0 T}\right)\nonumber \\
				&\quad + \sum_{i\in S_0} P\left(\frac{1}{\sqrt{T}} \left|\left(\boldsymbol{{\sigma}}_{11, i}^{\star}\right)^{-1} \boldsymbol{\widetilde{x}}_i'{\boldsymbol{\varepsilon}^{\star}}   - 
				\left(\boldsymbol{\sigma}^{\star}_{11, i}\right)^{-1}{\boldsymbol{{x}}^{\star}_i}'\boldsymbol{\varepsilon}^{\star}   
				\right|\geq\sqrt{T}\left|{\beta}_{i}\right|-\frac{\lambda\sqrt{s_0}}{8c_0 T}\right)\nonumber \\
				&\quad + \sum_{i\in S_0} P\left(\frac{1}{\sqrt{T}} \left|
				\left(\boldsymbol{\sigma}^{\star}_{11, i}\right)^{-1}{\boldsymbol{{x}}^{\star}_i}'\boldsymbol{\varepsilon}^{\star}   
				\right|\geq\sqrt{T}\left|{\beta}_{i}\right|-\frac{\lambda\sqrt{s_0}}{8c_0 T}\right) = A_1+A_2+A_3+A_4.\nonumber
			\end{align}
			Terms $A_{1}, \ldots, A_{4}$ are bounded following similar analysis as in Lemma \ref{Lemmathetaxu}, hence $A\leq A_{4} $.  Further,  notice that similarly with  $\boldsymbol{\widehat{{\Sigma}}}_{ij}$, $\forall \; i, j = 1,2$, $0<\lambda_{\min }\left(\boldsymbol{{{\Sigma}}}_{ij}^{\star}\right)\leq \lambda_{\max } \left(	\boldsymbol{{{\Sigma}}}_{ij}^{\star}\right)$.   Therefore by Lemma \ref{mds}, $\{\boldsymbol{c}_{t}\}=\{  \boldsymbol{x}_{t}^{\star}\varepsilon_{t}   \}$ is a m.d.s,  and by Lemma A3 of \cite{chudik2018one}, $$A_4\leq \sum_{i\in s_{0}}P\left(\frac{1}{\sqrt{T}} \left|
			{c}_{t, i}
			\right|\geq\frac{\left[\sqrt{T}\left|{\beta}_{i}\right|-\frac{\lambda\sqrt{s_0}}{16c_0 T}\right]}{C}\right)    = o\left(	s_{0}p^{-\frac{1}{2d\sqrt{s_0}}}		\right), $$  for some large enough constants $d,C>0$.   Similar analysis is conducted for $B$, concluding that 
			\begin{align}
				P\left(	\exists \lambda>0:\;\boldsymbol{\widehat{\beta}} = \boldsymbol{\beta}	\right)= o\left(	s_{0}p^{-\frac{1}{2d\sqrt{s_{0}}}}	\right). \label{signterm}
			\end{align}
			It remains to show \textcolor{red}{(21)}:   When $\widehat{S}_{0} = S_0$, $ \boldsymbol{\ddot{\beta}}_{S_0} = \boldsymbol{\widehat{\beta}}_{S_{0}} $, the latter can only differ when $\widehat{S}_{0} \neq  S_0$,  which is asymptotically negligible by \eqref{signterm}.
		\end{proof}
		\begin{proof} [\unskip\nopunct] \textbf{Proof of Theorem \textcolor{red}{3}. }
			We show that 
			\begin{align}
				\boldsymbol	{t}_{s}= \frac{\sqrt{T}	\left(\widehat{ \boldsymbol{b}} - \boldsymbol{\beta} \right) }{\sqrt{\widehat{{\boldsymbol\Theta}}_{i}\widehat{{\boldsymbol{\Sigma}}}_{xu}\widehat{{\boldsymbol\Theta}}'_{i}}}  \sim \mathcal{N}(\boldsymbol{0},\boldsymbol{I}_{p}),
			\end{align}
			where $\boldsymbol{\widehat{b}}$ is defined in \textcolor{red}{(26)} and $\widehat{{\boldsymbol\Theta}}_{i}$ is a  $1\times p$ vector from $\boldsymbol{\widehat{\Theta}}$ defined in  \textcolor{red}{(33)} of the main paper. By \textcolor{red}{(42)} we can write 
			\begin{align}
				\boldsymbol{t}_{s} = A+B ,\quad \text{where}
			\end{align} 
			\begin{align}
				{	A = \frac{\widehat{\boldsymbol{\Theta}}_{i}\widetilde{\boldsymbol{X}}'{ \widehat{\boldsymbol{L}}\boldsymbol{			{u}}}}{\sqrt{T\widehat{{\boldsymbol{\Theta}}}_{i}\widehat{{\boldsymbol{\Sigma}}}_{xu}\widehat{{\boldsymbol{\Theta}}}'_{i}}}, \quad B = -  \frac{\boldsymbol{{\delta}}}{\sqrt{\widehat{{\boldsymbol{\Theta}}}_{i}\widehat{{\boldsymbol{\Sigma}}}_{xu}\widehat{{\boldsymbol{\Theta}}}'_{i}}},\quad }
				\boldsymbol{\delta}= \sqrt{T} \left(\widehat{  \boldsymbol{{\Theta}}}\widehat{\boldsymbol{{\Sigma}} } - \boldsymbol{I}_{(p\times p)}\right) \left(\widehat{  \boldsymbol{{\beta}}} - \boldsymbol{\beta}\right)\nonumber,
			\end{align}
			where $ \widehat{\boldsymbol{L}}$ is defined in \eqref{L}. 
			It is sufficient to show that $A\sim \mathcal{N} \left(\boldsymbol{0},\boldsymbol{I}_p \right)$ and $B = o_{P}\left(1 \right)$.    We start by showing that $A\sim \mathcal{N} \left(\boldsymbol{0},\boldsymbol{I}_p \right)$. We denote 
			$ 
			\boldsymbol{\widehat{\Sigma}}_{xu} =(\boldsymbol{\widetilde{X}}'\boldsymbol{\widehat{L}}\boldsymbol{u})(\boldsymbol{\widetilde{X}}'\boldsymbol{\widehat{L}}\boldsymbol{u})'/T, 
			$ $\boldsymbol{\Sigma}_{xu}^{\star}  = (\boldsymbol{{X}}'\boldsymbol{{L}}\boldsymbol{u})(\boldsymbol{{X}}'\boldsymbol{{L}}\boldsymbol{u})'/T$.  To show that $A\sim \mathcal{N} \left(\boldsymbol{0},\boldsymbol{I}_p \right)$, we first need to show that 
			
			{
				\begin{align}	
					A' = \frac{\boldsymbol{\Theta}_{i}^{\star}{\boldsymbol{	X}^{\star}}'\boldsymbol{	\varepsilon}^{\star}}{\sqrt {T \boldsymbol{\Theta}^{\star}_{i}{{\boldsymbol{\Sigma}}}^{\star}_{xu}{\boldsymbol{\Theta}^{\star}_i}'}}\sim \mathcal{N} \left(\boldsymbol{0},\boldsymbol{I}_p \right), \; \text{and } A-A'=o_{P}(1).
				\end{align}
				where ${\boldsymbol{	X}^{\star}}'\boldsymbol{	\varepsilon}^{\star}={\boldsymbol{X}}'{ {\boldsymbol{L'Lu}}}.$
				As a first step we show that the nominator and denominator of $A$ is asymptotically equivalent to their corresponding quantities of $A'$.  Starting with the nominator, by Lemma \ref{Lemmathetaxu}.
				\begin{align}
					\frac{1}{\sqrt{T}}\left\Vert\widehat{\boldsymbol{\Theta}}_{i}\widetilde{\boldsymbol{X}}'{\boldsymbol{\widehat{\varepsilon}}} - \boldsymbol{\Theta}^{\star}_{i}{\boldsymbol{X}^{\star}}'\boldsymbol{\varepsilon}^{\star}\right\Vert_1 =o_P(1)\label{AApprox},
				\end{align}
				where $\widehat{\boldsymbol{\varepsilon} }= \boldsymbol{\widehat{L}}\boldsymbol{u}$, and ${\boldsymbol{\varepsilon} ^{\star}}= \boldsymbol{{L}}\boldsymbol{u}$ where $\boldsymbol{ L }$ is defined in \eqref{L}.
				Further, for the denominator, by Lemma \ref{identity}
				\small \begin{align}\label{thetApprox}
					\left|	\widehat{ \boldsymbol{\Theta}}_{i}\widehat{\boldsymbol{\Sigma}}_{xu}\widehat{ \boldsymbol{\Theta}}_{i}'-	
					\boldsymbol{ \Theta}^{\star}_{i}\boldsymbol{\Sigma}^{\star}_{xu}{\boldsymbol{ \Theta}^{\star}_{i}}'	\right| =o_P(1).
				\end{align}
				Hence, by  \eqref{AApprox}--\eqref{thetApprox}, $A-A'=o_P(1).$}
			
			{
				We  now show that $A'\sim \mathcal{N} \left(\boldsymbol{0},\boldsymbol{I}_p \right)$. 
				We remark that   $	\boldsymbol{\Theta}^{\star}_{i}{{\boldsymbol{\Sigma}}}^{\star}_{xu}{\boldsymbol{\Theta}^{\star}_i}$ is asymptotically bounded away from zero (positive definite), such that the following statement  holds:
				\begin{align}
					\boldsymbol{\Theta}^{\star}_{i}{{\boldsymbol{\Sigma}}}^{\star}_{xu}{\boldsymbol{\Theta}^{\star}_i}\geq \lambda_{\min}({{\boldsymbol{\Sigma}}}^{\star}_{xu})\Vert \boldsymbol{\Theta}\Vert_{2}^{2}\geq  \lambda_{\min}({{\boldsymbol{\Sigma}}}^{\star}_{xu})\lambda^{2}_{\min}(\boldsymbol{\Theta}^{\star})\geq \lambda_{\min}({{\boldsymbol{\Sigma}}}^{\star}_{xu})\lambda^{-2}_{\min}(\boldsymbol{\Theta}^{\star}),\label{eigval}
				\end{align}
				where $\lambda_{\min}({{\boldsymbol{\Sigma}}}^{\star}_{xu})$ the smallest eigenvalue of ${{\boldsymbol{\Sigma}}}^{\star}_{xu}$, and $\lambda_{\min}(\boldsymbol{\Theta}^{\star})$ the smallest eigenvalue of $\boldsymbol{\Theta}$, which obey   $0<\lambda_{\min}(\boldsymbol{\Theta}^{\star})\leq\lambda_{\max}(\boldsymbol{\Theta}^{\star})<\infty$.
				Then,  by consequence of  Assumption \textcolor{red}{3}, $\{{\boldsymbol{x}_{t}^{\star}}' \varepsilon^{\star}_{t}\}$ is a zero-mean stationary process, by consequence  $E \left[	({\boldsymbol{{x}}^{\star}_{1}}'\varepsilon^{\star}_{1})({\boldsymbol{{x}}^{\star}_{1}}'\varepsilon^{\star}_{1})'	\right] = \rm{Var}[{\boldsymbol{{x}}^{\star}_{1}}'\varepsilon^{\star}_{1}] >0$,  $E \left[{\boldsymbol{{x}}^{\star}_{1}}'\varepsilon^{\star}_{1}	\right]=0 $.
				Taking the expectation of $A'$ and $({A'})^{2}$ we obtain
				\begin{align}
					E\left[ 	\frac{{\boldsymbol{\Theta}}^{\star}_{i}  {\boldsymbol{x}_{t}^{\star}}' \varepsilon^{\star}_{t} }{\sqrt{T{{\boldsymbol{\Theta}}}^{\star}_{i}{{\boldsymbol{\Sigma}}}^{\star}_{xu}{{{\boldsymbol{\Theta}}}^{\star}_{i}}'}}\right] &= 0\label{condition24.62},\\
					E\left[  	\frac{{\boldsymbol{\Theta}}^{\star}_{i}  {\boldsymbol{x}_{t}^{\star}}' \varepsilon^{\star}_{t} }{\sqrt{T{{\boldsymbol{\Theta}}}^{\star}_{i}{{\boldsymbol{\Sigma}}}^{\star}_{xu}{{{\boldsymbol{\Theta}}}^{\star}_{i}}'}}
					\right]^{2} &=  E\left[   \frac{T^{-1}\left({\boldsymbol{\Theta}}^{\star}_{i}  {\boldsymbol{x}_{t}^{\star}}' \varepsilon^{\star}_{t}\right)'\left({\boldsymbol{\Theta}}^{\star}_{i}  {\boldsymbol{x}_{t}^{\star}}' \varepsilon^{\star}_{t}\right)} {T^{-1} \left({\boldsymbol{\Theta}}^{\star}_{i}  {\boldsymbol{x}_{t}^{\star}}' \varepsilon^{\star}_{t}\right)'\left({\boldsymbol{\Theta}}^{\star}_{i}  {\boldsymbol{x}_{t}^{\star}}' \varepsilon^{\star}_{t}\right)}
					\right] = 1\label{condition24.61}. 
				\end{align} 
				In view of Theorem 24.6 of \cite{davidson1994stochastic}, $A'$ is asymptotically standard normal, across $t=1,\ldots,T$.
			}
			{
				It remains to show that $B=o_{P}(1)$.  The denominators of $A,\, B$ are identical, so by \eqref{eigval}, the denominator of $B$ is asymptotically positive definite.   It suffices to show that 
				$
				\boldsymbol{\delta} = \sqrt{T}(\widehat{{\boldsymbol\Theta}}\boldsymbol{\widehat{{\Sigma}}}-\boldsymbol{I})(\boldsymbol{\widehat{\beta}}-\boldsymbol{\beta}) 
				$
				is asymptotically negligible.   As an implication of Proposition \ref{theta},  $\lVert \widehat{{\boldsymbol\Theta}}_{i}\boldsymbol{\widehat{{\Sigma}}}-\boldsymbol{e}_{i}\rVert_{\infty} = o_P(\lambda_i)$ and by Theorem \textcolor{red}{1}, $\lVert  \boldsymbol{\widehat{\beta}}-\boldsymbol{\beta}  \rVert_{1}=O_P(s_0 \log^{1/2}{p}T^{-1/2})$, therefore 
				\begin{align}
					\left\Vert  \boldsymbol{\delta}\right\Vert_{1} \leq \lVert{\boldsymbol{{\delta}} }\rVert_{\infty} =\left\lVert \sqrt{T}(\widehat{{\boldsymbol\Theta}}\boldsymbol{\widehat{{\Sigma}}}-\boldsymbol{I})(\boldsymbol{\widehat{\beta}}-\boldsymbol{\beta})\right\rVert_{\infty}\leq \sqrt{T}\underset{i= 1,\ldots,p}{\max}\left\lVert \widehat{{\boldsymbol\Theta}}_{i}\boldsymbol{\widehat{{\Sigma}}}-\boldsymbol{e}_{i}\right\rVert_{\infty}\left\lVert  \boldsymbol{\widehat{\beta}}-\boldsymbol{\beta}  \right\rVert_{1}= o_{P}(1),\nonumber 
				\end{align}	
				showing that $B=o_P(1)$. 
			}
		\end{proof}
		\section{Proofs of auxiliary  results}\label{secondary res}
		This section contains auxiliary technical lemmas used in the proofs of Section \ref{appendixb}.
		\begin{lemma}\label{mds}
			Let the processes $\{ \boldsymbol{x}_{t} \}$, $ \{ \varepsilon_{t}\} $ be series of r.v.'s with properties outlined in Assumption \textcolor{red}{1},   \textcolor{red}{2}  and Assumption \textcolor{red}{3.2} respectively and $1\leq t\leq T$.  Let $\mathcal{F}_{t-1}^{(1)}=\sigma\left(\left\{\varepsilon_{s}\right\}_{s=1}^{t-1},\left\{\boldsymbol{x}_{s}\right\}_{s=1}^{t-1}\right)$ and $\mathcal{F}_{t}^{(2)}=\sigma\left(\left\{\varepsilon_{s}\right\}_{s=1}^{t-1},\left\{\boldsymbol{x}_{s}\right\}_{s=1}^{t}\right)$.  Further, consider either that (i) $E\left(\varepsilon_{t} \mid \mathcal{F}_{t}^{(2)}\right)=0$ or (ii) $E\left(\boldsymbol{x}_{t} \varepsilon_{t}-\mu_{t} \mid \mathcal{F}_{t-1}^{(1)}\right)=0$, where $\mu_{t}=E\left(\boldsymbol{x}_{t} \varepsilon_{t}\right) =0.$    Then the $p$-dimensional series $\{\boldsymbol{x}_{t}\varepsilon_{t}\} $ is a martingale difference sequence (m.d.s.).				\end{lemma}
		\begin{proof}[\unskip\nopunct] 
			\textbf{Proof of Lemma \ref{mds}}.  
			Let $\mathcal{F}_{t-1} = \sigma\left(\{  \boldsymbol{x}_{s}\varepsilon_{s} \}^{t-1}_{s=1}\right)$, notice that under (i) and Assumption \textcolor{red}{2}, $E\left( \boldsymbol{x}_{t}\varepsilon_{t}|\mathcal{F}_{t-1} 	\right)= 0 $ and $E\left( E\left(	\varepsilon_{t}| \mathcal{F}_{t}^{(2)}	\right) \boldsymbol{x}_{t}| \mathcal{F}_{t}	 \right)= 0  $, hence  $\{\boldsymbol{x}_{t} \varepsilon_{t}\}$ is an m.d.s.
		\end{proof} 
		
		\begin{proposition}\label{theta}
			Let Assumption \textcolor{red}{3}  hold and let $\widehat{\boldsymbol{\Sigma}} = T^{-1} \widetilde{\boldsymbol{X}}'\widetilde{\boldsymbol{X}}$.  Then  $\widehat{{\boldsymbol{\Theta}}}$ is a good approximation of $\widehat{\boldsymbol{\Sigma}}$ uniformly for all $i = 1,\ldots,p$ if 
			\begin{align}
				\frac{\left \lVert  \boldsymbol{\widetilde{X}}'_{-i} \boldsymbol{\widetilde{X}}\widehat{{\boldsymbol\Theta}}_{i} \right \rVert_{\infty}}{T}
				=O_{P}\left(\sqrt\frac{\log{p}}{T}\right),\label{col_normXThet}
			\end{align}
			where  $\widehat{{\boldsymbol\Theta}} = \widehat{ \boldsymbol{T}}^{-2}\widehat{\boldsymbol{C}}$, $\widehat{\boldsymbol{T}}^{2}=\rm{diag}\left(\widehat{ \tau}^{2}_{1},\ldots,\widehat{ \tau}^{2}_{p} \right)$ and $\widehat{\boldsymbol{C}}$ as defined in \textcolor{red}{(32)}.
		\end{proposition}
		
		\begin{proof}[\unskip\nopunct] \textbf{Proof of Proposition \ref{theta}}
			We show that   $\widehat{{\boldsymbol{\Theta}}}$ is a good approximation of $\widehat{\boldsymbol{\Sigma}}^{-1}$,  for $\widehat{\boldsymbol{\Sigma}}= T^{-1} \left(\boldsymbol{\widetilde{X}}'\boldsymbol{\widetilde{X}} \right)$ the sample  variance-covariance matrix. Immediate application of the KKT condition on  $\widehat{{\boldsymbol{\gamma}}}_{i}^{2}$ gives
			\begin{align}
				&	\frac{1}{T}{\widetilde{\boldsymbol{X}}_{-i}' \left( \boldsymbol{\widetilde{\boldsymbol{x}}}_{i} -  \widetilde{\boldsymbol{X}}_{-i} {\widehat{\boldsymbol{\gamma}}_{i}} \right)}=\lambda_{i} \widehat{\boldsymbol{\eta}}_{i}. \label{eta} 
			\end{align}
			It suffices to show that 
			\begin{align}
				\frac{1}{T}	{\boldsymbol{X}_{-i}^{\star}}' \left( {{\boldsymbol{x}}}_{i}^{\star} - {\boldsymbol{X}}_{-i}^{\star} {\widehat{\boldsymbol{\gamma}}^{\star}}_{i}\right) + o_{P}\left(1\right)&= \left(\lambda_{i} \widehat{\boldsymbol{\eta}}_{i}- \lambda_{i} \widehat{\boldsymbol{\eta}}_{i}^{\star}\right) +  \lambda_{i} \widehat{\boldsymbol{\eta}}_{i}^{\star}, \label{etas}\\
				\frac{1}{T}	\widetilde{\boldsymbol{X}  }_{-i}'\left( \boldsymbol{\widetilde{\boldsymbol{x}}}_{i} -  \widetilde{\boldsymbol{X}}_{-i} {\widehat{\boldsymbol{\gamma}}^{\star}}_{i} \right) - 	\frac{1}{T}	{\boldsymbol{X}_{-i}^{\star}}' \left( \boldsymbol{{\boldsymbol{x}}}_{i} ^{\star}- {\boldsymbol{X}}_{-i}^{\star} {\widehat{\boldsymbol{\gamma}}_{i}}^{\star} \right)&= o_{P}\left(1\right)\label{etas2}, 
			\end{align}
			where \eqref{etas2}, the remainder term on \eqref{etas},  and  $\widehat{\boldsymbol{\eta}}_{i} = \text{sign}\left(\widehat{{{{\boldsymbol{\gamma}}}}}_{i}\right), \; \widehat{\boldsymbol{\eta}}_{i}^{\star} = \rm{sign}\left(\boldsymbol{\widehat{\gamma}}_{i}^{\star}	\right)$. Notice that 
			\[	\frac{{\boldsymbol{X}_{-i}^{\star}}' \left( {{\boldsymbol{x}}_{i}^{\star}} - {\boldsymbol{X}_{-i}^{\star}} {\widehat{\boldsymbol{\gamma}}_{i}}^{\star} \right)}{T} =\lambda_{i}\widehat{\boldsymbol{\eta}}_{i}^{\star},  \]
			are the KKT conditions of a node-wise regression, $\boldsymbol{x}_{i}^{\star}|\boldsymbol{X}^{\star}_{-i}$, and $\boldsymbol{\widehat{\gamma}}_{i}^{\star}$ the corresponding estimates, $\forall \; i=1,\ldots,p$.    Following similar arguments to Theorem \textcolor{red}{1}, $\lVert  \left(\boldsymbol{\widehat{\gamma}}_{i} - \boldsymbol{\widehat{\gamma}}^{\star}_{i}	\right)	\rVert_{1} = O_{P}\left(s_{0} \sqrt{T^{-1}\log p}\right)$, while $\lVert  \boldsymbol{\widehat{\gamma}}^{\star}_{i}\rVert_{1}$ will have the same properties as $\lVert  \boldsymbol{\widehat{\gamma}}_{i}\rVert_{1}$. 
			We proceed to show \eqref{etas2}, using the scalar representations of the processes involved,
			\begin{align}
				&\widetilde{\boldsymbol{X}}_{-i}' \left( {\widetilde{\boldsymbol{x}}}_{i} -  \widetilde{\boldsymbol{X}}_{-i} {\widehat{\boldsymbol{\gamma}}_{i}} \right)/T - 	{\boldsymbol{X}_{-i}^{\star}}' \left({{\boldsymbol{x}}_{i}^{\star}} - {\boldsymbol{X}}_{-i}^{\star} {\boldsymbol{\widehat{\gamma}}_{i}^{\star}} \right)/T  \nonumber\\
				&=\left( 	\sum_{t= 1}^{T} \boldsymbol{x}_{t,-i}  - \sum_{j=1}^{q}\widehat{\phi}_{j} \boldsymbol{x}_{t-j,-i} \right)\left[	\left(	\sum_{t= 1}^{T} {x}_{t,i}  - \sum_{j=1}^{q}\widehat{\phi}_{j}{x}_{t-j,i} \right)	\right.\nonumber\\
				&\quad \quad\quad\quad\quad\quad\quad\quad\quad\quad\quad\quad\quad
				\left.-\left( 	\sum_{t= 1}^{T} \boldsymbol{x}_{t,-i}  - \sum_{j=1}^{q}\widehat{\phi}_{j} \boldsymbol{x}_{t-j,-i} \right) \boldsymbol{\widehat{\gamma}}_{i} \right]\left/T\right.\nonumber \\
				&\quad -\left( 	\sum_{t= 1}^{T} \boldsymbol{x}_{t,-i}  - \sum_{j=1}^{q}{\phi}_{j} \boldsymbol{x}_{t-j,-i} \right)\left[	\left(	\sum_{t= 1}^{T} {x}_{t,i}  - \sum_{j=1}^{q}{\phi}_{j}{x}_{t-j,i} \right)	\right.\nonumber \\
				&\quad \quad\quad\quad\quad\quad \quad\quad\quad\quad\quad \quad\quad\quad
				\left.-\left( 	\sum_{t= 1}^{T} \boldsymbol{x}_{t,-i}  - \sum_{j=1}^{q}{\phi}_{j} \boldsymbol{x}_{t-j,-i} \right) \boldsymbol{\widehat{\gamma}}^{\star}_{i} \right]\left/T\right.\nonumber \\
				&\leq \frac{1}{T}	\sum_{t= 1}^{T} 		 \boldsymbol{x}_{t,-i}  \sum_{j=1}^{q}	 \left(\widehat{\phi}_{j}-\phi_{j}\right) x_{t-j, i} 	+ \frac{1}{T} \left(\boldsymbol{\widehat{\gamma}}_{i} - \boldsymbol{\widehat{\gamma}}^{\star}_{i}	\right)				\sum_{t= 1}^{T} 		 \boldsymbol{x}_{t,-i}	'	 \boldsymbol{x}_{t,-i} \nonumber \\
				&\quad +\frac{1}{T} \left(\boldsymbol{\widehat{\gamma}}_{i} - \boldsymbol{\widehat{\gamma}}^{\star}_{i}	\right)	\sum_{t= 1}^{T} 	 \boldsymbol{x}_{t,-i}	 \sum_{j=1}^{q}		 \boldsymbol{x}_{t-j,-i}  \left(\widehat{\phi}_{j}-\phi_{j}\right) +	\frac{1}{T}	\sum_{t= 1}^{T} x_{t,i} \sum_{j=1}^{q}	 \boldsymbol{x}_{t-j,-i}\left(\widehat{\phi}_{j}-\phi_{j}\right)	\nonumber \\
				&\quad +\frac{1}{T}\sum_{t= 1}^{T}  \sum_{j=1}^{q}  \boldsymbol{x}_{t-j,-i}{x}_{t-j,i}\left[\left( \widehat{\phi}_{j}- {\phi}_{j}  \right)^2 +\left(\widehat{\phi}_{j}- {\phi}_{j}\right)2\phi_{j}\right]\nonumber\\
				&\quad +\frac{1}{T} \sum_{t= 1}^{T} {x}_{t,i} \sum_{j=1}^{q}  \boldsymbol{x}_{t-j,-i}\left( \widehat{\phi}_{j}- {\phi}_{j}  \right)\nonumber\\
				&\quad  +\frac{1}{T}\left(\boldsymbol{\widehat{\gamma}}_{i} - \boldsymbol{\widehat{\gamma}}^{\star}_{i}	\right)			\sum_{t= 1}^{T}  \sum_{j=1}^{q}  \boldsymbol{x}_{t-j,-i}{x}_{t-j,-i}'\left[\left( \widehat{\phi}_{j}- {\phi}_{j}  \right)^2 +\left(\widehat{\phi}_{j}- {\phi}_{j}\right)2\phi_{j}\right]	\nonumber \\
				&= O_{P}\left(	 \frac{1}{T}\right) +  O_{P}\left(s_{0}\sqrt{\frac{\log{p}}{T}}\right)+O_{P}\left(s_{0}\frac{\sqrt{\log{p}}}{T^{3/2}}\right) +  O_{P}\left(	 \frac{1}{T}\right)    \nonumber \\
				&\quad + 		O_{P}\left(	 \frac{1}{T^{3/2}}\right)+ O_{P}\left(	 \frac{1}{T}\right)  + 		O_{P}\left(s_{0}{\frac{\sqrt{\log{p}}}{T^{3/2}}}\right).		\nonumber 
			\end{align}
			Notice that  $\widehat{{\boldsymbol{{\boldsymbol{\gamma}}}}}_{i}\lambda_{i}\widehat{\boldsymbol{\eta}}_{i}= \lambda_{i}\lVert\widehat{{\boldsymbol{{\boldsymbol{\gamma}}}}}_{i}\rVert_{1}$, and 	 $\widehat{{\boldsymbol{{\boldsymbol{\gamma}}}}}^{\star}_{i}\lambda_{i}\widehat{\boldsymbol{\eta}}^{\star}_{i}= \lambda_{i}\lVert\widehat{{\boldsymbol{{\boldsymbol{\gamma}}}}}^{\star}_{i}\rVert_{1}$, 
			\begin{align}
				\left(\boldsymbol{\widehat{\gamma}}_{i} - \boldsymbol{\widehat{\gamma}}^{\star}_{i}	\right)		= \left(\boldsymbol{\widehat{\gamma}}_{i} - \boldsymbol{{\gamma}}_{i}	\right)		+ \left(\boldsymbol{{\gamma}}_{i}-\boldsymbol{\widehat{\gamma}}^{\star}_{i} 	\right)	&\leq \lVert   \boldsymbol{\widehat{\gamma}}_{i} - \boldsymbol{{\gamma}}_{i}\rVert_{1} + \lVert \boldsymbol{{\gamma}}_{i}-\boldsymbol{\widehat{\gamma}}^{\star}_{i} \rVert_{1}
				=O_{P}\left(s_{i}T^{-1/2}\log^{1/2}{p}\right)\label{gammas}, 
			\end{align}
			where 	the first term obtains  its rate of convergence by Lemma \ref{consistentgamma} and  the second term has the same rate of convergence following similar arguments as in Lemma \ref{consistentgamma}.  Completing the proof of \eqref{etas2}.  
			Then we can show that 
			\begin{align}
				\left|\widehat{{\boldsymbol{{\boldsymbol{\gamma}}}}}_{i}\lambda_{i}\widehat{\boldsymbol{\eta}}_{i}- \widehat{{\boldsymbol{{\boldsymbol{\gamma}}}}}^{\star}_{i}\lambda_{i}\widehat{\boldsymbol{\eta}}^{\star}_{i}\right|& \leq \left|	\widehat{{\boldsymbol{{\boldsymbol{\gamma}}}}}_{i}\lambda_{i}\widehat{\boldsymbol{\eta}}_{i}- \widehat{{\boldsymbol{{\boldsymbol{\gamma}}}}}^{\star}_{i}\lambda_{i}\widehat{\boldsymbol{\eta}}_{i}\right| + \left|\widehat{{\boldsymbol{{\boldsymbol{\gamma}}}}}^{\star}_{i}\lambda_{i}\widehat{\boldsymbol{\eta}}_{i} - \widehat{{\boldsymbol{{\boldsymbol{\gamma}}}}}^{\star}_{i}\lambda_{i}\widehat{\boldsymbol{\eta}}^{\star}_{i}		\right|\\
				&\leq \lambda_{i} \left\Vert \widehat{\boldsymbol{\eta}}_{i}\right\Vert_{\infty} \left\Vert \boldsymbol{\widehat{\gamma}}_{i} - \boldsymbol{\widehat{\gamma}}^{\star}_{i}	\right\Vert_{1} + \lambda_{i} \left\Vert \widehat{\boldsymbol{\eta}}^{\star}_{i}\right\Vert_{\infty}\left\Vert\boldsymbol{\widehat{\gamma}}^{\star}_{i}	\right\Vert_{1}, \label{etasgammas} 
			\end{align}
			where $\lambda_{i}\asymp \sqrt{T^{-1}\log{p}}$, $\left\Vert \widehat{\boldsymbol{\eta}}^{\star}_{i}\right\Vert_{\infty} \leq 1$ as the sub-gradient of $\lVert \boldsymbol{\widehat{\gamma}}^{\star}_{i}\rVert_{1}$, $\left\Vert \widehat{\boldsymbol{\eta}}_{i}\right\Vert_{\infty} \leq 1$ as the sub-gradient of $\lVert \boldsymbol{\widehat{\gamma}}_{i}\rVert_{1}$, and $ \left\Vert \boldsymbol{\widehat{\gamma}}_{i} - \boldsymbol{\widehat{\gamma}}^{\star}_{i}	\right\Vert_{1} = O_{P}\left(s_{0} \sqrt{T^{-1}\log{p}}\right) $,  leading to $\left|\widehat{{\boldsymbol{{\boldsymbol{\gamma}}}}}_{i}\lambda_{i}\widehat{\boldsymbol{\eta}}_{i}- \widehat{{\boldsymbol{{\boldsymbol{\gamma}}}}}^{\star}_{i}\lambda_{i}\widehat{\boldsymbol{\eta}}^{\star}_{i}\right|=o_{P}\left(1\right)$ by \eqref{gammas}. 
			Considering the analysis on \eqref{etas2} and  \eqref{etasgammas}, 	\eqref{etas} becomes 
			\begin{equation}
				\boldsymbol{\widehat{\gamma}}^{\star}_{i}{{\boldsymbol{X}}^{\star}_{-i}}' \left( \boldsymbol{x}_{i}^{\star}-  {\boldsymbol{X}}^{\star}_{-i} \boldsymbol{\widehat{\gamma}}_{i}^{\star} \right)/T +o_{P}\left(1\right)
				= \lambda_{i}\lVert \boldsymbol{\widehat{\gamma}}_{i}^{\star}\rVert_{1} , \label{gammai}
			\end{equation}
			by plugging \eqref{gammai} into $(\widehat{\tau}^{\star}_{i})^{2} = T^{-1} \left \lVert \boldsymbol{x}^{\star}_{i}-
			{\boldsymbol X}^{\star}_{-i}{\boldsymbol{{\widehat{\gamma}}}}^{\star}_{i} \right \rVert_{2}^{2}+\lambda_{i}\lVert{\boldsymbol{{\widehat{\gamma}}}}^{\star}_{i}\rVert_{1}$, then,
			\begin{align}\label{tau2}
				(\widehat{\tau}^{\star}_{i})^{2} &= {T^{-1}} \left( \left(  \boldsymbol{{\boldsymbol{x}}}^{\star}_{i}-{\boldsymbol{X}}^{\star}_{-i} {\widehat{{\boldsymbol{{\boldsymbol{\gamma}}}}}^{\star}_{i}} \right)'\left(  \boldsymbol{{\boldsymbol{x}}}^{\star}_{i}-{\boldsymbol{X}}^{\star}_{-i} {\widehat{{\boldsymbol{{\boldsymbol{\gamma}}}}}^{\star}_{i}} \right)+			\boldsymbol{{\widehat{{{\gamma}}}}_{i}}	{{\boldsymbol{X}}^{\star}_{-i}}' \left( \boldsymbol{x}^{\star}_{i} -  	{\boldsymbol{X}}^{\star}_{-i} {\widehat{\boldsymbol{{\boldsymbol{\gamma}}}}_{i}}^{\star} \right)\right)\\
				&= {T^{-1}}\left( \boldsymbol{x}^{\star}_{i} -  	{\boldsymbol{X}}^{\star}_{-i} {\widehat{\boldsymbol{{\boldsymbol{\gamma}}}}_{i}}^{\star} \right) \left( \left( \boldsymbol{x}^{\star}_{i} -  	{\boldsymbol{X}}^{\star}_{-i} {\widehat{\boldsymbol{{\boldsymbol{\gamma}}}}_{i}}^{\star} \right)+    \widehat{{{{\boldsymbol{\gamma}}}}}_{i}^{\star}  {{\boldsymbol{X}}^{\star}_{-i}}' \right)  \label{taubound_1}\\
				&={T^{-1}} \left(  \boldsymbol{{\boldsymbol{x}}}^{\star}_{i}-\widetilde{\boldsymbol{X}}^{\star}_{-i}\boldsymbol{\widehat{\gamma}}_{i}^{\star} \right) \boldsymbol{x}^{\star}_{i} \label{taubound}.
			\end{align}
			Recall the definition of $\widehat{\boldsymbol{C}}$ in \textcolor{red}{(32)} and notice that  $\widehat{\boldsymbol{c}}_{i}$, is the $i$-th row of $\widehat{\boldsymbol{C}}$, we get that
			$ \boldsymbol{\widetilde{X}}\boldsymbol{\widehat{c}}_{i}=\left(  \boldsymbol{\widetilde{\boldsymbol{x}}}_{i}-\widetilde{\boldsymbol{X}}_{-i} {\widehat{{\boldsymbol{{\boldsymbol{\gamma}}}}}_{i}} \right)$ and from claims  \textcolor{red}{(31)}, \textcolor{red}{(33)}, $\widehat{ {\boldsymbol{\Theta}}}_{i} = \boldsymbol{\widehat{c}}_{i}/\widehat{\tau}_{i}^{2}$.  Equation \eqref{taubound} then becomes
			\begin{align}
				(\tau^{\star}_{i})^{2} &= {T^{-1}}{{\boldsymbol{X}}^{\star}_{-i}}'\boldsymbol{{X}}^{\star}\boldsymbol{\widehat{c}}^{\star}_{i} \Rightarrow
				{T^{-1}}{{\boldsymbol{X}}^{\star}_{-i}}'\boldsymbol{{X}}^{\star}\boldsymbol{{\Theta}}^{\star}_{i}=1  \label{xTheta},
			\end{align}
			where 	$\boldsymbol{{\Theta}}^{\star}_{i} = \boldsymbol{\widehat{c}}^{\star}_{i}/(\tau^{\star}_{i})^{2} $. 
			Then,  \eqref{xTheta} will hold if $\widehat{{\boldsymbol\Theta}}$ is a good approximation of  $\boldsymbol{\widehat{{\Sigma}}}^{-1}= \left( \boldsymbol{\widetilde{X}}'\boldsymbol{\widetilde{X}}/T\right)^{-1}$, in the sense that the approximation error $\left \lVert  \widehat{{\boldsymbol\Theta}}   \boldsymbol{\widehat{{\Sigma}}}-I  \right \rVert_{\infty} = o_{P}\left(1\right)$, which we now evaluate.
			
			For \eqref{col_normXThet} to hold, the following expression should hold as well.
			\begin{align}
				\left \lVert  \widehat{{\boldsymbol\Theta}}_{i}   \boldsymbol{\widehat{{\Sigma}}}-\boldsymbol{e}_{i} \right \rVert_{\infty}&\leq 	\left\Vert  \widehat{{\boldsymbol\Theta}}_{i}   \boldsymbol{\widehat{{\Sigma}}} -\widehat{{\boldsymbol\Theta}}_{i}     {\boldsymbol{\widehat{\Sigma}}^{\star} }  	\right\Vert_{\infty} + \left\Vert\widehat{{\boldsymbol\Theta}}_{i}     {\boldsymbol{\widehat{\Sigma}}^{\star} } -{{\boldsymbol\Theta}}^{\star}_{i}     {\boldsymbol{\widehat{\Sigma}}^{\star} }  \right\Vert_{\infty} +\left\Vert 	{{\boldsymbol\Theta}}^{\star}_{i}     {\boldsymbol{\widehat{\Sigma}}^{\star} }  -\boldsymbol{e}_{i}\right\Vert_{\infty}\\
				&=			B_1+B_2+B_3,\label{colnorm1}
			\end{align}
			where $\boldsymbol{\widehat{\Sigma}} = \left(\boldsymbol{\widetilde{x}}_{t}\boldsymbol{\widetilde{x}}_{t}'\right)/T,$ $\boldsymbol{\widehat{\Sigma}}^{\star} = \left(\boldsymbol{{x}}_{t}^{\star}{\boldsymbol{{x}}_{t}^{\star}}'\right)/T.$
			We wish to show that $B_{1}, B_{2}=o_{P}\left(1\right)$ and $B_{3}$ attains a bound such that the one in \eqref{col_normXThet}.
			We analyse $B_{1}$:
			\begin{align}
				B_{1}\leq  \left\Vert	   \boldsymbol{\widehat{{\Sigma}}}-   \boldsymbol{\widehat{\Sigma}}^{\star}    	\right\Vert_{\infty} 	\lVert \widehat{{\boldsymbol\Theta}}_{i} \rVert_{1}. 
			\end{align}
			We have that 
			\begin{align}
				\left\Vert				\boldsymbol{\widehat{{\Sigma}}}- \boldsymbol{\widehat{\Sigma}}^{\star} \right\Vert_{\infty}&\leq 
				\left\Vert	\frac{1}{T} \left(\boldsymbol{{X}}'\boldsymbol{\widehat{L}}'\boldsymbol{\widehat{L}}  \boldsymbol{{X}}  -   \boldsymbol{{X}}'\boldsymbol{{L}}'\boldsymbol{{L}}  \boldsymbol{{X}}  \right)\right\Vert_{\infty}\nonumber\\ 
				&\underset{(1), (2)}{\leq }  \frac{1}{T}\left\Vert \boldsymbol{{X}}'\boldsymbol{\widehat{L}}'  \left(	\boldsymbol{\widehat{L}} - \boldsymbol{L}	\right) \boldsymbol{{X}} 	 \right\Vert_{1} + \frac{1}{T}\left\Vert \boldsymbol{{X}} '\left(	\boldsymbol{\widehat{L}} - \boldsymbol{L}	\right)'\boldsymbol{{L}}\boldsymbol{{X}}\right\Vert_{1}= B_{11} +B_{12},		\label{Sigma}
			\end{align}
			where $(1)$ results from the norm inequality (8) of Chapter 8.5.2 of \cite{lutkepohl1997handbook} and $(2)$ results by applying the triangle inequality.
			For $i=1,\ldots, p, \; s=1,\ldots, T-q$, we analyse $B_{12}$: 
			\begin{align}
				B_{12} \leq \frac{1}{T} \max_{ i}\left|	\sum_{s=1}^{T-q}	 \boldsymbol{x}_{i}' (\boldsymbol{\widehat{\ell}}_{s}- \boldsymbol{\ell}_{s})'\boldsymbol{\ell}_{s} \boldsymbol{x}_{i}  	\right|  
				&\leq  
				\max_{i} \left\vert   \frac{1}{T}  \boldsymbol{x}_{i} '\boldsymbol{\ell}_{s}'\boldsymbol{x}_{i}\right\vert\left\Vert	\sum_{s=1}^{T-q}	(\boldsymbol{\widehat{\ell}}_{s}- \boldsymbol{\ell}_{s})\right\Vert_{1} = O_P(T^{-1/2}),\nonumber 
			\end{align}
			where $\boldsymbol{\ell}_s$, $\boldsymbol{\widehat{\ell}}_s$ two $1\times T$ row vectors of the matrices $\boldsymbol{L}$,  $\boldsymbol{\widehat{L}}$ respectively,   defined in \eqref{L}. The asymptotic rate of $B_{12}$ results by directly applying Corollary \textcolor{red}{1}.  Following a similar analysis to $B_{12}$, the term $B_{11}=O_P(T^{-1})$. 
			By Lemma \ref{Theta}, we have that $	\lVert \widehat{{\boldsymbol\Theta}}_{i} \rVert_{1}=O_{P}\left(\sqrt{s_{i}}\right)$, hence $B_{1}=O_{P}\left(\sqrt{s_i/T}\right)$.
			We analyse $B_{2}:$
			\begin{align}
				B_{2}&
				\underset{}{=} O_{P}\left(s_{i}\sqrt{\frac{\log{p}}{T}}	\right). 
			\end{align}
			The asymptotic rate of  $B_{2}$ is obtained by \eqref{ThetaCon}. Using $\boldsymbol{\widehat{\Sigma}}^{\star}=T^{-1} {\boldsymbol{X}^{\star}}'\boldsymbol{X}^{\star}$ and  expression \eqref{11}, we get that 
			$\Vert \boldsymbol{\widehat{\Sigma}}^{\star} \Vert_{\infty}= O_{P}\left(1\right).$
			We then continue the analysis for $B_{3}$, to derive the result. 		
			We examine the column norm of  $\Vert 	{{\boldsymbol\Theta}}^{\star}_{i}     {\boldsymbol{\widehat{\Sigma}}^{\star} }  -\boldsymbol{e}_{i}\Vert_{\infty}$, where $\boldsymbol{e}_{i}  $ is the $i^{th}$ column of the identity matrix $\boldsymbol{I}_{p\times p}$. By \eqref{xTheta},  $ \left \lVert T^{-1} {\boldsymbol{X}^{\star}}'_{-i} \boldsymbol{X}^{\star}{\boldsymbol{\Theta}}^{\star}_{i} \right \rVert_{\infty} =\Vert 	{{\boldsymbol\Theta}}^{\star}_{i}     {\boldsymbol{\widehat{\Sigma}}^{\star} }  -\boldsymbol{e}_{i}\Vert_{\infty}$, incorporating $\boldsymbol{{\Theta}}^{\star}_{i} = \boldsymbol{\widehat{c}}^{\star}_{i}/(\tau^{\star}_{i})^{2} $ into $\left \lVert T^{-1}  {\boldsymbol{X}^{\star}}'_{-i} \boldsymbol{X}^{\star}{\boldsymbol{\Theta}}^{\star}_{i} \right \rVert_{\infty} $, 
			\begin{align}
				\left \lVert  T^{-1}  {\boldsymbol{X}^{\star}}'_{-i} \boldsymbol{X}^{\star}{\boldsymbol{\Theta}}^{\star}_{i} \right \rVert_{\infty}
				= \left \lVert T^{-1} {\boldsymbol{X}^{\star}}'_{-i}  \boldsymbol{X}^{\star} \boldsymbol{\widehat{c}}^{\star}_{i}/(\tau^{\star}_{i})^{2} \right \rVert_{\infty} \leq \lambda_{i} \frac{\left \lVert    \widehat{\boldsymbol{\eta}}^{\star}_{i} \right \rVert _{\infty}}{(\tau^{\star}_{i})^{2}}\leq \frac{\lambda_{i}}{(\tau^{\star}_{i})^{2}},\label{bound XThet}
			\end{align}
			then \eqref{bound XThet} holds, since $ \left \lVert    \widehat{\boldsymbol{\eta}}^{\star}_{i} \right  \rVert _{\infty}:=\sup_{i} \left \lVert    \widehat{\boldsymbol{\eta}}^{\star}_{i} \right  \rVert_{1}\leq 1$.  Further, using \eqref{colnorm1} and since $ \left \lVert T^{-1} {\boldsymbol{X}^{\star}}'_{-i}  \boldsymbol{X}^{\star}{\boldsymbol{\Theta}}^{\star}_{i} \right \rVert_{\infty} =\Vert 	{{\boldsymbol\Theta}}^{\star}_{i}    \boldsymbol{\widehat    {\Sigma}}^{\star}  -\boldsymbol{e}_{i} \Vert_{\infty}$, we obtain 
			\begin{equation}\label{approximation}
				\left\Vert 	{{\boldsymbol\Theta}}^{\star}_{i}     {\boldsymbol{\widehat{\Sigma}}^{\star} }  -\boldsymbol{e}_{i}\right\Vert_{\infty}\leq \frac{\lambda_{i}}{(\tau^{\star}_{i})^{2} }+o_{P}\left(1\right),
			\end{equation}
			where $(\tau^{\star}_{i})^{2} = {\boldsymbol{x}_{i}^{\star}}'\left(\boldsymbol{x}^{\star}_{i}- \boldsymbol{X}_{-i}^{\star}\boldsymbol{\gamma}^{\star}_{i}\right)$.  It holds that  $\max_{i} \{1/(\tau^{\star}_{i})^{2}\} =O_P(1)$ by direct application of Corollary \ref{tauconsistent}, and given that $\lambda_{i}\asymp \sqrt{T^{-1}\log{p}}$, the result follows. 
		\end{proof}
		\begin{lemma}\label{consistentgamma}
			Let  Assumption \textcolor{red}{3}  hold.  If further, $\lambda_{i}\asymp \sqrt{\frac{\log p}{T}}$ and for $p>>T$, we have
			{\small			\begin{align}
					\left \lVert  \widetilde{\boldsymbol{X}}_{-i} \, (\widehat{{{\boldsymbol{{\boldsymbol{\gamma}}}}}}_{i} - {\boldsymbol{{\boldsymbol{\gamma}}}}_{i}) \right  \rVert_{2}^{2} =O_{P}\left(s_{i}\;T^{-1}\log p\right),& \quad \left \lVert  {\boldsymbol{X}}^{\star}_{-i} \, (\widehat{{{\boldsymbol{{\boldsymbol{\gamma}}}}}}^{\star}_{i} - {\boldsymbol{{\boldsymbol{\gamma}}}}_{i}) \right  \rVert_{2}^{2} =O_{P}\left(s_{i}\;T^{-1}\log p\right), \label{perror}\\
					\left	\lVert  \widehat{{{\boldsymbol{{\boldsymbol{\gamma}}}}}}_{i} - {\boldsymbol{{\boldsymbol{\gamma}}}}_{i} \right \rVert_{1}  = O_{P}(s_{i}\sqrt{T^{-1}\log p}),&\quad 	\left	\lVert  \widehat{{{\boldsymbol{{\boldsymbol{\gamma}}}}}}^{\star}_{i} - {\boldsymbol{{\boldsymbol{\gamma}}}}_{i} \right \rVert_{1}  = O_{P}(s_{i}\sqrt{T^{-1}\log p} ) \label{coeferror}. 
			\end{align}}
		\end{lemma}
		\begin{proof}[\unskip\nopunct] \textbf{Proof of Lemma \ref{consistentgamma}}
			\noindent The proof follows the same line of arguments as in Theorem \textcolor{red}{1}.
		\end{proof}
		\begin{corollary}
			\label{tauconsistent}
			Under Assumptions \textcolor{red}{3}, \textcolor{red}{4}, with row sparsity for $\boldsymbol{\Theta}$ bounded by $\max_{i} s_{i} = o(T/\log{p})$ and for a suitable choice of the regularisation parameter $\lambda_{i}\asymp \sqrt{\log{ p}/T}$ we have 
			\begin{equation}\label{tau}
				\max_{i}\left\{\frac{1}{{\widehat{\tau}_{i}}^{2}}\right\}=O_{P}(1),\quad  i=1,\ldots,p.
			\end{equation}
		\end{corollary}
		\begin{proof}[\unskip\nopunct] \textbf{Proof of Corollary \ref{consistentgamma}}
			
			We show that the errors of the node-wise regressions uniformly on $i$ are small.  First we make the following standard statements:  The compatibility condition holds uniformly for all node-wise regressions, and the corresponding compatibility constant is bounded away from zero.   Further, 
			\textcolor{black}{		$${1}/\tau_{i}^{2}=\boldsymbol{{\Theta}}^{\star}_{i,i}\geq \lambda^{2}_{\min}\left(\boldsymbol{ \Theta}^{\star} \right )>0, \quad \forall\; i=1,\ldots,p,$$ }and $\tau^{2}_{i}\leq E{\boldsymbol{{x}}_{i}^{\star}}'\boldsymbol{{x}}^{\star}_{i} ={\Sigma}_{i,i}^{\star}= O_{P}(1).$  Now, each node-wise regression has bounded prediction and estimation errors as in Lemma \ref{consistentgamma}, and we can then write 
			\begin{align}
				&\left[			\left \lVert \widetilde{\boldsymbol{x}}_{i}-\widetilde{\boldsymbol{X}}_{-i}  {{{\boldsymbol{{\widehat{\gamma}}}}}}_{i}\right \rVert_{2}^{2}/T -			\left \lVert {\boldsymbol{x}}^{\star}_{i}- {\boldsymbol{X}}^{\star}_{-i} {{{\boldsymbol{{\widehat{\gamma}}}}}}^{\star}_{i}\right \rVert_{2}^{2}/T\right] +	\left \lVert {\boldsymbol{x}}^{\star}_{i}-{\boldsymbol{X}}^{\star}_{-i}  {{{\boldsymbol{\widehat{{\gamma}}}}}}_{i}^{\star}\right \rVert_{2}^{2}/T  \nonumber\\
				&= A_1+\left \lVert {\boldsymbol{x}}^{\star}_{i}-{\boldsymbol{X}}^{\star}_{-i}  {{{\boldsymbol{\widehat{{\gamma}}}}}}_{i}^{\star}\right \rVert_{2}^{2}/T\nonumber \\
				&\underset{(1)}{=}	o_{p}\left(1\right) +
				\left \lVert {\boldsymbol{x}}^{\star}_{i}-{\boldsymbol{X}}^{\star}_{-i}  {{{\boldsymbol{\widehat{{\gamma}}}}}}_{i}^{\star}\right \rVert_{2}^{2}/T \nonumber\\
				&= 	o_{p}\left(1\right) +\left \lVert {\boldsymbol{x}}_{i}^{\star}-{\boldsymbol{X}}_{-i}^{\star} {\boldsymbol{\gamma}}_{i}^{\star} \right \rVert_{2}^{2}/T +   \left \lVert{\boldsymbol{X}}_{-i}^{\star}  (\widehat{{\boldsymbol{{\boldsymbol{\gamma}}}}}_{i}^{\star}-{\boldsymbol{{\boldsymbol{\gamma}}}}_{i}^{\star}) \right \rVert_{2}^{2}/T  +2\left( {\boldsymbol{x}}_{i}^{\star}- {\boldsymbol{X}}_{-i}^{\star}{{{\boldsymbol{{\boldsymbol{\gamma}}}}}}_{i}^{\star} \right)'\boldsymbol{{X}}_{-i}^{\star}\left(\widehat{{\boldsymbol{{\boldsymbol{\gamma}}}}}_{i}^{\star}-{\boldsymbol{{\boldsymbol{\gamma}}}}_{i}^{\star} \right) \nonumber\\
				&=o_{P}\left(1\right)+ \tau_{i}^{2} + O_{P}\left(s_{i}\log p / T \right) +O_{P} \left(\sqrt{s_{i}\log p/T} \right)+O_{P} \left(\sqrt{s_{i}\log p/T} \right)%
				= 	o_{p}\left(1\right) +\tau_{i}^{2}, \nonumber
			\end{align}
			where $		\boldsymbol{\gamma}_{i}^{\star}=\underset{\boldsymbol{\gamma_i^{\star}}\in \mathbb{R}^{p}}{\arg\min} \left \{{E}\left[(	{x}_{t,i}^{\star} -	\boldsymbol{x}_{-i,t}^{\star}\boldsymbol{\gamma}_{i}^{\star}	)^{2}\right]	\right	\}$, consequently, $\boldsymbol{\widetilde{\gamma}}_i$ can be defined following the same logic.   Above we used the fact that  $A_1=o_{P}(1)$, which we now prove. 	
			
			Notice that  $(1)$ results from the following expressions:  
			\begin{align}
				A_{1}&= 			\left \lVert \widetilde{\boldsymbol{x}}_{i}-\widetilde{\boldsymbol{X}}_{-i}  {{{\boldsymbol{{\widehat{\gamma}}}}}}_{i}\right \rVert_{2}^{2}/T -			\left \lVert {\boldsymbol{x}}^{\star}_{i}- {\boldsymbol{X}}^{\star}_{-i}{{{\boldsymbol{{\widehat{\gamma}}}}}}^{\star}_{i} \right \rVert_{2}^{2}/T \nonumber
				\\
				&=\widetilde{\boldsymbol{x}}_{i} ' \widetilde{\boldsymbol{x}}_{i}/T - 2 \widetilde{\boldsymbol{x}}_{i}'\widetilde{\boldsymbol{X}}_{-i}  {{{\boldsymbol{{\widehat{\gamma}}}}}}_{i}/T+ \left\Vert\widetilde{\boldsymbol{X}}_{-i}  {{{\boldsymbol{{\widehat{\gamma}}}}}}_{i}\right\Vert_{2}^2/T -  {\boldsymbol{x}_{i}^{\star}}' {\boldsymbol{x}}^{\star}_{i}/T
				\nonumber \\
				&\quad + 2 {\boldsymbol{x}^{\star}_{i}}' {\boldsymbol{X}}_{-i}^{\star}{{{\boldsymbol{{\widehat{\gamma}}}}}}^{\star}_{i} /T-  \left\Vert{\boldsymbol{X}}^{\star}_{-i}{{{\boldsymbol{{\widehat {\gamma}}}}}}^{\star}_{i} \right \rVert_{2}^{ 2}/T\nonumber\\
				&\leq \left\Vert  \boldsymbol{\widehat{\Sigma}} - \boldsymbol{\widehat{\Sigma}}^{\star}\right\Vert_{\infty} + \frac{1}{T}\left( 2 {{{\boldsymbol{{\boldsymbol{\gamma}}}}}^{\star}_{i}}' {\boldsymbol{X}^{\star}}_{-i} '{\boldsymbol{X}}^{\star}_{-i}{\boldsymbol{\widehat{\gamma}}^{\star}_{i}}+ 2{\boldsymbol{\upsilon}_i^{\star}}'{\boldsymbol{X}}^{\star}_{-i}{{{\boldsymbol{{\widehat{\gamma}}}}}}^{\star}_{i} -  2 {{{\boldsymbol{{\widetilde{\gamma}}}}}}_i '\boldsymbol{\widetilde{X}}_{-i}'\boldsymbol{\widetilde{X}}_{-i}\boldsymbol{\widehat{\gamma}}'_{i}+ 2\widetilde{\boldsymbol{\upsilon}}_i'\boldsymbol{\widetilde{X}}_{-i}{{{\boldsymbol{{\widehat{\gamma}}}}}}_{i} \right)\nonumber\\
				&\quad + \left(       \left\Vert\widetilde{\boldsymbol{X}}_{-i}  {{{\boldsymbol{{\widehat{\gamma}}}}}}_{i}\right\Vert_{2}^2/T -\left\Vert{\boldsymbol{X}}^{\star}_{-i}{{{\boldsymbol{{\widehat{\gamma}}}}}}^{\star}_{i} \right \rVert_{2}^{2}/T \right)= a_1+a_2+a_3\nonumber .
			\end{align}
			By \eqref{Sigma},  $a_1=O_{P}(T^{-1/2})$. 		For $1\leq i\leq p $ and $1\leq j\leq T$, we have that 
			\begin{align}
				a_{2}& = \left( 2 {{{\boldsymbol{{\boldsymbol{\gamma}}}}}^{\star}_{i}}' {\boldsymbol{X}^{\star}}_{-i} '{\boldsymbol{X}}^{\star}_{-i}{\boldsymbol{\widehat{\gamma}}^{\star}_{i}}+ 2{\boldsymbol{\upsilon}_i^{\star}}'{\boldsymbol{X}}^{\star}_{-i}{{{\boldsymbol{{\widehat{\gamma}}}}}}^{\star}_{i} -  2 {{{\boldsymbol{{\widetilde{\gamma}}}}}}_i '\boldsymbol{\widetilde{X}}_{-i}'\boldsymbol{\widetilde{X}}_{-i}\boldsymbol{\widehat{\gamma}}'_{i}+ 2\widetilde{\boldsymbol{\upsilon}}_i'\boldsymbol{\widetilde{X}}_{-i}{{{\boldsymbol{{\widehat{\gamma}}}}}}_{i} \right)/T\\
				&\leq \frac{1}{T}\left[\left\Vert	\boldsymbol{{\widetilde{X}}}_{-i}	\right\Vert_2^2 \left\Vert {{{\boldsymbol{{\widetilde{\gamma}}}}}}_{i} \right\Vert_{2} \left\Vert \left({{{\boldsymbol{{\widehat{\gamma}}}}}}_{i}-{{{\boldsymbol{{\widetilde{\gamma}}}}}}_{i}\right)\right\Vert_2+  \left\Vert 	\boldsymbol{{\widetilde{\upsilon}}}_{i}	'\boldsymbol{{\widetilde{X}}}_{-i}	\right\Vert_2\left\Vert {{{\boldsymbol{{\widetilde{\gamma}}}}}}_{i} \right\Vert_2 \left\Vert \left({{{\boldsymbol{{\widehat{\gamma}}}}}}_{i}-{{{\boldsymbol{{\widetilde{\gamma}}}}}}_{i}\right)\right\Vert_2 \right.\nonumber\\
				&\quad \left.- \left\Vert	{\boldsymbol{{X}}_{-i}^{\star}}\right\Vert_{2}^{2}\left\Vert {{{\boldsymbol{{\boldsymbol{\gamma}}}}}}^{\star}_{i} \right\Vert_{2} \left\Vert \left({{{\boldsymbol{{\widehat{\gamma}}}}}}_{i}^{\star}-{{{\boldsymbol{{{\gamma}}}}}}_{i}^{\star}\right)\right\Vert_2 + \left\Vert{\boldsymbol{\upsilon}_i^{\star}}'{{\boldsymbol{X}^{\star}}}_{-i}\right\Vert_2\left\Vert {{{\boldsymbol{{\widetilde{\gamma}}}}}}_{i} \right\Vert_2 \left\Vert \left({{{\boldsymbol{{\widehat{\gamma}}}}}}_{i}-{{{\boldsymbol{{\widetilde{\gamma}}}}}}_{i}\right)\right\Vert_2 \right.\nonumber\\
				&\quad \left.- \left\Vert	{\boldsymbol{{X}}_{-i}^{\star}}\right\Vert_{2}^{2}\left\Vert {{{\boldsymbol{{\boldsymbol{\gamma}}}}}}^{\star}_{i} \right\Vert_{2} \left\Vert \left({{{\boldsymbol{{\widehat{\gamma}}}}}}_{i}^{\star}-{{{\boldsymbol{{{\gamma}}}}}}_{i}^{\star}\right)\right\Vert_2\right]\nonumber \\
				&\leq\frac{1}{T}\left[  \left\Vert	\boldsymbol{{\widetilde{X}}}_{-i}	\right\Vert_2^2 O_{P}\left(\sqrt{s_i}\right)O_P\left(s_i\sqrt{\log{p}/T}\right)
				+\left\Vert{\boldsymbol{\widetilde{\upsilon}}_{i}}'\boldsymbol{{\widetilde{X}}}_{-i}\right\Vert_2 O_{P}\left(\sqrt{s_i}\right)O_P\left(s_i\sqrt{\log{p}/T}\right)\right.\nonumber\\
				& \left.+\left\Vert	\boldsymbol{{{X}}}_{-i}^{\star}	\right\Vert_2^2O_{P}\left(\sqrt{s_i}\right)O_P\left(s_i\sqrt{\log{p}/T}\right) + \left\Vert{{\boldsymbol{X}^{\star}}'}_{-i}\boldsymbol{\upsilon}_i^{\star}\right\Vert_2  O_{P}\left(\sqrt{s_i}\right)O_P\left(s_i\sqrt{\log{p}/T}\right)\right]\nonumber\\
				& = a_{21} + a_{22} + a_{23}+a_{24}\nonumber .
			\end{align}
			Using the  reverse triangle inequality we have, 
			{\small \begin{align}
					\left| \left\Vert	\boldsymbol{{\widetilde{X}}}_{-i}	\right\Vert_2^2 - \left\Vert	\boldsymbol{{{X}}}_{-i}^{\star}	\right\Vert_2^2 \right|O_P\left(s^{3/2}_i\sqrt{\frac{\log{p}}{T}}\right) /T
					&\leq \left\Vert\boldsymbol{{\widetilde{X}}}_{-i} - 	\boldsymbol{{{X}}}_{-i}^{\star}\right\Vert_{2}^2 O_P\left(s^{3/2}_i\sqrt{\frac{\log{p}}{T}}\right) /T\nonumber \\
					&\leq \left\Vert 	\boldsymbol{\widehat{L}}\boldsymbol{X}_{-i}/T - 	\boldsymbol{{L}}\boldsymbol{X}_{-i}/T\right\Vert_{2}^2O_P\left(s^{3/2}_i\sqrt{\frac{\log{p}}{T}}\right) \nonumber \\
					&\leq \left\Vert  \left( \boldsymbol{\widehat{L}}-	\boldsymbol{{L}} \right)	\boldsymbol{X}_{-i} /T\right\Vert_{2}^2 O_P\left(s^{3/2}_i\sqrt{\frac{\log{p}}{T}}\right) \nonumber \\ &\leq \left\Vert 	\boldsymbol{X}_{-i} '	\boldsymbol{X}_{-i} /T\right\Vert_{\infty} \left\Vert  \boldsymbol{\widehat{L}}-	\boldsymbol{{L}}  \right\Vert_{1} O_P\left(s^{3/2}_i\sqrt{\frac{\log{p}}{T}}\right) .
			\end{align}}
			Using similar arguments to prove \eqref{Sigma}, we have that $a_{21}- a_{23}= O_P(s^{3/2}_i\sqrt{{\log{p}}}/T)$.   Following a similar line of arguments and utilising the definitions of $\boldsymbol{\upsilon}_i^{\star} =E[ \boldsymbol{x}_i^{\star}-\boldsymbol{X}_i^{\star}\boldsymbol{\gamma}_i^{\star}]$ and $\widetilde{\boldsymbol{\upsilon}}_i=E[ \boldsymbol{\widetilde{x}}_i-\boldsymbol{\widetilde{X}}_i\boldsymbol{\widetilde{\gamma}}_i] $, we conclude that $a_{2}= O_P(s^{3/2}_i\sqrt{{\log{p}}}/T)\vee O_P(s^{3/2}_i\sqrt{{\log{p}}/T})=  O_P(s^{3/2}_i\sqrt{{\log{p}}/T}). $
			For $a_3$ and  using the reverse triangle inequality we have,
			\begin{align}
				a_{3}&\leq 	\left\vert      \left\Vert\widetilde{\boldsymbol{X}}_{-i}  {{{\boldsymbol{{\widehat{\gamma}}}}}}_{i}\right\Vert_{2}^2/T -\left\Vert {\boldsymbol{X}}^{\star}_{-i}{{{\boldsymbol{{\widehat{\gamma}}}}}}^{\star}_{i} \right \rVert_{2}^{2}/T \right\vert\nonumber\\
				& \leq 	\left\Vert   \widetilde{\boldsymbol{X}}_{-i}  {{{\boldsymbol{{\widehat{\gamma}}}}}}_{i}/T -{\boldsymbol{X}}^{\star}_{-i}{{{\boldsymbol{{\widehat{\gamma}}}}}}^{\star}_{i} /T \right\Vert^{2}_2\nonumber \\
				&= \left\Vert\left(	\boldsymbol{\widehat{L}}\boldsymbol{X}_{-i} \boldsymbol{\widehat{\gamma}}_{i} -	\boldsymbol{\widehat{L}}\boldsymbol{X}_{-i} \boldsymbol{{\gamma}}_{i} \right)   +  \left(		\boldsymbol{\widehat{L}}\boldsymbol{X}_{-i} \boldsymbol{{\gamma}}_{i} - 	\boldsymbol{{L}}\boldsymbol{X}_{-i} \boldsymbol{{\gamma}}_{i} 		\right)  +
				\left(	\boldsymbol{{L}}\boldsymbol{X}_{-i} \boldsymbol{{\gamma}}_{i} 	-	\boldsymbol{{L}}\boldsymbol{X}_{-i}  {{{\boldsymbol{{\boldsymbol{\gamma}}}}}}^{\star}_{i}\right)\right\Vert_{2}^{2}/T \nonumber \\
				&\leq \left(\left\Vert	\boldsymbol{\widehat{L}}\boldsymbol{X}_{-i} \boldsymbol{\widehat{\gamma}}_{i} -	\boldsymbol{\widehat{L}}\boldsymbol{X}_{-i} \boldsymbol{{\gamma}}_{i} \right\Vert   +  \left\Vert	\boldsymbol{\widehat{L}}\boldsymbol{X}_{-i} \boldsymbol{{\gamma}}_{i} - 	\boldsymbol{{L}}\boldsymbol{X}_{-i} \boldsymbol{{\gamma}}_{i} 		\right\Vert  +
				\left\Vert	\boldsymbol{{L}}\boldsymbol{X}_{-i} \boldsymbol{{\gamma}}_{i} 	-	\boldsymbol{{L}}\boldsymbol{X}_{-i}  {{{\boldsymbol{{\boldsymbol{\gamma}}}}}}^{\star}_{i}\right\Vert\right)^{2}/T\nonumber \\
				&= a_{31}+a_{32}+a_{33}\label{as}.
			\end{align}
			To show that $a_3=o_P(1)$, we  first show that $a_{31}, a_{32}, a_{33}=o_{P}(1)$.  We start with $a_{31}$:
			\begin{align}
				a_{31}&= \left\Vert	\boldsymbol{\widehat{L}}\boldsymbol{X}_{-i} \boldsymbol{\widehat{\gamma}}_{i} -	\boldsymbol{\widehat{L}}\boldsymbol{X}_{-i} \boldsymbol{\widetilde{\gamma}}_{i} \pm 	\boldsymbol{{L}}\boldsymbol{X}_{-i} \boldsymbol{\widehat{\gamma}}_{i}  \pm 	\boldsymbol{{L}}\boldsymbol{X}_{-i} \boldsymbol{\widetilde{\gamma}}_{i} \right\Vert /T\nonumber\\
				&\leq \left\Vert	\boldsymbol{\widehat{L}} -	\boldsymbol{{L}}  \right\Vert_1\left\Vert \boldsymbol{X}_{-i}\right\Vert_2\left\Vert \boldsymbol{\widetilde{\gamma}}_{i}\right\Vert_2 \left\Vert\left(\boldsymbol{\widehat{\gamma}}_{i}-\boldsymbol{\widetilde{\gamma}}_{i}\right)   \right\Vert_2 /T+  \left\Vert		\boldsymbol{\widehat{L}} -	\boldsymbol{{L}}  \right\Vert_1\left\Vert \boldsymbol{X}_{-i} \right\Vert_2\left\Vert\boldsymbol{\widetilde{\gamma}}_{i}	\right\Vert_2 /T\nonumber\\
				&		+  \left\Vert	\boldsymbol{{L}}  \boldsymbol{X}_{-i} \right\Vert_2	\left\Vert \boldsymbol{\widetilde{\gamma}}_{i}\right\Vert_2		\left\Vert\left(\boldsymbol{\widehat{\gamma}}_{i} -\boldsymbol{\widetilde{\gamma}}_{i} \right)		\right\Vert_2/T\nonumber\\
				&\underset{(1)}{=} O_P(T^{-1/2})O(1) O_P\left(s_i T^{-1/2}{\log^{1/2}{p}}\right)\vee  O_P(T^{-1/2})O(1) O_(\sqrt{s_{i}}) \nonumber\\
				&\quad  \vee O(1)O(\sqrt{s_i}) O_P\left(s_i T^{-1/2}{\log^{1/2}{p}}\right)= O_P\left(\sqrt{{s_i}/{T}}\right) \label{a3},
			\end{align}
			where 		$(1)$ results by following similar arguments as the ones to show \eqref{Sigma}, further the rate of $\Vert{\widetilde{\boldsymbol{\gamma}}} _i\Vert_2$ and $\left\Vert\left(\boldsymbol{\widehat{\gamma}}_{i} -\boldsymbol{\widetilde{\gamma}}_{i} \right) 			\right\Vert_2$, is obtained  by implication of Lemma  \ref{consistentgamma}. Regarding $\Vert{\widetilde{\boldsymbol{\gamma}}} _i\Vert_1$, we can write
			{		\begin{align}\left \lVert{\boldsymbol{{\widetilde{\gamma}}}}_{i} \right \rVert_{1}\leq \lambda_{i} \left( \left \lVert{\boldsymbol{{\widetilde{\gamma}}}}_{i}\right \rVert_{1} +\left \lVert\widehat{\boldsymbol{\gamma}}_{i}-{\boldsymbol{{\widetilde{\gamma}}}}_{i} \right \rVert_{1} \right) = \lambda_{i} \,O(\sqrt{s_{i}})+\lambda_{i} \, O_{P}(\sqrt{s_{i}\log p/T})\label{gammastar}.  \end{align}}
			Lastly $a_{32}, a_{33}$ follow similar analysis to $a_{31}$. Hence, we conclude  that $a_{31},a_{32}, a_{33}=o_{P}(1)$, showing that  $A_{1}=o_{P}(1)$.   Showing a similar result as in \eqref{gammastar} for $\Vert\boldsymbol{\gamma}_{i}^{\star}\Vert_1$, we proceed to show  \eqref{tau}
			\textcolor{black}{	\begin{align}
					\max_{i}\left	\{1/\widehat{\tau}^{2}\right	\} &= \max_{i}\left\{	\left[(\tau_{i})^{2}  + O_{P}\left( \sqrt{s_{i}/T}\right) + O_{P}\left(s_{i}\log p / T \right) +O_{P} \left(\sqrt{s_{i}\log p/T}\right)	\right]^{-1}\right \} 		\nonumber \\
					&\leq   \max_{i}\left\{  \left[\left({{\tau}_{i}}\right)^{2} +o_{P}\left(1\right)\right]^{-1}\right\}  ,   \nonumber  
			\end{align}}where $(\tau_{i})^{2} \leq E{\boldsymbol{{x}}_{i}^{\star}}'\boldsymbol{{x}}^{\star}_{i} ={\Sigma}_{i,i}^{\star}= O_{P}(1),$  completing  the proof. 
		\end{proof}
		
		\begin{lemma}\label{Lemmatau}
			Suppose Assumption \textcolor{red}{3} holds, then 
			\begin{align}
				|  \widehat{\tau}_{i}^{2} - {\tau}_{i}^{2} | &=  O_{P}\left(\sqrt{\frac{s_{i}\log{p}}{T}}\right)\\
				\left|	\frac{1}{{\widehat{\tau}_{i}}^{2}} -\frac{1}{{{\tau}_{i}}^{2}}\ \right| &= O_{P}\left(\sqrt{\frac{s_{i}\log{p}}{T}}\right)\label{T^{-1}au}
			\end{align}
		\end{lemma}
		\begin{proof}[\unskip\nopunct] \textbf{Proof of Lemma \ref{Lemmatau}. }
			
			Before we proceed to the analysis, it is useful to  point out that, from the KKT conditions of the node-wise regressions of $\boldsymbol{\widetilde{x}}_{i}|\boldsymbol{\widetilde{X}}_{-i}$ we have $ \widehat{ \tau}^{2}_{i} = \widetilde{\boldsymbol{x}}_{i}'(\widetilde{\boldsymbol{x}}_{i} -\widetilde{\boldsymbol{X}}_{-i}{{\boldsymbol{{\widehat{\gamma}}}}_{i}})$ with residuals
			$\widehat{\boldsymbol{\upsilon}}_{i}=(\widetilde{\boldsymbol{x}}_{i} - \widetilde{\boldsymbol{X}}_{-i}    {{\boldsymbol{{\widehat{\gamma}}}}_{i}}) $, further on the same note, the KKT conditions of the node-wise regressions of $\boldsymbol{{x}}^{\star}_{i}|\boldsymbol{{X}}^{\star}_{-i}$ we have $ (\widehat{ \tau}^{\star}_{i})^{2} = {{\boldsymbol{x}}^{\star}_{i}}'({\boldsymbol{x}}^{\star}_{i} - {\boldsymbol{X}}^{\star}_{-i}{{\boldsymbol{{\widehat{\gamma}}}}_{i}}^{\star})$ with residuals
			$\widehat{\boldsymbol{\upsilon}}^{\star}_{i}=({\boldsymbol{x}}^{\star}_{i} - {\boldsymbol{X}}^{\star}_{-i}{{\boldsymbol{{\widehat{\gamma}}}}_{i}}^{\star}) $ and $\tau^{2}_{i} = E\left[\left({\boldsymbol{x}}^{\star}_{i}	- \sum_{i\neq k}\gamma_{i,k}{\boldsymbol{X}}^{\star}_{-i}		\right)^{2}\right]$, with population residuals, $	E\left(\upsilon^{\star}_{i}\right) = E\left[  \boldsymbol{x}^{\star}_{i}-{\boldsymbol{X}^{\star}_{-i}}\boldsymbol\gamma^{\star}_{i}\right]$. 
			
			We first consider $|  \widehat{\tau}_{i}^{2} - {\tau}_{i}^{2} |$. 
			Then we have that
			\begin{align}
				\left|\widehat{ \tau}^{2}_{i}- { \tau}^{2}_{i}  \right| &\leq		\left|\widehat{ \tau}^{2}_{i}-(\widehat{ \tau}^{\star}_{i})^{2}  \right| + \left| (\widehat{ \tau}^{\star}_{i})^{2} - { \tau}^{2}_{i}  \ \right| = B_{1}+B_{2} .\nonumber 
			\end{align}
			The first part of the Lemma follows if $B_1, \; B_2=o_{P}\left(1\right)$, we proceed to 
			analyse $B_{1}$, where $1\leq i\leq p, \; 1\leq j\leq q$:
			\begin{align}
				B_{1}				&=  {T^{-1}}\left|	\widetilde{\boldsymbol{x}}_{i}'\widehat{\boldsymbol{\boldsymbol{\upsilon}}}_{i} - 	{{\boldsymbol{x}}^{\star}_{i}}'\widehat{\boldsymbol\upsilon}^{\star}_{i}\right|		\nonumber\\
				&= T^{-1}\left|	(	\boldsymbol{\widehat{L}}_{j}	\boldsymbol{x}_{i})'\left(\boldsymbol{\widehat{L}}_{j}	\boldsymbol{x}_{i} - \boldsymbol{\widehat{L}}_{j}	\boldsymbol{X}_{-i}	\boldsymbol{\widehat{\gamma}}_{i}	\right)	- 	(\boldsymbol{{L}}_{j}	\boldsymbol{x}_{i})'\left(\boldsymbol{{L}}_{j}	\boldsymbol{x}_{i} - \boldsymbol{{L}}_{j}	\boldsymbol{X}_{-i}	\boldsymbol{\widehat{\gamma}}_{i}^{\star}	\right)		\right|\nonumber \\
				&\leq T^{-1}	\left\vert\boldsymbol{x}_{i}'		\boldsymbol{\widehat{L}}_{j}	'	\boldsymbol{\widehat{L}}_{j}\boldsymbol{X}_{-i}\boldsymbol{\widehat{\gamma}}_{i}	 - \boldsymbol{x}_{i}'	{\boldsymbol{{L}}_{j}}'\boldsymbol{{L}}_{j}\boldsymbol{X}_{-i}		{	\boldsymbol{\widehat{\gamma}}_{i}^{\star}} \right\vert +T^{-1} \left\Vert\boldsymbol{x}_{i}'	\left( 	\boldsymbol{\widehat{L}}_{j}	'	\boldsymbol{\widehat{L}}_{j}	 - {\boldsymbol{{L}}_{j}}'\boldsymbol{{L}}_{j}\right)	\boldsymbol{x}_{i}\right\Vert_1\nonumber \\
				&= b_{1} + b_{2}\nonumber. 
			\end{align}
			Following a similar analysis as  in \eqref{as}, \eqref{Sigma},  it can be shown that $b_{2}= O_P\left(	T^{-1/2}	\right)$.  We analyse $b_{2}$
			{		\begin{align}
					b_{1} &=  	\left\vert\boldsymbol{x}_{i}'		\boldsymbol{\widehat{L}}_{j}	'	\boldsymbol{\widehat{L}}_{j}\boldsymbol{X}_{-i}\boldsymbol{\widehat{\gamma}}_{i}'	 - \boldsymbol{x}_{i}'	{\boldsymbol{{L}}_{j}}'\boldsymbol{{L}}_{j}\boldsymbol{X}_{-i}		{	\boldsymbol{\widehat{\gamma}}_{i}^{\star}} \pm  \boldsymbol{x}_{i}'	{\boldsymbol{{L}}_{j}}'\boldsymbol{{L}}_{j}\boldsymbol{X}_{-i}		{	\boldsymbol{\widehat{\gamma}}_{i}}	\pm 	\boldsymbol{x}_{i}'	{\boldsymbol{{L}}_{j}}'\boldsymbol{{L}}_{j}\boldsymbol{X}_{-i}	\boldsymbol{{\gamma}}_{i}^{\star}	\right\vert /T \nonumber \\
					&= \left\vert		\boldsymbol{x}_{i}'	\left(	\boldsymbol{\widehat{L}}_{j}	'	\boldsymbol{\widehat{L}}_{j}	 - 		{\boldsymbol{{L}}_{j}}'\boldsymbol{{L}}_{j}\right)	\boldsymbol{X}_{-i}\boldsymbol{\widehat{\gamma}}_{i}	\right\vert/T   
					+   \left\vert		\boldsymbol{x}_{i}'	{\boldsymbol{{L}}_{j}}'\boldsymbol{{L}}_{j}\boldsymbol{X}_{-i}	\left(	{	\boldsymbol{\widehat{\gamma}}^{\star}_{i}}	- \boldsymbol{{\gamma}}_{i}	\right)	\right\vert/T\nonumber\\		
					&\quad 	+  \left\vert		\boldsymbol{x}_{i}'	{\boldsymbol{{L}}_{j}}'\boldsymbol{{L}}_{j}\boldsymbol{X}_{-i}	\left(	{	\boldsymbol{\widehat{\gamma}}_{i}}	- \boldsymbol{{\gamma}}_{i}	\right)		\right\vert/T		= b_{11}+b_{12}+b_{13}. \nonumber 
			\end{align}		}
			Further, $b_{11}$ can be analysed such that 
			{	\begin{align}
					b_{11}& = \left\vert		\boldsymbol{x}_{i}'	\left(	\boldsymbol{\widehat{L}}_{j}	'	\boldsymbol{\widehat{L}}_{j}	 - 		{\boldsymbol{{L}}_{j}}'\boldsymbol{{L}}_{j}\right)	\boldsymbol{X}_{-i}\boldsymbol{\widehat{\gamma}}_{i}	\right\vert/T   \nonumber\\
					&\leq\frac{1}{T} \left\Vert 		\boldsymbol{x}_{i}'	\left(	\boldsymbol{\widehat{L}}_{j}	'	\boldsymbol{\widehat{L}}_{j}	 - 		{\boldsymbol{{L}}_{j}}'\boldsymbol{{L}}_{j}\right)	\boldsymbol{X}_{-i} \right\Vert_1  \left\Vert \boldsymbol{\widehat{\gamma}}_{i} - \boldsymbol{{\gamma}}_{i}\right\Vert_1 \label{terms2}\\
					&\quad +\frac{1}{T} \left\Vert 		\boldsymbol{x}_{i}'	\left(	\boldsymbol{\widehat{L}}_{j}	'	\boldsymbol{\widehat{L}}_{j}	 - 		{\boldsymbol{{L}}_{j}}'\boldsymbol{{L}}_{j}\right)	\boldsymbol{X}_{-i} \right\Vert_1 \left\Vert  \boldsymbol{{\gamma}}_{i}\right\Vert_1\label{terms1} \\
					&\underset{(1)}{=} O_P\left(T^{-1/2}\right) O_{P}\left(s_i\sqrt{\frac{\log{p}}T}\right) +  O_P\left(T^{-1/2}\right) O_{P}\left(\sqrt{s_i}\right)\nonumber.
			\end{align}}
			The first and third term in  \eqref{terms2}--\eqref{terms1} can be analysed following similar analysis as \eqref{Sigma}, the second term is obtained by directly applying Corollary \ref{consistentgamma} and the last term is an implication of Corollary \ref{consistentgamma}, hence $b_{11}= O_{P}\left(s_i\frac{\sqrt{{\log{p}}}}{T}\right)$. 
			Following the analysis  in \eqref{a3} and by direct application of Corollary \ref{consistentgamma}, we conclude that $b_{12}, b_{13} =O_{P}\left(s_i\sqrt{{\log{p}}/T}\right) $.   	Concluding, $B_1= b_1\vee b_2= O_P\left(s_i\sqrt{\log{p}/T}\right),$ and $B_2$ follows the same analysis with $B_1$.   Therefore the following holds,
			\[	\left|\widehat{ \tau}^{2}_{i}- { \tau}^{2}_{i}  \right| =O_P\left(	\sqrt{s_{i} \log{p}T^{-1}}	\right).\]
			To show \eqref{T^{-1}au}, 	note that $\tau_{i}^2= \frac{1}{\Theta_{i,i}} \geq \frac{1}{\lambda_{\max}\left(\boldsymbol{ \Theta}\right)} = \lambda_{\min}\left( {\boldsymbol{\Sigma}} \right)$, for all $i=1, \ldots, p$. Recall that $\lambda_{min}\left(\boldsymbol{\Sigma} \right)>0$, thus 
			\[
			\widehat{ \tau}_i^2 \leq  \left|\widehat{ \tau}_i^2 - (\widehat{ \tau}^{\star}_{i})^{2} \right|+ \left|(\widehat{ \tau}^{\star}_{i})^{2} - { \tau}_i^2 \right|+ { \tau}_i^2   
			= { \tau}_i^2  + o_{P}(1) >0,
			\]
			which taken together with Corollary \ref{tauconsistent}, imply 
			\begin{align}
				\left|	\frac{1}{{\widehat{\tau}_{i}}^{2}} -\frac{1}{{{\tau}_{i}}^{2}} \right| \leq \left|	\frac{1}{{\widehat{\tau}_{i}}^{2}} - \frac{1}{ (\widehat{ \tau}^{\star}_{i})^{2}} \right|+ \left|	 \frac{1}{ (\widehat{ \tau}^{\star}_{i})^{2}} - \frac{1}{{{\tau}_{i}}^{2}} \right|
				&= \frac{|\widehat{ \tau}_{i}^{2} - (\widehat{ \tau}^{\star}_{i})^{2} | }{{\widehat{\tau}_{i}}^{2}  (\widehat{ \tau}^{\star}_{i})^{2}} + \frac{| (\widehat{ \tau}^{\star}_{i})^{2}- 	{\tau}_{i}^{2} | }{{ 	 (\widehat{ \tau}^{\star}_{i})^{2}			{\tau}_{i}^{2} } } 
				\nonumber \\
				&= O_{P}\left(\sqrt{\frac{s_{i}\log{p}}{T}}\right).\nonumber 
			\end{align}
			Completing the proof.
		\end{proof}
		\begin{lemma}\label{Theta}
			Suppose Assumptions \textcolor{red}{2 -- 4} hold, and that $\lambda_i \asymp \sqrt{\log{p}/T}$, then the following statements hold
			\begin{align}
				\lVert \widehat{{\boldsymbol\Theta}}_{i} -{{\boldsymbol\Theta}}_{i} \rVert_{1} &= O_{P}\left(	s_{i}\sqrt{\frac{\log p}{T}}		\right)\label{thetapop1}\\
				\lVert \widehat{{\boldsymbol\Theta}}_{i} -{{\boldsymbol\Theta}}_{i} \rVert_{2} &= O_{P}\left(	\sqrt{s_{i}\frac{\log p}{T}}		\right)\\
				\lVert \widehat{{\boldsymbol\Theta}}_{i} \rVert_{1} &= O_{P}\left(	\sqrt{s_{i}}		\right)	\\
				\lVert {{\boldsymbol\Theta}}_{i} \rVert_{1} &= O\left(	\sqrt{s_{i}}		\right)	\label{thetapop}
			\end{align}
			where $s_{i} :=|\{j\neq i:\;\Theta_{i,j}\neq 0\}| $.
		\end{lemma}
		
		\begin{proof}[\unskip\nopunct] \textbf{Proof of Lemma \ref{Theta}}
			In this proof, we largely follow \cite{kock2016oracle}. By definition $\widehat{ {\boldsymbol{\Theta}}}_{i} = \boldsymbol{\widehat{c}}_{i}/\widehat{\tau}_{i}^{2}$, similar arguments hold for ${ {\boldsymbol{\Theta}}}_{i}^{\star} $, then
			\begin{align}
				\lVert \widehat{{\boldsymbol\Theta}}_{i} - {{\boldsymbol\Theta}}_{i} \rVert_{1} & \leq 				\lVert \widehat{{\boldsymbol\Theta}}_{i} - {{\boldsymbol{{\Theta}}}}_{i} ^{\star}\rVert_{1} +		\lVert {{\boldsymbol\Theta}}^{\star}_{i} - {{\boldsymbol\Theta}}_{i} \rVert_{1} = C_{1}+C_{2}. \label{thetass}
			\end{align}
			Starting with $C_{1}, $
			\begin{align}
				C_{1}&= \left \lVert \frac{ \boldsymbol{\widehat{c}}_{i}}{\widehat{\tau}_{i}^{2}}	-		\frac{ \boldsymbol{{c}}_{i}^{\star}}{ (\widehat{ \tau}^{\star}_{i})^{2}}	\right\rVert_{1}
				= \left \lVert \frac{ 1- \widehat{{\boldsymbol{{\boldsymbol{\gamma}}}}}_{i}}{\widehat{\tau}_{i}^{2}}	-		\frac{ 1-{\boldsymbol{{\widehat{\gamma}}}}^{\star}_{i}}{ (\widehat{ \tau}^{\star}_{i})^{2}}	\right\rVert_{1}\nonumber\\
				&=  \left \lVert \frac{1}{{\widehat{\tau}_{i}}^{2}} -\frac{1}{ (\widehat{ \tau}^{\star}_{i})^{2}}     + \frac{ {{\boldsymbol{{\widehat{\gamma}}}}}^{\star}_{i}}{\widehat{\tau}_{i}^{2}} - \frac{ {{\boldsymbol{{\widehat{\gamma}}}}}^{\star}_{i}}{\widehat{\tau}_{i}^{2}} +\frac{ {\boldsymbol{{\widehat{\gamma}}}}_{i}^{\star}}{ (\widehat{ \tau}^{\star}_{i})^{2}} - \frac{  \widehat{{\boldsymbol{{{\gamma}}}}}_{i}}{\widehat{\tau}_{i}^{2}} 
				\right\rVert_{1}\nonumber\\
				&\leq\left \lVert  \frac{1}{{\widehat{\tau}_{i}}^{2}} -\frac{1}{ (\widehat{ \tau}^{\star}_{i})^{2}}  \right  \rVert_{1} + \left\lVert{\boldsymbol{{\widehat{\gamma}}}}^{\star}_{i} \right\rVert_{1}	 \left \lVert  \frac{1}{{\widehat{\tau}_{i}}^{2}} -\frac{1}{ (\widehat{ \tau}^{\star}_{i})^{2}}  \right  \rVert_{1}    + \left\lVert\widehat{{\boldsymbol{{\boldsymbol{\gamma}}}}}_{i}-{\boldsymbol{{\widehat{\gamma}}}}^{\star}_{i} \right\rVert_{1} \left \lVert  \frac{1}{{\widehat{\tau}_{i}}^{2}} \right\rVert_{1}\nonumber\\
				&	= O_{P}\left(\sqrt{\frac{s_{i}\log{p}}{T}}\right)+ O(\sqrt{s_{j}}) \, O_{P}\left(	\sqrt{\frac{s_{i} \log p}{T}}		\right)  + O_{P}\left( s_{i} \sqrt{\frac{\log{p}}{T}}\right) \, O_{P}(1)  \nonumber\\
				&=O_{P}\left(s_{i}\sqrt{\frac{\log{p}}{T}}\right). \label{ThetaCon} 
			\end{align}
			The first term of \eqref{ThetaCon} is a direct result of  Lemma \ref{Lemmatau}, the second results from \eqref{gammastar}, and Lemma \ref{Lemmatau} and the third term results from \eqref{gammas}, and direct application of Corollary \ref{taubound}.
			Following similar arguments to \eqref{ThetaCon}, we can show that $\lVert \widehat{{\boldsymbol\Theta}}_{i} -{{\boldsymbol\Theta}}_{i} \rVert_{2} = O_{P}(\sqrt{	T^{-1}s_{i}{\log p}}		)$.
			By definition of $	\lVert \widehat{{\boldsymbol\Theta}}_{i} \rVert_{1}$,  $ \widehat{{\boldsymbol{\Theta}}}_{i} = \widehat{{\tau_{i}}}^{-2}\widehat{\boldsymbol{c}}_{i}$ (similar arguments hold for ${\boldsymbol{\Theta}}^{\star}_{i}$),  we have 
			\begin{align}
				\lVert{{\boldsymbol{\Theta}}}_{i} \rVert_{1} &= 	\lVert  {{\tau_{i}}}^{-2}{\boldsymbol{c}}_{i}\rVert_{1} = \left\lVert 	\frac{ 1-{\boldsymbol{{\boldsymbol{\gamma}}}}_{i}}{{\tau}_{i}^{2}}\right	\rVert_{1}
				\leq \left\lVert     \frac{1}{{{\tau}_{i}^{2}}}     \right\rVert_{1} - \lVert {\boldsymbol{{\boldsymbol{\gamma}}}}_{i}  \rVert_{1} \left\lVert     \frac{1}{{{\tau}_{i}^{2}}}     \right\rVert_{1} = O(\sqrt{s_{i}}) \nonumber
			\end{align}
			\begin{align}
				\lVert \widehat{{\boldsymbol\Theta}}_{i} \rVert_{1}&\leq 	\lVert\widehat{{\boldsymbol{\Theta}}}_{i}  - {{\boldsymbol{\Theta}}}^{\star}_{i} \rVert_{1} +  \lVert{{\boldsymbol{\Theta}}}^{\star}_{i} \rVert_{1} \nonumber\\
				& 	\leq		O_{P}\left(s_{i}\sqrt{\frac{\log{p}}{T}}\right)+ \left\lVert     \frac{1}{(\widehat{ \tau}^{\star}_{i})^{2}}     \right\rVert_{\infty} - \lVert {{\boldsymbol{{\widehat{\gamma}}}}}^{\star}_{i} -{\boldsymbol{{\boldsymbol{\gamma}}}}_{i} \rVert_{1} \left\lVert     \frac{1}{{(\widehat{ \tau}^{\star}_{i})^{2}}}     \right\rVert_{\infty} -\left\lVert  {\boldsymbol{{\boldsymbol{\gamma}}}}_{i}  \right\rVert_{1}  \left\lVert     \frac{1}{(\widehat{ \tau}^{\star}_{i})^{2}}     \right\rVert_{\infty} \nonumber \\
				&=  O_{P}(1) + O_{P}\left(s_{i}\sqrt{\frac{\log{p}}{T}}\right) O_{P}(1)  + O(\sqrt{s_{i}})O_{P}(1) =O_{P}(\sqrt{s_{i}}). \label{Theistar} 
			\end{align}
			The result is based on a direct application of Corollary \ref{tauconsistent}, from which  we get that $ \left \lVert  \frac{1}{{{\tau}_{i}}^{2}} \right\rVert_{\infty}=O(1)$ and  $ \left \lVert  \frac{1}{{(\widehat{\tau}_{i}^{\star})}^{2}} \right\rVert_{\infty}=O_{P}(1)$, from Lemma \ref{consistentgamma} we get $  \lVert \boldsymbol{\widehat{\gamma}}^{\star}_{i} -{\boldsymbol{{\boldsymbol{{\boldsymbol{\gamma}}}}}}_{i} \rVert_{1} = O_{P}(s_{i}\sqrt{T^{-1}\log p})$, and by following similar arguments as \eqref{gammastar}, we have $\lVert  {\boldsymbol{{\boldsymbol{\gamma}}}}_{i} \rVert_{1} =O(\sqrt{s_{i}})$.	 
			
			Using a similar line of arguments with $C_{1}$,   we have that $C_{2}=  O_{P}\left(	s_{i}\sqrt{{T^{-1}\log p}}		\right)$, completing the proof. 
		\end{proof}

		\begin{lemma}\label{Lemmathetaxu}
			Under Assumption \textcolor{red}{3}, we have  for  $ i=1,\ldots,p$,
			\begin{align}
				\frac{1}{\sqrt{T}}\left\Vert\widehat{\boldsymbol{\Theta}}_{i}\widetilde{\boldsymbol{X}}'{ \widehat{\boldsymbol{L}}\boldsymbol{u}} - \boldsymbol{\Theta}_{i}{\boldsymbol{X}}'{ {\boldsymbol{L}}\boldsymbol{u}} \right\Vert_1=o_{P}\left(1\right)\label{thetaxu1},
			\end{align}
			where $ \widehat{ \boldsymbol{\Theta}}_{i}$ is defined in \textcolor{red}{(33)} of the main paper, while $\boldsymbol{ \Theta}_{i}$ is its sample counterpart,   and $\boldsymbol{\widetilde{X}} = \boldsymbol{\widehat{L}}\boldsymbol{X}$ is defined in \eqref{L}. 
		\end{lemma}
		
		\begin{proof}[\unskip\nopunct] \textbf{Proof of Lemma \ref{Lemmathetaxu} }
			We use the following decomposition of the left hand side of \eqref{thetaxu1}:
			\textcolor{black}{\small \begin{align}
					\frac{1}{\sqrt{T}}\left\Vert\widehat{\boldsymbol{\Theta}}_{i}\widetilde{\boldsymbol{X}}'{ \widehat{\boldsymbol{L}}\boldsymbol{u}} - \boldsymbol{\Theta}_{i}{\boldsymbol{X}}'{ \boldsymbol{{L}}{\boldsymbol{u}}} \right\Vert_1 
					&\leq\frac{1}{\sqrt{T}} \left[\left\Vert 	\widehat{\boldsymbol{\Theta}}_{i}\widetilde{\boldsymbol{X}}'{ \widehat{\boldsymbol{L}}\boldsymbol{u}} -  {\boldsymbol{\Theta}}^{\star}_{i}\widetilde{\boldsymbol{X}}'{ \widehat{\boldsymbol{L}}\boldsymbol{u}}	\right\Vert_1 
					+ \left \Vert  {\boldsymbol{\Theta}}^{\star}_{i}\widetilde{\boldsymbol{X}}'{ \widehat{\boldsymbol{L}}\boldsymbol{u}}	- {\boldsymbol{\Theta}}^{\star}_{i}{\boldsymbol{X}^{\star}}'{ \widehat{\boldsymbol{L}}\boldsymbol{u}}			\right\Vert_1 \right.\nonumber\\
					&\left.\quad + \left \Vert     {\boldsymbol{\Theta}}^{\star}_{i}{\boldsymbol{X}^{\star}}'{ \widehat{\boldsymbol{L}}\boldsymbol{u}}	 -  {\boldsymbol{\Theta}}^{\star}_{i}{\boldsymbol{X}^{\star}}'{ {\boldsymbol{L}}\boldsymbol{u}}	\right\Vert_{1}
					+ \left \Vert     {\boldsymbol{\Theta}}^{\star}_{i}{\boldsymbol{X}^{\star}}'{ {\boldsymbol{L}}\boldsymbol{u}}			 -  {\boldsymbol{\Theta}}_{i}{\boldsymbol{X}^{\star}}'{ {\boldsymbol{L}}\boldsymbol{u}}\right\Vert_{1}\right.\nonumber\\
					& \left. \quad +\left \Vert     {\boldsymbol{\Theta}}_{i}{\boldsymbol{X}^{\star}}'{ {\boldsymbol{L}}\boldsymbol{u}}	 -  {\boldsymbol{\Theta}}_{i}{\boldsymbol{X}}'{ {\boldsymbol{L}}\boldsymbol{u}}	\right\Vert_{1}  \right]= \sum_{k=1}^5 D_k.
			\end{align}}
			It suffices to show the following :
			\small\begin{align}
				D_{1} &= O_{P}\left(\frac{s_i\sqrt{\log{p}}}{T}\right),\; D_{2}= O_P\left(\sqrt\frac{s_i}{T}\right) , \; D_{3}=O_P\left({\left(\frac{s_{i}}{T}\right)}^{3/2}\sqrt{\log{p}}\right)\nonumber\\
				D_4&= O_{P}\left(\sqrt\frac{s_i\log{p}}{T}\right) ,\;  D_5= O\left(\sqrt{\frac{s_i}{T}}\right). 
				\label{suffTheta}
			\end{align}
			We consider each term, starting with $D_{1}$:
			\begin{align}
				D_{1}
				&\leq  \frac{1}{\sqrt{T}}\left\Vert \widehat{\boldsymbol{\Theta}}_{i}-  {\boldsymbol{\Theta}}^{\star}_{i}\right\Vert_{1} \left[   \left\Vert \widetilde{\boldsymbol{X}}'{ \widehat{\boldsymbol{L}}\boldsymbol{u}}- {\boldsymbol{X}^{\star}}'{ \widehat{\boldsymbol{L}}\boldsymbol{u}}  \right\Vert_1 + 	  \left\Vert {\boldsymbol{X}^{\star}}'{ \widehat{\boldsymbol{L}}\boldsymbol{u}}   - {\boldsymbol{X}^{\star}}'{ {\boldsymbol{L}}\boldsymbol{u}} \right\Vert_1+ \left\Vert {\boldsymbol{X}^{\star}}'{ {\boldsymbol{L}}\boldsymbol{u}}  \right\Vert_1	\right]\nonumber\\
				&= \left\Vert \widehat{\boldsymbol{\Theta}}_{i}-  {\boldsymbol{\Theta}}^{\star}_{i}\right\Vert_{1}\left[D_{11}+D_{12}+D_{13}\right]. \nonumber
			\end{align}
			By \eqref{Theistar},   $\Vert \widehat{\boldsymbol{\Theta}}_{i}-  {\boldsymbol{\Theta}}^{\star}_{i}\Vert_{1} = O_P\left(s_{i}\sqrt{\frac{\log{p}}{T}}\right)$, as for term $D_{11}$, we have that 
			\begin{align}	\textcolor{black}{ \frac{1}{\sqrt{T}}\left\Vert \widetilde{\boldsymbol{X}}'{ \widehat{\boldsymbol{L}}\boldsymbol{u}}- {\boldsymbol{X}^{\star}}'{ \widehat{\boldsymbol{L}}\boldsymbol{u}} \right \Vert_1  }&\leq  \frac{1}{\sqrt{T}}\left\vert \sum_{t=1}^{T} \sum_{j=1}^{q}(\widehat\phi_j- \phi_j ) \sum_{i=1}^p x_{t-j,i}\right\vert\nonumber\\
				&\quad \quad \times  \left \vert  \sum_{t=1}^{T}\left[u_t + \sum_{j=1}^{q} \left[(\widehat\phi_j- \phi_j ) \sum_{i=1}^p x_{t-j,i}( \widehat{\beta}_i- \beta_i) \right] + \sum_{j=1}^{q}\phi_j x_{t-j,i}\right]\right\vert		\nonumber \\
				& \leq \frac{1}{\sqrt{T}} \left\vert \sum_{t=1}^{T} \sum_{j=1}^{q}(\widehat\phi_j- \phi_j ) \sum_{i=1}^p x_{t-j,i}u_t \right\vert\nonumber\\
				&\quad  + \frac{1}{\sqrt{T}} \left\vert \sum_{t=1}^{T} \sum_{j=1}^{q}(\widehat\phi_j- \phi_j )^2\sum_{i=1}^p x_{t-j,i}^2 \right\vert \left\vert \sum_{i=1}^p( \widehat{\beta}_i- \beta_i) \right\vert\nonumber \\
				&\quad +  \frac{1}{\sqrt{T}}\left\vert  \sum_{t=1}^{T} \sum_{j=1}^{q}(\widehat\phi_j- \phi_j )\phi_j \sum_{i=1}^p x_{t-j,i} ^2\right\vert = d_{1} + (d_2\times d_3 )+ d_{4} 	\label{til}.
			\end{align}
			By Corollary \textcolor{red}{1} and \eqref{11} $d_1= O_P(T^{-1/2})$, $d_{2} = O_P({T^{-1/2}} )$, and by Lemma \textcolor{red}{1}, $d_{3}= O_{P}(s_{0}\sqrt{\frac{\log{p}}{T}})$, and by \eqref{11},  $d_{4} = O_P(1)$, hence  $D_{11} =O_P({T^{-1/2}} )$.  We proceed to analyse $D_{12}$:
			\begin{align}
				\textcolor{black}	{\frac{1}{\sqrt{T}}	  \left\Vert {\boldsymbol{X}^{\star}}'{ \widehat{\boldsymbol{L}}\boldsymbol{u}} - {\boldsymbol{X}^{\star}}'{ {\boldsymbol{L}}\boldsymbol{u}}\right\Vert_1 }
				&\leq \frac{1}{\sqrt{T}}  \left\Vert   {\boldsymbol{X}}'{\boldsymbol{L}}'{\boldsymbol{X}} \right\Vert_{\infty}  \left\Vert \boldsymbol{	\widehat{\beta}}- \boldsymbol{	 \beta} \right\Vert_1  =  O_P(s_{0}\sqrt{\log{p}}) \nonumber ,
				\label{ustar}   
			\end{align}
			hence $D_{12}=  O_P(s_{0}\sqrt{\frac{\log{p}}{T}})$. 
			We continue  with $D_{13}$, by \eqref{11}, $D_{13}=  O_P(s_{0}\sqrt{\frac{\log{p}}{T}})$.  By consequence, $D_1= O_P(s_{0}\sqrt{\log{p}/T})$, showing  the first part of \eqref{suffTheta}. For the second part of \eqref{suffTheta}, by  \eqref{Theistar} and \eqref{til}
			$
			D_2\leq \Vert 	\boldsymbol{\Theta}_{i}^{\star}	\Vert_{1}   \Vert {\boldsymbol{X}^{\star}}'{ \widehat{\boldsymbol{u}}}   - {\boldsymbol{X}^{\star}}'\boldsymbol{u}\Vert_1 = O_P(\sqrt{\frac{s_{i}}{T}}) .
			$
			For $D_3$,  by   \eqref{Theistar} and  by \eqref{ustar} $D_3=O_P({(\frac{s_{i}}{T})}^{3/2}\sqrt{\log{p}})$.   For  $D_4$
			\textcolor{black}{	\begin{align}
					\left \Vert     {\boldsymbol{\Theta}}^{\star}_{i}{\boldsymbol{X}^{\star}}'{ {\boldsymbol{L}}\boldsymbol{u}}	 -  {\boldsymbol{\Theta}}_{i}{\boldsymbol{X}^{\star}}'{ {\boldsymbol{L}}\boldsymbol{u}}	\right\Vert_{1} \leq \left\Vert {\boldsymbol{\Theta}}^{\star}_{i}-  {\boldsymbol{\Theta}}_{i}\right\Vert_{1}\left \Vert  {\boldsymbol{X}^{\star}}'{ {\boldsymbol{L}}\boldsymbol{u}}		\right\Vert_{1} .
			\end{align}}
			By the arguments of \eqref{thetass}, $\left\Vert {\boldsymbol{\Theta}}^{\star}_{i}-  {\boldsymbol{\Theta}}_{i}\right\Vert_{1}= O_{P}(s_i\sqrt{\log{p}/T}) $ and by Lemma \textcolor{red}{1} $\left \Vert  {\boldsymbol{X}^{\star}}'{ {\boldsymbol{u}}}		\right\Vert_{1} \leq pT^{-1/2}\left \Vert  {\boldsymbol{X}^{\star}}'{ {\boldsymbol{u}}}		\right\Vert_{\infty} \leq \frac{\sqrt{T}\lambda_{0}}{2}$, for some $\lambda_{0}=\sqrt{T^{-1}\log{p}}$.  Noting that \eqref{11} holds,  along with the latter,${T}^{-1/2} \left \Vert  {\boldsymbol{X}^{\star}}'{ {\boldsymbol{L}}\boldsymbol{u}}	\right\Vert_{1}=O_P(1). $  Lastly,  for $D_4$
			\textcolor{black}{	\begin{align}
					T^{-1/2}\left \Vert     {\boldsymbol{\Theta}}_{i}{\boldsymbol{X}^{\star}}'{ {\boldsymbol{L}}\boldsymbol{u}}	 -  {\boldsymbol{\Theta}}_{i}{\boldsymbol{X}}'{ {\boldsymbol{L}}\boldsymbol{u}}	\right\Vert_{1}  \leq T^{-1/2} \left\Vert  {\boldsymbol{\Theta}}_{i}\right\Vert_1  \left\Vert( \boldsymbol{X}^{\star} -\boldsymbol{X})'{ {\boldsymbol{L}}\boldsymbol{u}} \right\Vert_1
			\end{align}}
			By Lemma \ref{Theta}, $\left\Vert  {\boldsymbol{\Theta}}_{i}\right\Vert_1 = O(\sqrt{s}_i)$, and 
			{\begin{align}
					T^{-1/2}\left\Vert( \boldsymbol{X}^{\star} -\boldsymbol{X})'{ {\boldsymbol{L}}\boldsymbol{u}} \right\Vert_1 \leq 
					T^{-1/2}    \left\Vert{ {\boldsymbol{L}}\boldsymbol{u}}\right\Vert_1 .
			\end{align} }
			Note that Concluding that $\boldsymbol{X} $ corresponds to the design matrix when ${\phi}_j$, for all $j=1,\ldots, q$, is known, which coincides with the case of $\boldsymbol{X}^{\star}$, where the design contains the infeasible estimates of $\phi_j$. Therefore    $D_5 = O(\sqrt{s_iT^{-1}})$, and the result follows. 
		\end{proof}
		
		\begin{lemma}\label{identity}
			Under Assumption \textcolor{red}{3}, we have  $\forall\; i\in\{1,\ldots,p\}$,
			\begin{align}
				\left|	\widehat{\boldsymbol{ \Theta}}_{i}\widehat{\boldsymbol{\Sigma}}_{xu}\widehat{ \boldsymbol{\Theta}}_{i}'-	\boldsymbol{ \Theta}_{i} \boldsymbol{\Sigma}_{xu}{ \boldsymbol{\Theta}}_{i}'	\right|=o_{p}(1), \label{thetas}
			\end{align}
			where $	\widehat{ \boldsymbol{\Theta}}_{i},\; \widehat{\boldsymbol{\Sigma}}_{xu}$ are defined in  Section   \textcolor{red}{4} of the main paper.
		\end{lemma}
		
		\begin{proof}[\unskip\nopunct] \textbf{Proof of Lemma \ref{identity}. }
			To show the result in \eqref{thetas}, we  use the following decomposition:
			\begin{align}
				\left|	\widehat{ \boldsymbol{\Theta}}_{i}\widehat{\boldsymbol{\Sigma}}_{xu}\widehat{ \boldsymbol{\Theta}}_{i}'-	\boldsymbol{ \Theta}_{i}\boldsymbol{\Sigma}_{xu}\boldsymbol{ \Theta}_{i}'	\right| &\leq 
				\left|\widehat{\boldsymbol{\Theta}}_{i}\widehat{\boldsymbol{\Sigma}}_{xu} \widehat{\boldsymbol{\Theta}}_{i}'-\widehat{\boldsymbol{\Theta}}_{i} \boldsymbol{\Sigma}_{xu} \widehat{\boldsymbol{\Theta}}_{i}'\right|+\left|\widehat{\boldsymbol{\Theta}}_{i}\boldsymbol{\Sigma}_{xu} \widehat{\boldsymbol{\Theta}}_{i}'-\boldsymbol{\Theta}_{i}\boldsymbol{\Sigma}_{xu} \boldsymbol{\Theta}_{i}'\right| \nonumber \\
				&\leq 			\left|\widehat{\boldsymbol{\Theta}}_{i} \widehat{\boldsymbol{\Sigma}}_{xu} \widehat{\boldsymbol{\Theta}}_{i}'-\widehat{\boldsymbol{\Theta}}_{i} {\boldsymbol{\Sigma}}^{\star}_{xu} \boldsymbol{\widehat{\Theta}}_{i}'\right| + \left|\boldsymbol{\widehat{\Theta}}_{i} {\boldsymbol{\Sigma}}^{\star}_{xu} \boldsymbol{\widehat{\Theta}}_{i}'-\boldsymbol{\widehat{\Theta}}_{i}\boldsymbol{\Sigma}_{xu} \boldsymbol{\widehat{\Theta}}_{i}'\right|\nonumber\\
				&		\quad 	+\left|\boldsymbol{\widehat{\Theta}}_{i} \boldsymbol{\Sigma}_{xu} \boldsymbol{\widehat{\Theta}}_{i}'-\boldsymbol{\Theta}_{i} \boldsymbol{\Sigma}_{xu} \boldsymbol{\Theta}_{i}'\right| 
				= F_{1}+F_{2}+F_{3} \nonumber.
			\end{align}
			It is sufficient to show the following:
			\begin{align}
				F_{1}=o_{P}(1),\quad F_{2}=o_{P}(1), \quad F_{3}=o_{P}(1) \label{Fs}.
			\end{align}
			We consider each term, starting with $F_1$:
			\begin{align}
				F_{1}&=\left|\widehat{\boldsymbol{\Theta}}_{i}\widehat{\boldsymbol{\Sigma}}_{xu} \widehat{\boldsymbol{\Theta}}_{i}'-\widehat{\boldsymbol{\Theta}}_{i} \boldsymbol{\Sigma}_{xu} \widehat{\boldsymbol{\Theta}}_{i}'\right|
				\leq \left\Vert			\widehat{\boldsymbol{\Sigma}}_{xu} -{\boldsymbol{\Sigma}}^{\star}_{xu} \right\Vert_{\infty} \Vert \widehat{\boldsymbol{\Theta}}_{i}  \Vert^{2}_{1}. \label{proff1}
			\end{align}
			Using the definition $\widehat{u}_{t} = u_{t}+ \widetilde{\boldsymbol{x}}_{t}\left(	\boldsymbol{\beta} - \widehat{ \boldsymbol{\beta}}	\right)$,  we  analyse the term $ \left\Vert			\widehat{\boldsymbol{\Sigma}}_{xu} -{\boldsymbol{\Sigma}}^{\star}_{xu} \right\Vert_{\infty} $ as:
			\begin{align}
				\left\Vert	\widehat{\boldsymbol{\Sigma}}_{xu} -{\boldsymbol{\Sigma}}^{\star}_{xu}\right\Vert_{\infty}
				&\leq \left\Vert \frac{1}{T} \sum_{t=1}^{T}\left( \widetilde{\boldsymbol{x}}_{t} - {\boldsymbol{x}}^{\star}_{t}\right) u_{t}\right\Vert^{2}_{2} + \left\Vert \frac{1}{T} \sum_{t=1}^{T} \widetilde{\boldsymbol{x}}_{t} \left(\widehat{\boldsymbol\beta} - \boldsymbol{\beta}\right) \right\Vert_{2}^{2}\nonumber\\
				&+\left\Vert \frac{2}{T} \sum_{t=1}^{T} \widetilde{\boldsymbol{x}}_{t}u_{t} \right\Vert_{1}\left\Vert \widehat{\boldsymbol\beta} -\boldsymbol{\beta}\right\Vert_{1}\nonumber\\
				&\leq \left\Vert \frac{1}{T} \sum_{t=1}^{T}\left[\sum_{j=1}^{q}\left( \widehat{\phi}_{j}-{\phi}_{j}\right)\boldsymbol{x}_{t-j}\right] u_{t}\right\Vert^{2}_{2} + \left\Vert \frac{1}{T} \sum_{t=1}^{T} \widetilde{\boldsymbol{x}}_{t} \left(\widehat{\boldsymbol\beta} - \boldsymbol{\beta}\right) \right\Vert_{2}^{2}\nonumber\\
				&+\left\Vert \frac{2}{T} \sum_{t=1}^{T} \widetilde{\boldsymbol{x}}_{t}u_{t} \right\Vert_{1}\left\Vert \widehat{\boldsymbol\beta} -\boldsymbol{\beta}\right\Vert_{1}
				=  G_{1}+G_{2}+G_{3}\nonumber .
			\end{align}
			Notice that $ T^{-1}\sum_{t=1}^{T}\left[\sum_{j=1}^{q}\left( \widehat{\phi}_{j}-{\phi}_{j}\right)\boldsymbol{x}_{t-j}\right] u_{t}$ is bounded in probability by the same arguments used in $(I)$ of \eqref{expoall} and as a consequence  $G_{1}=o_{P}(1)$.  
			$G_{2}$ attains a non-asymptotic bound by direct application of Theorem \textcolor{red}{1}, hence $G_{2}=o_{P}(1)$.  The term $G_{3}$ is bounded in probability in Theorem \textcolor{red}{1}, hence  $G_{3} = o_{P}(1)$, therefore $ \left\Vert			\widehat{\boldsymbol{\Sigma}}_{xu} -{\boldsymbol{\Sigma}}^{\star}_{xu} \right\Vert_{\infty}=o_{p}(1)$. Then, $ \Vert \widehat{\boldsymbol{\Theta}}_{i}  \Vert^{2}_{1}= O_{P}(s_{i})$ by direct application of Lemma \ref{Theta}, which completes the analysis for $F_{1}$.
			We continue with $F_2$:
			\begin{align}
				F_{2} =\left|\boldsymbol{\widehat{\Theta}}_{i}{\boldsymbol{\Sigma}}^{\star}_{xu} \boldsymbol{\widehat{\Theta}}_{i}'-\boldsymbol{\widehat{\Theta}}_{i} \boldsymbol{\Sigma}_{xu} \boldsymbol{\widehat{\Theta}}_{i}'\right| \leq \left\Vert	{\boldsymbol{\Sigma}}^{\star}_{xu} -{\boldsymbol{\Sigma}}_{xu} \right\Vert_{\infty} \Vert \widehat{\boldsymbol{\Theta}}_{i}  \Vert^2_{1},
			\end{align}
			where $\left\Vert{\boldsymbol{\Sigma}}^{\star}_{xu} -{\boldsymbol{\Sigma}}_{xu} \right\Vert_{\infty}= \boldsymbol{ \Delta}^{\star}_{\boldsymbol{\Sigma}}$, 
			\begin{align}
				\boldsymbol{ \Delta}^{\star}_{\boldsymbol{\Sigma}}&\leq  \left| \frac{1}{T} \sum_{t=1}^{T}\left({\boldsymbol{x}}^{\star}_{t} u_{t}\right)'\left({\boldsymbol{x}}^{\star}_{t}u_{t}\right)-E\left[\frac{1}{T} \sum_{t=1}^{T}\left(\widetilde{\boldsymbol{x}}_{t} u_{t}\right)'\left(\widetilde{\boldsymbol{x}}_{t} u_{t}\right)\right]\right|\nonumber\\
				&= \left| \frac{1}{T} \sum_{t=1}^{T}\left( \left( \boldsymbol{{x}}_{t}- \sum_{j=1}^{q} \phi_{j}\boldsymbol{x}_{t-j}	\right) \left(u_{t}-\sum_{j=1}^{q} \phi_{j}u_{t-j} \right)\right)'\right.\nonumber\\
				&\quad \quad \quad \quad \left.\left( \left( \boldsymbol{{x}}_{t}- \sum_{j=1}^{q} \phi_{j}\boldsymbol{x}_{t-j}	\right) \left(u_{t}-\sum_{j=1}^{q} \phi_{j}u_{t-j} \right)\right)   \right.  \nonumber\\
				&\left.-E\left[\sum_{t=1}^{T} \left(\left( \boldsymbol{{x}}_{t}- \sum_{j=1}^{q} \phi_{j}\boldsymbol{x}_{t-j}	\right) \left(u_{t}-\sum_{j=1}^{q} \phi_{j}u_{t-j} \right)\right)'\right.\right. \nonumber\\
				&\quad \quad \quad \quad \left. \left. \left( \left( \boldsymbol{{x}}_{t}- \sum_{j=1}^{q} \phi_{j}\boldsymbol{x}_{t-j}	\right) \left(u_{t}-\sum_{j=1}^{q} \phi_{j}u_{t-j} \right)\right)\right] \right|\nonumber.
			\end{align}

			Set  $  \boldsymbol{ \xi}_t= \left( \boldsymbol{{x}}_{t}- \sum_{j=1}^{q} \phi_{j}\boldsymbol{x}_{t-j}	\right) \left(u_{t}-\sum_{j=1}^{q} \phi_{j}u_{t-j} \right)$, and its expectation $$E\left(\boldsymbol{\xi}_{t}\right) = E\left[ \left( \boldsymbol{{x}}_{t}- \sum_{j=1}^{q} \phi_{j}\boldsymbol{x}_{t-j}	\right) \left(u_{t}-\sum_{j=1}^{q} \phi_{j}u_{t-j}\right)\right].$$    By Theorem 14.1 of \cite{davidson1994stochastic}, and as a product of $\alpha$-mixing series, $\{ \boldsymbol{\xi}_t\}$, is also a $p$-dimensional  $\alpha$-mixing series  with similar properties as  \textcolor{red}{(5)}--\textcolor{red}{(6)}.   We then write
			\begin{align}
				\boldsymbol{ \Delta}^{\star}_{\boldsymbol{\Sigma}}&\leq \left\vert \frac{1}{T} \sum_{t=1}^{T}\boldsymbol{\xi}_{t}'\boldsymbol{\xi}_{t} -\sum_{t=1}^{T} E\left[\frac{1}{T}\sum_{t=1}^{T} \boldsymbol{\xi}_{t}'\right] E\left[	\frac{1}{T}\sum_{t=1}^{T} \	\boldsymbol{\xi}_{t}\right] \right\vert\nonumber\\
				&\leq \frac{1}{T}  \left \Vert  \sum_{t=1}^{T} \boldsymbol{\xi}_{t} - E\left(\boldsymbol{\xi}_{t}\right)    \right\Vert_{2}^{2}  +  \frac{2}{T}  \left\Vert E\left(\boldsymbol{\xi}_{t}\right)   \right\Vert^{2}_2\leq  \frac{1}{T}  \left \Vert  \sum_{t=1}^{T} \boldsymbol{\xi}_{t} - E\left(\boldsymbol{\xi}_{t}\right)    \right\Vert_{2}^{2}   + \frac{2}{T}  E \left\Vert \boldsymbol{\xi}_{t}\right\Vert_{2}^{2} \label{expectation} \\
				&			= J_{1}+J_{2}.\nonumber
			\end{align}
			Note that  the term $J_{1}$ includes  an $\alpha$-mixing series and can be  bounded following similar arguments as in Lemma A1 of  \cite{DGK}, $J_{2}$ is bounded by Assumption \textcolor{red}{1}, which implies that  second moments exist and are bounded.  Therefore,  $\left \Vert {\boldsymbol{\Sigma}}^{\star}_{xu} -{\boldsymbol{\Sigma}}_{xu}\right\Vert_{\infty} = o_{P}(1)$, while $\Vert \widehat{\boldsymbol{\Theta}}_{i}  \Vert^2_{1}=O_{P}({s_{i}})$ by direct application of Lemma \ref{Theta}, which  completes the analysis for $F_{2}$.
			
			We continue with $F_{3}$: 
			\begin{align}
				\left|\boldsymbol{\widehat{\Theta}}_{i}\boldsymbol{\Sigma}_{xu} \boldsymbol{\widehat{\Theta}}_{i}'-\boldsymbol{\Theta}_{i} \boldsymbol{\Sigma}_{xu} \boldsymbol{\Theta}_{i}'\right| \leq\left\|\boldsymbol{\Sigma}_{xu}\right\|_{\infty}\left\|\widehat{\boldsymbol{\Theta}}_{i}-\boldsymbol{\Theta}_{i}\right\|_{1}^{2}+2\left\|\boldsymbol{\Sigma}_{xu} \boldsymbol{\Theta}_{i}\right\|_{2}\left\|\widehat{\boldsymbol{\Theta}}_{i}-\boldsymbol{\Theta}_{i}\right\|_{2}, \nonumber
			\end{align}
			where, by Lemma \ref{Theta} we have that,  $$\left\|\widehat{\boldsymbol{\Theta}}_{i}-\boldsymbol{\Theta}_{i}\right\|_{1}=O_{P}\left(s_{i} \sqrt{T^{-1}\log{p} }\right), \; \text{and} \;  \left\|\widehat{\boldsymbol{\Theta}}_{i}-\boldsymbol{\Theta}_{i}\right\|_{2}=O_{P}\left(\sqrt{s_{i}T^{-1} \log{p} }\right).$$ 
			Note that  $\boldsymbol{\Sigma}_{xu}$ and $\boldsymbol{\Theta}$ are symmetric positive definite matrices, that satisfy the following properties:
			\begin{align}
				0<\lambda_{\min}(\boldsymbol{\Sigma})&\leq \lambda_{\max}\left(\boldsymbol{\Sigma}\right),\quad
				0<\lambda_{\min}(\boldsymbol{\Theta})\leq \lambda_{\max}\left(\boldsymbol{\Theta}\right)\nonumber,
			\end{align} 
			where $\lambda_{\min}(\boldsymbol{\Sigma}),  \lambda_{\max}(\boldsymbol{\Sigma})$ are the smallest and largest eigenvalues of $\boldsymbol{{\Sigma}}$ respectively and  $\lambda_{\min}(\boldsymbol{\Theta}), \lambda_{\max}(\boldsymbol{\Theta})$ are the smallest and largest eigenvalues of $\boldsymbol{{\Theta}}$ respectively.    Denote $E[u_{t}^{2}]:=\operatorname{Var}(u)=\sigma_{u}^{2}$ which is a scalar and $0<\sigma_{u}^{2}<\infty$.
			Then $\boldsymbol{\Sigma}_{xu}:=\boldsymbol{\Sigma }\sigma_{u}^{2}$.  Hence we obtain the following statements: 
			\begin{align}
				\left\|\boldsymbol{\Sigma}_{xu}\right\|_{\infty} \leq\left\|\boldsymbol{\Sigma}_{xu}\right\|_{2}&=\lambda_{\max }\left(\boldsymbol{\Sigma}_{xu}\right)=\sigma_{u}^{2} \lambda_{\max }(\boldsymbol{\Sigma})=O_{P}(1) ,\\
				\left\|\boldsymbol{\Sigma}_{xu} \boldsymbol{\Theta}_{i}\right\|_{\infty} &\leq\left\|\boldsymbol{\Sigma}_{xu} \right\|_{2}\left\|\boldsymbol{\Theta}_{i}\right\|_{\infty} \leq O_{P}(1)\|\Theta\|_{2}\nonumber \\
				&=O_{P}(1) \lambda_{\max }(\boldsymbol{\Theta})=\frac{O_{P}(1) }{ \lambda_{\min }(\boldsymbol{\Sigma})}=O_{P}(1),
			\end{align}
			which completes the proof.
		\end{proof}
		
		{\begin{lemma}\label{vcov}
				Under Assumptions \textcolor{red}{3},  \textcolor{red}{4}, the following holds:
				\begin{align}
					\nonumber 
					P\left(\max_{0<i,k\leq p}	\left|	(\widetilde{	\boldsymbol{X}}'\widetilde{	\boldsymbol{X}} )_{i,k}- E\left(	\widetilde{	\boldsymbol{X}}'\widetilde{	\boldsymbol{X}}\right)_{i,k}	\right|	>  \nu T \right) &\leq  pc_{0}\exp\left(	-c_{1}\left(\frac{\nu T^{1/2}}{3\delta_{1}\delta_{2}}	\right)^{2}	\right) \to 0
				\end{align}
				where  $\widehat{\boldsymbol{\Sigma}}= T^{-1}{\widetilde{\boldsymbol{X}}'\widetilde{	\boldsymbol{X}}}$, and  $\boldsymbol{{\Sigma}}= E({\boldsymbol{\widetilde{X}}}'{	\widetilde{\boldsymbol{X}}})$, for some large enough positive and finite constants,  $c_1,\nu>0$ and $\delta_{1}, \delta_{2}>0$ independent from $T$ and $p$.
		\end{lemma}}
		\begin{proof} [\unskip\nopunct] \textbf{Proof of Lemma \ref{vcov}}
			Notice that, for some $i,k\in[1,p]$
			\begin{align}
				\left\vert(\widetilde{	\boldsymbol{X}}'\widetilde{	\boldsymbol{X}} )_{i,k}- E\left(	\widetilde{	\boldsymbol{X}}'\widetilde{	\boldsymbol{X}}\right)_{i,k}	\right\vert
				&\leq 	\	\left|	({	\boldsymbol{X}}'\widehat{\boldsymbol{L}}'\widehat{\boldsymbol{L}}{	\boldsymbol{X}} )_{i,k}- ({	\boldsymbol{X}}'{\boldsymbol{L}}'{\boldsymbol{L}}{	\boldsymbol{X}} )_{i,k} \right\vert\nonumber\\
				&\quad  +\left\vert  ({	\boldsymbol{X}}'{\boldsymbol{L}}'{\boldsymbol{L}}{	\boldsymbol{X}} )_{i,k} -  E\left(	{	\boldsymbol{X}}'{L}'{{L}}{	\boldsymbol{X}}\right)_{i,k}	\right\vert\nonumber \\
				&\quad +\left|
				E\left(	{	\boldsymbol{X}}'{\boldsymbol{L}}'{\boldsymbol{L}}{	\boldsymbol{X}}\right)_{i,k} - E\left(	{	\boldsymbol{X}}'\widehat{\boldsymbol{L}}'\widehat{\boldsymbol{L}}{	\boldsymbol{X}}\right)_{i,k}	\right|	=A + B+C\nonumber.
			\end{align}
			By \eqref{Sigma}, $A=O_{P}(T^{-1/2})$, we analyse $B$:
			\begin{align}
				B &\leq \left|\boldsymbol{x}_{i}'\widehat{\boldsymbol{L}} -  E(\boldsymbol{x_{i}})'\widehat{\boldsymbol{L}}	\right|		  \left|\boldsymbol{x}_{k}'\widehat{\boldsymbol{L}} -  E(\boldsymbol{x}_{k})'\widehat{\boldsymbol{L}}	\right|
				+\left|E(\boldsymbol{x}_{i}'\boldsymbol{L}) \right|\left|\boldsymbol{x}_{k}'\widehat{\boldsymbol{L}} -E(\boldsymbol{x_{k}})'{\boldsymbol{L}} 	\right| \nonumber\\
				&\quad +\left| E(\boldsymbol{x}_{k}'\boldsymbol{L})\right| \left|\boldsymbol{x}_{i}'\widehat{\boldsymbol{L}} - E(\boldsymbol{x_{i}})'{\boldsymbol{L}} 	\right|\nonumber\\
				&= \left|  \left(\boldsymbol{x}_{i}'(\widehat{\boldsymbol{L}}- \boldsymbol{L}) - E\boldsymbol{x}_{i}\right)	+ \boldsymbol{x}_{i}'\boldsymbol{L}-E\boldsymbol{x}_{i}'\widehat{\boldsymbol{L}}				\right|
				\times \left| \left(\boldsymbol{x}_{k}'(\widehat{\boldsymbol{L}}- \boldsymbol{L})  - E\boldsymbol{x}_{k}\right)	+ \boldsymbol{x}_k'\boldsymbol{L}-E\boldsymbol{x}_{k}'\widehat{\boldsymbol{L}}			\right|\nonumber\\
				&\quad +\left|E(\boldsymbol{x}_{i}'\boldsymbol{L}) \right|\left| (\boldsymbol{x_{k}} - E\boldsymbol{x}_{k})' (\widehat{\boldsymbol{L}}- \boldsymbol{L})	+ \boldsymbol{x}_k'\boldsymbol{L}  -  E\boldsymbol{x}_{k}'\widehat{\boldsymbol{L}} \right| 
				\nonumber\\
				&\quad +\left|E(\boldsymbol{x}_{k}'\boldsymbol{L}) \right|\left|  (\boldsymbol{x_{i}} - E\boldsymbol{x}_{i})'(\widehat{\boldsymbol{L}}- \boldsymbol{L})	+ \boldsymbol{x}_{i}'\boldsymbol{L}  -  E\boldsymbol{x}_{i}'\widehat{\boldsymbol{L}} \right| \nonumber,
			\end{align}
			\noindent where $\boldsymbol{L}$ and  $\widehat{\boldsymbol{L}}$ are defined in \eqref{L}. 
			Under Assumption \textcolor{red}{3}, $\boldsymbol{x}_{i,k}$ is a thin tailed $\alpha$-mixing sequence with properties defined in \textcolor{red}{(6)} of the main paper, with $\max_{i,k}E|\boldsymbol{x}_{i,k}|\leq d<\infty$, for some $d>0$, we have 
			\begin{align}
				\left|	\widehat{\sigma}_{i,k} - \sigma_{ik}	\right| &\leq \left|  \left(\boldsymbol{x}_{i} - E\boldsymbol{x}_{i}\right)'(\widehat{\boldsymbol{L}}- \boldsymbol{L})	+ \boldsymbol{x}_{i}'\boldsymbol{L}-E\boldsymbol{x}_{i}'\widehat{\boldsymbol{L}}				\right|\nonumber\\
				&\quad \times \left|\left(\boldsymbol{x}_{k} - E\boldsymbol{x}_{k}\right)' (\widehat{\boldsymbol{L}}- \boldsymbol{L}) 	+ \boldsymbol{x}_{k}\boldsymbol{L}-\widehat{\boldsymbol{L}}E\boldsymbol{x}_{k}				\right|\nonumber\\
				&\quad+d\left|  (\boldsymbol{x_{k}} - E\boldsymbol{x}_{k})'(\widehat{\boldsymbol{L}}- \boldsymbol{L})	+ \boldsymbol{x}_k'\boldsymbol{L}  -  E\boldsymbol{x}_{k}'\widehat{\boldsymbol{L}} \right| \nonumber\\
				&\quad +d\left|(\boldsymbol{x_{i}} - E\boldsymbol{x}_{i})' (\widehat{\boldsymbol{L}}- \boldsymbol{L}) 	+ \boldsymbol{x}_{i}' \boldsymbol{L} -  E\boldsymbol{x}_{i}'\widehat{\boldsymbol{L}}\right| .\nonumber
			\end{align}
			By Assumption \textcolor{red}{3}, $\{\boldsymbol{x}_{t}\}$ is a   $p$-dimensional ergodic $\alpha$-mixing sequences, and consequently by Theorem 14.1 of \cite{davidson1994stochastic}, $\boldsymbol{x}_{i}'\boldsymbol{L}, \boldsymbol{x}_{k}'\boldsymbol{L} $ are $\alpha$-mixing series satisfying Assumption \textcolor{red}{3}.  Therefore, 
			\begin{align}
				P\left(\max_{0<i,k\leq p}\left|	\widehat{\sigma}_{i,k} - \sigma_{ik}	\right| >tT\right) &\leq\sum_{i,k}
				P \left(	\left|  \left(\boldsymbol{x}_{i} - E\boldsymbol{x}_{i}\right)'(\widehat{\boldsymbol{L}}- \boldsymbol{L})	+ \boldsymbol{x}_{i}'\boldsymbol{L}-E\boldsymbol{x}_{i}'\widehat{\boldsymbol{L}}			\right|\right.\nonumber\nonumber\\
				&\left.\quad  \times \left| \left(\boldsymbol{x}_{k} - E\boldsymbol{x}_{k}\right)' (\widehat{\boldsymbol{L}}- \boldsymbol{L}) 	+ \boldsymbol{x}_{k}\boldsymbol{L}-\widehat{\boldsymbol{L}}E\boldsymbol{x}_{k}			\right|>tT/3	\right)\nonumber\\
				&\quad +P\left(	d\left|   (\boldsymbol{x_{k}} - E\boldsymbol{x}_{k})'(\widehat{\boldsymbol{L}}- \boldsymbol{L})	+ \boldsymbol{x}_k'\boldsymbol{L}  -  E\boldsymbol{x}_{k}'\widehat{\boldsymbol{L}} \right| >tT/3	\right)\nonumber\\
				&\quad +P\left(d\left| (\boldsymbol{x_{i}} - E\boldsymbol{x}_{i})' (\widehat{\boldsymbol{L}}- \boldsymbol{L}) 	+ \boldsymbol{x}_{i}' \boldsymbol{L} -  E\boldsymbol{x}_{i}'\widehat{\boldsymbol{L}} \right| >tT/3\right). \nonumber
			\end{align}
			Notice that 
			\begin{align}
				& \left|  \left(\boldsymbol{x}_{i} - E\boldsymbol{x}_{i}\right)'\left(\widehat{\boldsymbol{L}}- \boldsymbol{L}\right)	+ \boldsymbol{x}_{i}'\boldsymbol{L}-E\boldsymbol{x}_{i}'\widehat{\boldsymbol{L}}				\right|\times \left|\left(\boldsymbol{x}_{k} - E\boldsymbol{x}_{k}\right)' \left(\widehat{\boldsymbol{L}}- \boldsymbol{L}\right)	+ \boldsymbol{x}_{k}\boldsymbol{L}-\widehat{\boldsymbol{L}}E\boldsymbol{x}_{k}				\right|\nonumber\\
				&\leq 	\left|  \left(\boldsymbol{x}_{i} - E\boldsymbol{x}_{i}\right)'\left(\widehat{\boldsymbol{L}}- \boldsymbol{L}\right)		\right|\times \left| \left(\boldsymbol{x}_{k} - E\boldsymbol{x}_{k}\right)'\left(\widehat{\boldsymbol{L}}- \boldsymbol{L}\right)			\right|\nonumber \\
				&=	\left| \left(\boldsymbol{x}_{i} - E\boldsymbol{x}_{i}\right)		\right|\times \left|  \left(\boldsymbol{x}_{k} - E\boldsymbol{x}_{k}\right)'				\right|\|	\widehat{\boldsymbol{L}}- \boldsymbol{L} 	\|^{2}_{2},\nonumber\\
				&d\left|(\boldsymbol{x_{k}} - E\boldsymbol{x}_{k})' \left(\widehat{\boldsymbol{L}}- \boldsymbol{L}\right) 	+ \boldsymbol{x}_k'\boldsymbol{L}  -E\boldsymbol{x}_{k}'  \widehat{\boldsymbol{L}} \right|\leq d\left|  (\boldsymbol{x_{k}} - E\boldsymbol{x}_{k})'\left(\widehat{\boldsymbol{L}}- \boldsymbol{L}\right)	\right| \nonumber\\
				&\text{and}\nonumber\\
				&d\left| (\boldsymbol{x_{i}} - E\boldsymbol{x}_{i})'\left(\widehat{\boldsymbol{L}}- \boldsymbol{L}\right)+ \boldsymbol{x}_{i}'\boldsymbol{L}  -  E\boldsymbol{x}_{i}'\widehat{\boldsymbol{L}} \right| \leq d\left|  (\boldsymbol{x_{i}} - E\boldsymbol{x}_{i})'\left(\widehat{\boldsymbol{L}}- \boldsymbol{L}\right)\right|.  \nonumber
			\end{align}
			Then we can write,
			{\small \begin{align}
					P\left(\max_{0<i,k\leq p}T^{-1}\left|	\widehat{\sigma}_{i,k} - \sigma_{i,k}	\right| >\nu \right) &\leq\sum_{i,k}
					P \left(T^{-1}\left| \left(\boldsymbol{x}_{i} - E\boldsymbol{x}_{i}\right)		\right|\times \left|  \left(\boldsymbol{x}_{k} - E\boldsymbol{x}_{k}\right)'	\right|\left \|	\widehat{\boldsymbol{L}}- \boldsymbol{L} \right	\|^{2}	>\frac{\nu }{3}		\right)\nonumber\\
					&\quad +\sum_{i,k}P\left(	dT^{-1}\left| \left(\boldsymbol{x_{k}} - E\boldsymbol{x}_{k}\right)'\left(\widehat{\boldsymbol{L}} - \boldsymbol{L}\right)	\right|  >\frac{\nu }{3}		\right)\nonumber\\
					&\quad +\sum_{i,k}P\left(dT^{-1}\left|  \left(\boldsymbol{x_{i}} - E\boldsymbol{x}_{i}\right)'\left(\widehat{\boldsymbol{L}} - \boldsymbol{L}\right) \right|>\frac{\nu }{3}	\right).\label{probXX}
			\end{align}}
			Since $\left(\boldsymbol{x}_{i} - E\boldsymbol{x}_{i}\right)$ is $\alpha$-mixing with properties similar to \textcolor{red}{(5)}, \textcolor{red}{(6)} and $\left(\widehat{\boldsymbol{L}} - \boldsymbol{L}\right)= O_{P}({T}^{-1/2})$ by direct application of Corollary \textcolor{red}{1}.    The three terms are bounded using Lemma A1 of the Online Supplement of \cite{DGK}.   To show that \eqref{probXX} holds, it suffices to show that 
			{\small \begin{align}\label{cov}
					\sum_{i,k}P \left(T^{-1}\left| \left(\boldsymbol{x}_{i} - E\boldsymbol{x}_{i}\right)		\right|\times \left|  \left(\boldsymbol{x}_{k} - E\boldsymbol{x}_{k}\right)				\right|  \left \|	\widehat{\boldsymbol{L}}- \boldsymbol{L} \right	\|^{2}>\frac{\nu }{3}		\right)
					&\leq  pc_{0}\exp\left(	-c_{1}\left(\frac{\nu T^{1/2}}{3\delta_{1}\delta_{2}}	\right)^{2}	\right) \to 0\\
					\sum_{k}P\left(	dT^{-1}\left| (\boldsymbol{x_{k}} - E\boldsymbol{x}_{k})'(\widehat{\boldsymbol{L}}- \boldsymbol{L}) \right|  >\frac{\nu }{3}	\right) & \leq pc_{0}\exp\left(	-c_{2}\left(\frac{\nu T^{1/2}}{3 d_1e_{1}}	\right)^{2}	\right) \to 0
					\label{127}\\
					\sum_{i}P\left(	dT^{-1}\left| (\boldsymbol{x_{i}} - E\boldsymbol{x}_{i})'(\widehat{\boldsymbol{L}}- \boldsymbol{L}) 	\right|  >\frac{\nu }{3}		\right)  &\leq pc_{0}\exp\left(	-c_{3}\left(\frac{\nu T^{1/2}}{3 d_2e_{2}}	\right)^{2}	\right) \to 0
					\label{2},
			\end{align}}
			for some finite constants $ c_{1}, \delta_{1},\delta_{2},   d_{1}, d_{2}, e_{1},e_{2}>0$ independent of $T,p$, where $c_{1}, \delta_1,\delta_2, d_{1}, d_{2}$ are large constants.  
			By Lemma A11 , Equation  B.61 of \cite{chudik2018one} we have that 
			\begin{align}
				&	\sum_{i,k}P \left(T^{-1}\left| \left(\boldsymbol{x}_{i} - E\boldsymbol{x}_{i}\right)		\right|\times \left|  \left(\boldsymbol{x}_{k} - E\boldsymbol{x}_{k}\right)				\right|  \times \left \|	\widehat{\boldsymbol{L}}- \boldsymbol{L} \right	\|^{2}>\frac{\nu }{3}		\right)\label{cov1}\\
				&	\leq\sum_{i}P \left(T^{-1}\left| \left(\boldsymbol{x}_{i} - E\boldsymbol{x}_{i}\right)		\right|>\nu /(3\delta_{1}\delta_{2}) 	\right) +\sum_{k}P \left(\left|  \left(\boldsymbol{x}_{k} - E\boldsymbol{x}_{k}\right)				\right|>\delta_{1} 	\right)\nonumber\\
				&\quad +P \left(\left \|\widehat{\boldsymbol{L}} - \boldsymbol{L} \right	\|^{2}>\delta_{2} 	\right)= 	S_{1}+S_{2}+S_{3},  \nonumber 
			\end{align}
			where, analysing $S_{1}$ we obtain 
			\begin{align}
				\sum_{i}P  \left(T^{-1/2}\left| \left(\boldsymbol{x}_{i} - E\boldsymbol{x}_{i}\right)		\right|>\frac{\nu T^{1/2}}{3\delta_{1}\delta_{2}}	\right)&\leq p c_{0}\left\{ \exp\left(	-c_{1}\left(\frac{\nu T^{1/2}}{3\delta_{1}\delta_{2}}\right)^{2}	\right)  \right.\nonumber\\
				&\quad\quad\quad  \left. + \exp\left(-c_{2}\left( \frac{\nu T^{1/2}}{3\delta_{1}\delta_{2}\log^{2}T}\right)^{{\boldsymbol{\gamma}}_{1}} 	\right)		\right\}  \nonumber\\
				&\leq  pc_{0}\exp\left(	-c_{1}\left(\frac{\nu T^{1/2}}{3\delta_{1}\delta_{2}}	\right)^{2}	\right), \label{firstTermcov}
			\end{align}
			for some $\nu>0$.  
			For $S_{2}$,   let $\delta_{1}\geq \nu/3$, then $S_{2}\leq pc_* \exp\left(-c_{2}9\log{p}T^{-2}\right)\leq S_{1}, $ for some $c_{*}>0$. For $S_{3}$, notice that $\boldsymbol{ \widehat{L} },\; \text{and}\; \boldsymbol{ L }$ are non-singular matrices defined in \eqref{L}, then 
			by  Corollary \ref{probabilisticRates} we have that 
			$$S_{3}\leq P \left(  \max_{s=1,\ldots,T} \left|\boldsymbol{\widehat{\ell}}_{s}-  \boldsymbol{\ell}_{s} \right|	^{2}>\delta_{2} 	\right) = o(1), \; \delta>0.$$  
			\eqref{1}--\eqref{2} can be bounded using similar arguments to \eqref{cov1}.    
			Finally, by Assumption \textcolor{red}{3} and following a similar analysis with term $B$,  by Corollary \ref{probabilisticRates} term $C=o(1),$ for some $\delta>0$.
			Then, for some    large enough finite constant  $\nu>0$
			\begin{align}
				\sum_{k}P\left(	dT^{-1}\left| (\boldsymbol{x_{k}} - E\boldsymbol{x}_{k})'(\widehat{\boldsymbol{L}}- \boldsymbol{L}) \right|  >\frac{\nu }{3}	\right) & \leq pc_{0}\exp\left(	-c_{2}\left(\frac{\nu T^{1/2}}{3 d_2e_{2}}	\right)^{2}	\right) \to 0
				\\
				\sum_{i}P\left(	dT^{-1}\left| (\boldsymbol{x_{i}} - E\boldsymbol{x}_{i})'(\widehat{\boldsymbol{L}}- \boldsymbol{L}) 	\right|  >\frac{\nu }{3}		\right)  &\leq pc_{0}\exp\left(	-c_{3}\left(\frac{\nu T^{1/2}}{3 d_3e_{3}}	\right)^{2}	\right) \to 0
			\end{align}
			Notice that \begin{align} pc_{0}\exp\left(	-c_{1}\left(\frac{\nu T^{1/2}}{3\delta_{1}\delta_{2}}	\right)^{2}	\right) &\geq pc_{0}\exp\left(	-c_{2}\left(\frac{\nu T^{1/2}}{3 d_2e_{2}}	\right)^{2}	\right) \nonumber\\
				& \approx  pc_{0}\exp\left(	-c_{3}\left(\frac{\nu T^{1/2}}{3 d_3e_{3}}	\right)^{2}	\right) \to 0, 
			\end{align}
			for some positive and finite constants $ \nu, \delta_{1}, \delta_{2}, d_2,d_3, e_2, e_{3}$, and  $T, p\to \infty$.  Therefore,  the dominant rate corresponds to \eqref{cov}, completing the proof.
		\end{proof}

		\section{Simulation Study Supplement}\label{simulations}
		In the main paper we consider the experimental design in \textcolor{red}{(46)}, studying the small sample properties of our methodology compared  to methods previously seen in the literature, e.g. \emph{Lasso, debiased Lasso} where the cardinality of the non-sparse set is $s_{0}=|S_0|=3$.   
		In this supplement we examine the case where the cardinality of  the active set is increased to $7$ non-sparse elements, i.e. $s_{0}=7$. Notice that the selection of the regularisation parameter corresponding to this simulation study, follows  Section  \textcolor{red}{5} of the main paper, with $k=10$.

		The first panel of Table \ref{rmse2} reports the ratio of the average root mean squared error (RMSE) of the \emph{Lasso} estimator  over the RMSE of the \emph{GLS Lasso}. The second panel of  Table \ref{rmse2}, reports the ratio of the average   RMSE of the \emph{debiased Lasso} estimator over 
		the RMSE of \emph{debiased \emph{GLS} Lasso}.   Entries larger than 1 indicate superiority of the competing model (\emph{GLS} \emph{Lasso}).   Highlighted are the entries corresponding to the RMSE of the   {\emph{GLS} Lasso} in Panel I and  the \emph{debiased \emph{GLS} Lasso} in Panel II.  
		Table \ref{Avg_cov_07} below reports average coverage rates, lengths of CI, and size-adjusted power of the \emph{debiased} estimators. 
		The results in Table \ref{rmse2} suggest similar patterns with Table \textcolor{red}{1} of the main paper.

		As the results in Table \textcolor{red}{2} of the main paper suggest,  when autocorrelation is present,  our method outperforms the \emph{debiased Lasso}, while in the cases of $\phi=0,$ i.e. $ u_t\sim \; i.i.d.$ both the \emph{debiased Lasso} and \emph{debiased GLS Lasso} report asymptotically similar results.  Furthermore, regarding all sample sizes, $T$, and $p=100$, the average coverage rate of the \emph{debiased Lasso} appears to approach 95\% for all $\phi>0$, outperforming the \emph{debiased Lasso} which covers the true parameter on average 20\% less than  our  method,  as the level of autocorrelation increases from $\phi=0.5$ to $\phi =0.8$.   The reported size ($1-\text{AvgCov}\; S_0^{c}$) for our method appears less than the nominal rate $5\%$ in the aforementioned cases.  This pattern  follows the cases where the number of covariates, $p$, is  significantly larger compared to the sample size, while when $p=100, \; T=500$ and  $\phi=0.9$, our method report size equal to the nominal rate, and better in performance than the \emph{debiased Lasso} which in part appears more correctly sized than in cases of $p>>T$.    Additionally, the \emph{debiased Lasso } performs poorly under larger $p,T$, which indicates that as the sample size and number of covariates diverge, under the cases of strong dependence, i.e. at least $\phi>0.5 $, the \emph{debiased Lasso}  undercovers the true parameters with a high frequency and is significantly more biased (see,  for example,   Panel II of Table \ref{rmse2}) than our method. 
		
		Finally,   in Table \ref{power_07} we also report size-adjusted power (refer to this term in  Section \textcolor{red}{6} of the main paper) for both the \emph{debiased} models under the same sparsity level, $s_0=7$.    The results here,  suggest that our method reports power closer to 95\% as  $p,T$ and $\phi$ increase, while $|\phi|<1$.  These results are in line with the results of  Table \textcolor{red}{2} of the main paper.   It is evident that the size-adjustment uncovers the underlying behaviour of the methods, given that  \emph{debiased Lasso} exhibits significant size distortions as  $p$ and $\phi$ increase as a pair and/or individually. 
		
		These findings, are in line with our theoretical results, while the behaviour of the \emph{debiased} estimators in higher degrees of sparsity are in line with the empirical results of \cite{van2014asymptotically}, suggesting that  inference with the \emph{debiased Lasso} has its limit when the problem is not sufficiently sparse. 
		\subsection{Simulations with $t$-distributed errors and/or covariates}\label{t_errors}
		In the main paper we consider the model in \textcolor{red}{(46)} with $\boldsymbol{x}_t\sim \rm{i.i.d.}\; \mathcal{N}(0,1)$, $ \varepsilon_{t}\sim  \rm{i.i.d.}\;  \mathcal{N}(0,1)$ and study the small sample properties of our methodology compared  to methods previously seen in the literature, e.g. \emph{Lasso, debiased Lasso}. Here we use simulations to examine the impact of a violation in Assumption \textcolor{red}{1} and consequently  Assumption \textcolor{red}{3}  on the empirical coverage, and length of the proposed estimator and \emph{Debiased Lasso}. 
		
		We consider the following three  data generating processes that involve $t$-distributed errors and/or covariates:
		\begin{enumerate}
			\item[DGP 1:] $	 y_t=\boldsymbol{x}_{t}'\boldsymbol{\beta} + u_t, \;
			u_t={\phi} u_{t-1 } + \varepsilon_{t},$ $\boldsymbol{x}_t\sim  \rm{i.i.d.}\;  \mathcal{N}(0,1)$ and $\varepsilon_{t}\sim t_{d}$, 
			\item[DGP 2:]  $y_t=	 \boldsymbol{x}_{t}'\boldsymbol{\beta} + u_t, \; 
			u_t={\phi} u_{t-1 } + \varepsilon_{t},$ $\boldsymbol{x}_t\sim t_{d}$ and $\varepsilon_{t}\sim t_d $,
			\item[DGP 3:] $	y_t= \boldsymbol{x}_{t}'\boldsymbol{\beta} + u_t, \;
			u_t={\phi} u_{t-1 } + \varepsilon_{t},$ $\boldsymbol{x}_t\sim t_{d}$ and $\varepsilon_{t}\sim  \rm{i.i.d.}\;  \mathcal{N}(0,1)$,
		\end{enumerate}
		where $t=1, \ldots, T$ and $d\in\{ 4, 8, 16 \}$. 	We report these results in Tables \ref{t_4}--\ref{set2_t_16}, where the reported quantities have been   described in detail in  Section \textcolor{red}{6} of the main paper.  
		
		In DGP 1, we consider the case where only $\varepsilon_{t}\sim t_{d}$ and $\boldsymbol{x}_t\sim  \rm{i.i.d.}\;  \mathcal{N}(0,1)$.  In this DGP we have a clear violation of Assumptions \textcolor{red}{1}, \textcolor{red}{3} when $d=4$, for $\varepsilon_{t}$.
		In this case, we observe some mild undercoverage of the true parameter, see Table \ref{t_4}, which decreases as $d$ increases, see Tables \ref{t_8}  and \ref{t_16}. Overall the results (see Tables \ref{t_8}--\ref{t_16}) are similar to the results we obtain in the main paper.
		
		In DGP 2  we consider the case of where both the covariates and the error term, $\boldsymbol{x}_{t}\sim t_{d}$, $\varepsilon_{t}\sim t_{d}$.  In this DGP we have a clear violation of Assumptions \textcolor{red}{1}, \textcolor{red}{3}, when $d=4$, which leads to massive undercoverage of the true parameter from all methods, see, for example, Table \ref{set1_t_4}.    Similar to the observations in DGP 1, as $d$ increases, see Tables \ref{set1_t_8}, \ref{set1_t_16} we move closer to our  theoretical  framework, hence coverage rates resemble the results reported  in Table \textcolor{red}{2}.  
		
		Finally, in DGP 3, for $d=4$, we observe a similar pattern to DGP 2 for the same case.  More specifically, in Table \ref{set2_t_4}, departing from the Assumption of thin-tailed distribution in  $\{\boldsymbol{	x}_t\}$, affects the frequency of covering the true parameter significantly more than assuming a heavy-tailed $\{ \varepsilon_{t} \}$,  for example in DGP 1.   Assuming heavy-tailed $\boldsymbol{x}_t$, changes the rate at which \textcolor{red}{(12)} goes to zero (see Lemma \ref{vcov}), hence the results in such case are not indicative, as they are outside of the scope of our framework.  In Tables \ref{set2_t_8}, \ref{set2_t_16} where $d=8$ and $d=16$ respectively, the average coverage rates resemble the results in Tables \ref{set1_t_8}, \ref{set1_t_16}, discussed above.

		In terms of RMSE,  for all DGPs, 
		\emph{GLS}-based methods outperforms \emph{Lasso}-based ones in the majority of the cases where $\phi >0$, similarly to the results of the main paper.  These results are available upon request.  
		
		\begin{table}[h!]
			
			\small{
				\caption{Average coverage rates, lengths of $CI$, size-adjusted power and size of \emph{debiased} estimates throughout 1000 replications of model \textcolor{red}{(46)}, $\boldsymbol{x}_t\sim  \rm{i.i.d.}\;  \mathcal{N}(0,1)$ and $\varepsilon_{t}\sim  \rm{i.i.d.}\;  \mathcal{N}(0,1)$, $s_0= 7.$}
				\label{Avg_cov_07}
				\resizebox{\linewidth}{!}{%
\begin{tabular}{lllllllllllllll}
						\toprule
						&                            & p/T & \multicolumn{4}{c}{100}       & \multicolumn{4}{c}{200}       & \multicolumn{4}{l}{500}       \\ \midrule 
						&
						
						&	$\phi$ &
						\multicolumn{1}{c}{0} &
						\multicolumn{1}{c}{0.5} &
						\multicolumn{1}{c}{0.8} &
						\multicolumn{1}{c}{0.9} &
						\multicolumn{1}{c}{0} &
						\multicolumn{1}{c}{0.5} &
						\multicolumn{1}{c}{0.8} &
						\multicolumn{1}{c}{0.9} &
						\multicolumn{1}{c}{0} &
						\multicolumn{1}{c}{0.5} &
						\multicolumn{1}{c}{0.8} &
						\multicolumn{1}{c}{0.9} \\ \midrule
						\emph{Debiased Lasso} & AvgCov $S_0$               & 100 & 0.896 & 0.864 & 0.779 & 0.716 & 0.933 & 0.903 & 0.807 & 0.722 & 0.959 & 0.936 & 0.841 & 0.729 \\
						& AvgCov $S_0^{c}$ &     & 0.955 & 0.934 & 0.860 & 0.802 & 0.964 & 0.941 & 0.858 & 0.770 & 0.977 & 0.957 & 0.864 & 0.758 \\
						& AvgLength                  &     & 0.423 & 0.426 & 0.447 & 0.482 & 0.321 & 0.323 & 0.333 & 0.352 & 0.221 & 0.221 & 0.224 & 0.230 \\
						& AvgLength$^{c}$  &     & 0.423 & 0.426 & 0.447 & 0.481 & 0.321 & 0.323 & 0.333 & 0.352 & 0.221 & 0.221 & 0.224 & 0.230 \\
						\emph{Debiased	GLS }  & AvgCov $S_0$               &     & 0.891 & 0.907 & 0.926 & 0.926 & 0.932 & 0.952 & 0.970 & 0.973 & 0.959 & 0.976 & 0.988 & 0.989 \\
						& AvgCov $S_0^{c}$ &     & 0.956 & 0.967 & 0.978 & 0.980 & 0.965 & 0.978 & 0.989 & 0.990 & 0.977 & 0.988 & 0.994 & 0.996 \\
\bottomrule
					\end{tabular}%
				}
				\label{power_07}
				\resizebox{\linewidth}{!}{%
					\begin{tabular}{llllllllllllll}
						\multicolumn{14}{c}{Size adjusted power}                                                                                           \\ \toprule 
						& p/T & \multicolumn{4}{c}{100}       & \multicolumn{4}{c}{200}       & \multicolumn{4}{c}{500}       \\ \midrule
						
						$\phi$ &
						&
						\multicolumn{1}{c}{0} &
						\multicolumn{1}{c}{0.5} &
						\multicolumn{1}{c}{0.8} &
						\multicolumn{1}{c}{0.9} &
						\multicolumn{1}{c}{0} &
						\multicolumn{1}{c}{0.5} &
						\multicolumn{1}{c}{0.8} &
						\multicolumn{1}{c}{0.9} &
						\multicolumn{1}{c}{0} &
						\multicolumn{1}{c}{0.5} &
						\multicolumn{1}{c}{0.8} &
						\multicolumn{1}{c}{0.9} \\ \midrule
						\emph{Debiased GLS}   & 100 & 0.734 & 0.751 & 0.766 & 0.770 & 0.837 & 0.837 & 0.476 & 0.854 & 0.891 & 0.905 & 0.918 & 0.921 \\
						\emph{Debiased Lasso} &     & 0.739 & 0.708 & 0.604 & 0.521 & 0.826 & 0.801 & 0.719 & 0.637 & 0.890 & 0.878 & 0.828 & 0.773 \\
						\emph{Debiased GLS }  & 200 & 0.746 & 0.753 & 0.753 & 0.755 & 0.845 & 0.858 & 0.873 & 0.876 & 0.905 & 0.909 & 0.917 & 0.922 \\
						\emph{Debiased Lasso} &     & 0.744 & 0.707 & 0.609 & 0.512 & 0.847 & 0.827 & 0.755 & 0.678 & 0.905 & 0.890 & 0.846 & 0.789 \\
						\emph{Debiased GLS  } & 500 & 0.744 & 0.742 & 0.728 & 0.725 & 0.843 & 0.855 & 0.868 & 0.866 & 0.905 & 0.909 & 0.917 & 0.922 \\
						\emph{Debiased Lasso} &
						&
						0.745 &
						0.711 &
						0.584 &
						0.483 &
						0.821 &
						0.819 &
						0.745 &
						0.689 &
						0.905 &
						0.890 &
						0.846 &
						0.797 \\ \bottomrule\\
					\end{tabular}%
			}}
		\end{table}
		\begin{table}[h!]
			\caption{Average relative RMSE throughout 1000 replications of model \textcolor{red}{(46)}, $\boldsymbol{x}_t\sim  \rm{i.i.d.}\;  \mathcal{N}(0,1)$ and $\varepsilon_{t}\sim  \rm{i.i.d.}\;  \mathcal{N}(0,1)$, $s_0= 7.$}
			\label{rmse2}
			\resizebox{\linewidth}{!}{%
				\begin{tabular}{llllllllllllll}
					\toprule
					& $p/T$ & \multicolumn{4}{c}{100}       & \multicolumn{4}{c}{200}       & \multicolumn{4}{c}{500}       \\ \midrule
					$\phi$         &       & 0     & 0.5   & 0.8   & 0.9   & 0     & 0.5   & 0.8   & 0.9   & 0     & 0.5   & 0.8   & 0.9   \\ \midrule
					\multicolumn{14}{l}{Panel I}                                                                                      \\ \midrule\midrule     
					\emph{Lasso}/GLS \emph{Lasso}            & 100   & 0.990 & 1.183 & 1.707 & 2.096 & 0.997 & 1.198 & 1.821 & 2.439 & 0.999 & 1.284 & 1.921 & 2.496 \\
					\rowcolor[HTML]{C0C0C0} 
					\emph{GLS Lasso}      &       & 0.058 & 0.055 & 0.051 & 0.050 & 0.041 & 0.037 & 0.033 & 0.033 & 0.025 & 0.023 & 0.021 & 0.020 \\ \midrule
					\emph{Lasso}/GLS \emph{Lasso}       & 200   & 0.996 & 1.201 & 1.749 & 2.212 & 0.997 & 1.203 & 1.901 & 2.500 & 1.000 & 1.251 & 1.762 & 2.489 \\
					\rowcolor[HTML]{C0C0C0} 
					\emph{GLS Lasso}      &       & 0.047 & 0.045 & 0.043 & 0.043 & 0.031 & 0.029 & 0.026 & 0.026 & 0.021 & 0.019 & 0.017 & 0.016 \\ \midrule
					\emph{Lasso}/GLS \emph{Lasso}       & 500   & 0.997 & 1.180 & 1.736 & 2.249 & 0.996 & 1.228 & 1.923 & 2.558 & 0.999 & 1.207 & 1.859 & 2.715 \\
\bottomrule
				\end{tabular}%
			}
		\end{table}

		\begin{table}[h!]
			\centering
			\caption{Average Coverage rates and confidence interval lengths throughout 1000  replications of DGP 1, $\boldsymbol{x}_t\sim  \rm{i.i.d.}\;  \mathcal{N}(0,1)$ and $\varepsilon_{t}\sim t_{d}$, $d=4$, $s_0=3$.}
			\resizebox{\linewidth}{!}{%
				\begin{tabular}{@{}llllllllllll@{}}
					\toprule
					&  & p/T & \multicolumn{1}{c}{100} &  &  & \multicolumn{1}{c}{200} &  &  & 500 &  &  \\ \midrule
					$\phi$ &  &  & \multicolumn{1}{c}{0} & \multicolumn{1}{c}{0.5} & \multicolumn{1}{c}{0.8} & \multicolumn{1}{c}{0} & \multicolumn{1}{c}{0.5} & \multicolumn{1}{c}{0.8} & \multicolumn{1}{c}{0} & \multicolumn{1}{c}{0.5} & \multicolumn{1}{c}{0.8} \\ \midrule
					\emph{		Debiased 	Lasso} & AvgCov $S_0$ & 50 & 0.686 & 0.605 & 0.663 & 0.579 & 0.688 & 0.568 & 0.739 & 0.787 & 0.647 \\
					& AvgCov $S_0^{c}$ &  & 0.724 & 0.651 & 0.717 & 0.612 & 0.702 & 0.576 & 0.753 & 0.802 & 0.671 \\
					& AvgLength &  & 0.343 & 0.400 & 0.243 & 0.271 & 0.156 & 0.166 & 0.153 & 0.152 & 0.158 \\
					& AvgLength$^{c}$ &  & 0.343 & 0.399 & 0.243 & 0.271 & 0.156 & 0.166 & 0.153 & 0.152 & 0.158 \\
					\emph{Debiased GLS}  & AvgCov $S_0$ &  & 0.730 & 0.776 & 0.767 & 0.812 & 0.782 & 0.827 & 0.738 & 0.866 & 0.905 \\
					& AvgCov $S_0^{c}$ &  & 0.798 & 0.843 & 0.804 & 0.850 & 0.805 & 0.854 & 0.753 & 0.887 & 0.923 \\
\end{tabular}%
			}
			
			\label{t_4}
		\end{table}

		\begin{table}[h!]
			\caption{Average Coverage rates and confidence interval lengths throughout 1000  replications of DGP 1, $\boldsymbol{x}_t\sim  \rm{i.i.d.}\;  \mathcal{N}(0,1)$ and $\varepsilon_{t}\sim t_{d}$, $d=8$, $s_0=3$. }
			\resizebox{\linewidth}{!}{%
				\begin{tabular}{@{}llllllllllll@{}}
					\toprule
					&  & p/T & \multicolumn{1}{c}{100} &  &  & \multicolumn{1}{c}{200} &  &  & 500 &  &  \\ \midrule
					$\phi$ &  &  & \multicolumn{1}{c}{0} & \multicolumn{1}{c}{0.5} & \multicolumn{1}{c}{0.8} & \multicolumn{1}{c}{0} & \multicolumn{1}{c}{0.5} & \multicolumn{1}{c}{0.8} & \multicolumn{1}{c}{0} & \multicolumn{1}{c}{0.5} & \multicolumn{1}{c}{0.8} \\ \midrule
					\emph{Debiased Lasso}& AvgCov$S_0$ & 50 & 0.731 & 0.645 & 0.750 & 0.636 & 0.756 & 0.618 & 0.739 & 0.787 & 0.647 \\
					& AvgCov $S_0^{c}$ &  & 0.782 & 0.691 & 0.778 & 0.669 & 0.778 & 0.644 & 0.753 & 0.802 & 0.671 \\
					& AvgLength &  & 0.326 & 0.363 & 0.234 & 0.254 & 0.153 & 0.160 & 0.153 & 0.152 & 0.158 \\
					& AvgLength$^{c}$ &  & 0.325 & 0.362 & 0.234 & 0.254 & 0.153 & 0.160 & 0.153 & 0.152 & 0.158 \\
					\emph{Debiased GLS}& AvgCov $S_0$ &  & 0.786 & 0.828 & 0.824 & 0.866 & 0.845 & 0.887 & 0.738 & 0.866 & 0.905 \\
					& AvgCov $S_0^{c}$ &  & 0.855 & 0.893 & 0.860 & 0.900 & 0.866 & 0.905 & 0.753 & 0.887 & 0.923 \\
\end{tabular}%
			}
			\label{t_8}
			
		\end{table}

		\begin{table}[h!]
			\centering 
			\caption{Average Coverage rates and confidence interval lengths throughout 1000 replications of DGP 1, $\boldsymbol{x}_t\sim  \rm{i.i.d.}\;  \mathcal{N}(0,1)$ and $\varepsilon_{t}\sim t_{d}$, $d={16}$, $s_0=3$.}
			\resizebox{\linewidth}{!}{%
				\begin{tabular}{@{}llllllllllll@{}}
					\toprule
					&  & p/T & \multicolumn{1}{c}{100} &  &  & \multicolumn{1}{c}{200} &  &  & 500 &  &  \\ \midrule
					$\phi$ &  &  & \multicolumn{1}{c}{0} & \multicolumn{1}{c}{0.5} & \multicolumn{1}{c}{0.8} & \multicolumn{1}{c}{0} & \multicolumn{1}{c}{0.5} & \multicolumn{1}{c}{0.8} & \multicolumn{1}{c}{0} & \multicolumn{1}{c}{0.5} & \multicolumn{1}{c}{0.8} \\ \midrule
					\emph{Debiased Lasso}& AvgCov$S_0$ & 50 & 0.742 & 0.659 & 0.765 & 0.653 & 0.787 & 0.647 & 0.739 & 0.787 & 0.647 \\
					& AvgCov $S_0^{c}$ &  & 0.803 & 0.711 & 0.803 & 0.690 & 0.802 & 0.671 & 0.753 & 0.802 & 0.671 \\
					& AvgLength &  & 0.320 & 0.351 & 0.231 & 0.248 & 0.152 & 0.158 & 0.153 & 0.152 & 0.158 \\
					& AvgLength$^{c}$ &  & 0.319 & 0.351 & 0.231 & 0.247 & 0.152 & 0.158 & 0.153 & 0.152 & 0.158 \\
					\emph{Debiased GLS}& AvgCov $S_0$ &  & 0.799 & 0.838 & 0.836 & 0.879 & 0.866 & 0.905 & 0.738 & 0.866 & 0.905 \\
					& AvgCov $S_0^{c}$ &  & 0.872 & 0.906 & 0.883 & 0.916 & 0.887 & 0.923 & 0.753 & 0.887 & 0.923 \\
\end{tabular}%
			}
			
			\label{t_16}
			
		\end{table}

		\begin{table}[h!]
			\caption{Average Coverage rates and confidence interval lengths throughout 1000 replications of DGP 2,  $\boldsymbol{x}_t\sim t_{d}$ and $\varepsilon_{t}\sim t_d $, $d={4}$, $s_0=3$.}
			\resizebox{\linewidth}{!}{%
				\begin{tabular}{@{}llllllllllll@{}}
					\toprule
					&                  & p/T & \multicolumn{1}{c}{100} &       &       & \multicolumn{1}{c}{200} &       &       & 500   &       &       \\ \midrule
					$\phi$ &
					&
					&
					\multicolumn{1}{c}{0} &
					\multicolumn{1}{c}{0.5} &
					\multicolumn{1}{c}{0.8} &
					\multicolumn{1}{c}{0} &
					\multicolumn{1}{c}{0.5} &
					\multicolumn{1}{c}{0.8} &
					\multicolumn{1}{c}{0} &
					\multicolumn{1}{c}{0.5} &
					\multicolumn{1}{c}{0.8} \\ \midrule
					\emph{Debiased Lasso}& AvgCov$S_0$     & 50  & 0.613                   & 0.599 & 0.572 & 0.441                   & 0.448 & 0.452 & 0.264 & 0.274 & 0.282 \\
					& AvgCov $S_0^{c}$ &     & 0.858                   & 0.826 & 0.734 & 0.867                   & 0.832 & 0.722 & 0.871 & 0.829 & 0.704 \\
					& AvgLength        &     & 0.464                   & 0.473 & 0.514 & 0.328                   & 0.332 & 0.350 & 0.215 & 0.216 & 0.222 \\
					& AvgLength$^{c}$  &     & 0.459                   & 0.468 & 0.508 & 0.327                   & 0.331 & 0.349 & 0.214 & 0.215 & 0.221 \\
					\emph{Debiased GLS}& AvgCov $S_0$     &     & 0.612                   & 0.644 & 0.667 & 0.450                   & 0.469 & 0.470 & 0.264 & 0.267 & 0.267 \\
					& AvgCov $S_0^{c}$ &     & 0.859                   & 0.887 & 0.915 & 0.868                   & 0.899 & 0.929 & 0.871 & 0.905 & 0.935 \\
\bottomrule 
			\end{tabular}}
			
			\label{set1_t_4}
		\end{table}
		
		\begin{table}[h!]
			\caption{Average Coverage rates and confidence interval lengths throughout 1000 replications of DGP 2,  $\boldsymbol{x}_t\sim t_{d}$ and $\varepsilon_{t}\sim t_d $, $d={8}$, $s_0=3$.}
			\resizebox{\linewidth}{!}{%
				\begin{tabular}{@{}llllllllllll@{}}
					\toprule
					&                  & p/T & \multicolumn{1}{c}{100} &       &       & \multicolumn{1}{c}{200} &       &       & 500   &       &       \\ \midrule
					$\phi$ &
					&
					&
					\multicolumn{1}{c}{0} &
					\multicolumn{1}{c}{0.5} &
					\multicolumn{1}{c}{0.8} &
					\multicolumn{1}{c}{0} &
					\multicolumn{1}{c}{0.5} &
					\multicolumn{1}{c}{0.8} &
					\multicolumn{1}{c}{0} &
					\multicolumn{1}{c}{0.5} &
					\multicolumn{1}{c}{0.8} \\ \midrule
					\emph{Debiased Lasso}& AvgCov$S_0$     & 50  & 0.771                   & 0.740 & 0.669 & 0.705                   & 0.680 & 0.611 & 0.563 & 0.546 & 0.502 \\
					& AvgCov $S_0^{c}$ &     & 0.853                   & 0.820 & 0.734 & 0.868                   & 0.829 & 0.721 & 0.871 & 0.828 & 0.700 \\
					& AvgLength        &     & 0.356                   & 0.362 & 0.395 & 0.264                   & 0.266 & 0.282 & 0.174 & 0.175 & 0.180 \\
					& AvgLength$^{c}$  &     & 0.355                   & 0.362 & 0.394 & 0.264                   & 0.266 & 0.282 & 0.174 & 0.175 & 0.180 \\
					\emph{Debiased GLS}& AvgCov $S_0$     &     & 0.770                   & 0.800 & 0.836 & 0.708                   & 0.732 & 0.759 & 0.562 & 0.593 & 0.608 \\
					& AvgCov $S_0^{c}$ &     & 0.855                   & 0.887 & 0.917 & 0.869                   & 0.901 & 0.931 & 0.871 & 0.905 & 0.935 \\
\end{tabular}
			}
			\label{set1_t_8}
		\end{table}
		
		\begin{table}[h!]
			\caption{Average Coverage rates and confidence interval lengths throughout 1000 replications of DGP 2,  $\boldsymbol{x}_t\sim t_{d}$ and $\varepsilon_{t}\sim t_d $, $d={16}$, $s_0=3$}
			\resizebox{\linewidth}{!}{%
				\begin{tabular}{@{}llllllllllll@{}}
					\toprule
					&                  & p/T & \multicolumn{1}{c}{100} &       &       & \multicolumn{1}{c}{200} &       &       & 500   &       &       \\ \midrule
					$\phi$ &
					&
					&
					\multicolumn{1}{c}{0} &
					\multicolumn{1}{c}{0.5} &
					\multicolumn{1}{c}{0.8} &
					\multicolumn{1}{c}{0} &
					\multicolumn{1}{c}{0.5} &
					\multicolumn{1}{c}{0.8} &
					\multicolumn{1}{c}{0} &
					\multicolumn{1}{c}{0.5} &
					\multicolumn{1}{c}{0.8} \\ \midrule
					\emph{Debiased Lasso}& AvgCov$S_0$     & 50  & 0.804                   & 0.770 & 0.679 & 0.800                   & 0.764 & 0.672 & 0.773 & 0.738 & 0.633 \\
					& AvgCov $S_0^{c}$ &     & 0.853                   & 0.819 & 0.729 & 0.867                   & 0.828 & 0.717 & 0.871 & 0.826 & 0.697 \\
					& AvgLength        &     & 0.328                   & 0.334 & 0.364 & 0.243                   & 0.246 & 0.261 & 0.161 & 0.162 & 0.167 \\
					& AvgLength$^{c}$  &     & 0.328                   & 0.334 & 0.363 & 0.243                   & 0.246 & 0.261 & 0.161 & 0.162 & 0.167 \\
					\emph{Debiased GLS}& AvgCov $S_0$     &     & 0.798                   & 0.829 & 0.867 & 0.801                   & 0.833 & 0.877 & 0.775 & 0.804 & 0.843 \\
					& AvgCov $S_0^{c}$ &     & 0.855                   & 0.887 & 0.917 & 0.868                   & 0.899 & 0.931 & 0.872 & 0.905 & 0.936 \\
\end{tabular}
			}
			\label{set1_t_16}
			
		\end{table}
		
		\begin{table}[h!]
			\caption{Average Coverage rates and confidence interval lengths throughout 1000 replications of DGP 3, $\boldsymbol{x}_t\sim t_{d}$ and $\varepsilon_{t}\sim  \rm{i.i.d.}\;  \mathcal{N}(0,1)$, $d=4$, $s_0=3.$}
			\resizebox{\linewidth}{!}{%
				\begin{tabular}{@{}llllllllllll@{}}
					\toprule
					&                  & p/T & \multicolumn{1}{c}{100} &       &       & \multicolumn{1}{c}{200} &       &       & 500   &       &       \\ \midrule
					$\phi$ &
					&
					&
					\multicolumn{1}{c}{0} &
					\multicolumn{1}{c}{0.5} &
					\multicolumn{1}{c}{0.8} &
					\multicolumn{1}{c}{0} &
					\multicolumn{1}{c}{0.5} &
					\multicolumn{1}{c}{0.8} &
					\multicolumn{1}{c}{0} &
					\multicolumn{1}{c}{0.5} &
					\multicolumn{1}{c}{0.8} \\ \midrule
					\emph{Debiased Lasso}& AvgCov$S_0$     & 50  & 0.623                   & 0.616 & 0.590 & 0.452                   & 0.448 & 0.450 & 0.257 & 0.270 & 0.280 \\
					& AvgCov $S_0^{c}$ &     & 0.931                   & 0.907 & 0.833 & 0.937                   & 0.911 & 0.827 & 0.943 & 0.920 & 0.824 \\
					& AvgLength        &     & 0.437                   & 0.441 & 0.459 & 0.320                   & 0.322 & 0.330 & 0.213 & 0.213 & 0.216 \\
					& AvgLength$^{c}$  &     & 0.435                   & 0.438 & 0.457 & 0.319                   & 0.321 & 0.329 & 0.213 & 0.213 & 0.215 \\
					\emph{Debiased GLS}& AvgCov $S_0$     &     & 0.628                   & 0.638 & 0.665 & 0.459                   & 0.456 & 0.465 & 0.257 & 0.261 & 0.265 \\
					& AvgCov $S_0^{c}$ &     & 0.931                   & 0.948 & 0.964 & 0.937                   & 0.956 & 0.971 & 0.944 & 0.960 & 0.974 \\
\end{tabular}
			}
			\label{set2_t_4}
		\end{table}

		\begin{table}[h!]
			\caption{Average Coverage rates and confidence interval lengths throughout 1000 replications of DGP 3, $\boldsymbol{x}_t\sim t_{d}$ and $\varepsilon_{t}\sim  \rm{i.i.d.}\;  \mathcal{N}(0,1)$, $d=8$, $s_0=3.$}
			\resizebox{\linewidth}{!}{%
				\begin{tabular}{@{}llllllllllll@{}}
					\toprule
					&                  & p/T & \multicolumn{1}{c}{100} &       &       & \multicolumn{1}{c}{200} &       &       & 500   &       &       \\ \midrule
					$\phi$ &
					&
					&
					\multicolumn{1}{c}{0} &
					\multicolumn{1}{c}{0.5} &
					\multicolumn{1}{c}{0.8} &
					\multicolumn{1}{c}{0} &
					\multicolumn{1}{c}{0.5} &
					\multicolumn{1}{c}{0.8} &
					\multicolumn{1}{c}{0} &
					\multicolumn{1}{c}{0.5} &
					\multicolumn{1}{c}{0.8} \\ \midrule
					\emph{Debiased Lasso}& AvgCov$S_0$     & 50  & 0.786                   & 0.759 & 0.684 & 0.736                   & 0.708 & 0.636 & 0.582 & 0.570 & 0.524 \\
					& AvgCov $S_0^{c}$ &     & 0.891                   & 0.862 & 0.776 & 0.899                   & 0.866 & 0.760 & 0.908 & 0.873 & 0.751 \\
					& AvgLength        &     & 0.355                   & 0.360 & 0.384 & 0.260                   & 0.262 & 0.274 & 0.173 & 0.174 & 0.178 \\
					& AvgLength$^{c}$  &     & 0.355                   & 0.360 & 0.384 & 0.259                   & 0.262 & 0.273 & 0.173 & 0.174 & 0.177 \\
					\emph{Debiased GLS}& AvgCov $S_0$     &     & 0.787                   & 0.815 & 0.856 & 0.741                   & 0.757 & 0.786 & 0.585 & 0.600 & 0.621 \\
					& AvgCov $S_0^{c}$ &     & 0.894                   & 0.918 & 0.942 & 0.900                   & 0.927 & 0.952 & 0.909 & 0.935 & 0.956 \\
\end{tabular}
			}
			\label{set2_t_8}
		\end{table}

		\begin{table}[h!]
			
			\caption{Average Coverage rates and confidence interval lengths throughout 1000 replications of DGP 3, $\boldsymbol{x}_t\sim t_{d}$ and $\varepsilon_{t}\sim  \rm{i.i.d.}\;  \mathcal{N}(0,1)$, $d=16$, $s_0=3.$}
			\resizebox{\linewidth}{!}{%
				\begin{tabular}{@{}llllllllllll@{}}
					\toprule
					&                  & p/T & \multicolumn{1}{c}{100} &       &       & \multicolumn{1}{c}{200} &       &       & 500   &       &       \\ \midrule
					$\phi$ &
					&
					&
					\multicolumn{1}{c}{0} &
					\multicolumn{1}{c}{0.5} &
					\multicolumn{1}{c}{0.8} &
					\multicolumn{1}{c}{0} &
					\multicolumn{1}{c}{0.5} &
					\multicolumn{1}{c}{0.8} &
					\multicolumn{1}{c}{0} &
					\multicolumn{1}{c}{0.5} &
					\multicolumn{1}{c}{0.8} \\ \midrule
					\emph{Debiased Lasso}& AvgCov$S_0$     & 50  & 0.821                   & 0.789 & 0.695 & 0.828                   & 0.787 & 0.694 & 0.792 & 0.758 & 0.632 \\
					& AvgCov $S_0^{c}$ &     & 0.873                   & 0.842 & 0.753 & 0.880                   & 0.844 & 0.735 & 0.890 & 0.851 & 0.724 \\
					& AvgLength        &     & 0.331                   & 0.336 & 0.363 & 0.242                   & 0.245 & 0.257 & 0.161 & 0.162 & 0.165 \\
					& AvgLength$^{c}$  &     & 0.330                   & 0.336 & 0.362 & 0.242                   & 0.244 & 0.257 & 0.161 & 0.161 & 0.165 \\
					\emph{Debiased GLS}& AvgCov $S_0$     &     & 0.815                   & 0.846 & 0.883 & 0.827                   & 0.869 & 0.904 & 0.794 & 0.816 & 0.846 \\
					& AvgCov $S_0^{c}$ &     & 0.875                   & 0.903 & 0.931 & 0.881                   & 0.911 & 0.941 & 0.891 & 0.921 & 0.947 \\
\end{tabular}
			}
			\label{set2_t_16}
		\end{table}
		
	\end{document}